\documentclass[12pt]{article}

\usepackage[tmargin=1in,body={6in, 8.5in},left=1.2in,right=1.2in]{geometry} 
\usepackage[labelfont={bf,small},textfont={small},aboveskip=15pt,belowskip=12pt]{caption}
\usepackage[small]{titlesec}
\usepackage{cite}

\usepackage{amsmath,amssymb}
\usepackage{epsfig}

\newcommand{\Dslash}{\!\not\!\! D}
\newcommand{\dslash}{\!\not\! \partial}
\newcommand{\eps}{\epsilon}
\newcommand{\pslash}{\!\not\! p}

\newcommand{\refcite}{\citen}

\begin{document}

\thispagestyle{empty}

\vspace*{1cm}
\begin{center}
\Large The Higgs as a Composite Nambu-Goldstone Boson
\end{center}

\vspace{1cm}
\begin{center}
{\large R. Contino}
\vspace{0.5cm}

{Dipartimento di Fisica, Universit\`a La Sapienza,\\
and INFN, Roma, Italy\\[0.2cm]
roberto.contino@roma1.infn.it
}
\end{center}

\vspace{1cm}
\begin{abstract}
This is an introduction to theories where the Higgs is a composite Nambu-Goldstone boson
of a new strongly-interacting dynamics not much above the weak scale.  
A general discussion is presented  based on the pattern of global symmetries at low energy, and
the analogy with the QCD pion is analyzed.  The last part of the lectures shows how a composite Higgs can emerge
as the hologram of a 5-dimensional gauge field.
\end{abstract}

\vspace{3.5cm}
\begin{center}
Lectures presented at the 2009 TASI Summer School, \\ 
`Physics of the Large and the Small' \\
University of Colorado, Boulder,  June 1-29, 2009
\end{center}

\newpage 

\tableofcontents

\newpage

\section{Introduction}

More than a century of experimental results and theoretical progress
has led us to the formulation of an extremely elegant and compact  theory of the fundamental interactions among
particles.  Its success in reproducing a huge amount
of experimental data, spanning several orders of magnitude in energy, is impressive.
Despite their profoundly different manifestations on macroscopic scales, the electromagnetic, weak and
strong forces are all described within the same mathematical framework of gauge theories.
The electromagnetic and weak interactions are associated to the same $SU(2)_L\times U(1)_Y$ gauge invariance 
at short distances, although only electromagnetism is experienced as a long-range force.
The rest of  the electroweak symmetry is hidden at large distances or low energies, 
\textit{i.e.} it is spontaneously broken by the vacuum.
As a matter of fact,  despite the abundance of experimental information,  we do not know much about
the dynamics responsible for such spontaneous breaking.

The  Standard Model (SM) of particle physics gives an extremely economical formulation of the electroweak
symmetry breaking (EWSB) in terms of only one new fundamental degree of freedom: the Higgs boson.
It does not explain, however, the dynamical origin of the symmetry breaking, nor why the Higgs boson should be light, as
required to comply with the LEP data.  
An old and still attractive idea is that the Higgs boson might be a bound state
of a new strongly-interacting dynamics not much above the weak scale.
Its being composite would solve the SM hierarchy problem, as quantum corrections to its mass are now saturated 
at the compositeness scale.  
Significant theoretical progress on the construction of these theories has recently come from the intriguing connection
between gravity in higher-dimensional curved spacetimes and strongly-coupled gauge theories.
Fully realistic models have been built where some longstanding problems of the original constructions are solved.

These lectures are aimed to give an introduction to composite Higgs theories. We start in Section~\ref{sec:2paradigms} 
by recalling why new dynamics is expected to emerge at the TeV scale in the electroweak  sector.  
We then present the two paradigms for such new dynamics: the weakly-interacting Higgs model,
and strongly-coupled Technicolor theories.
The idea of the Higgs as a composite pseudo Nambu-Goldstone (pNG) boson is introduced in Section~\ref{sec:CHM},
as an interpolation between these two scenarios.  We illustrate one explicit example of symmetry breaking pattern, $SO(5)\to SO(4)$,
and make use of symmetry arguments to derive the expression of the Higgs potential in terms of form factors.
The same approach is then followed to describe the electromagnetic potential of the  pion in QCD,
and the analogy with the composite Higgs is analyzed in detail.
The constraints from electroweak precision tests and  from flavor-changing neutral currents (FCNC) are then discussed at length,
and the concept of partial compositeness introduced.
Section~\ref{sec:holoHiggs}  shows how a composite pNG boson can emerge as the fifth component of
a gauge field living in  a 5-dimensional spacetime.   The basic features of this kind of theories are illustrated by means 
of a simple abelian model in a flat extra dimension. We discuss how  the bulk of the fifth dimension gives
a model of the 4-dimensional strong dynamics
which is perturbative and thus calculable. As an important application we compute the form factors that parametrize the  couplings
of the composite Higgs and obtain an analytic expression for its potential.
We conclude with a few  words on the phenomenology of composite Higgs models.

In selecting the above topics I had necessarily to omit some other important ones, as for example
warped extra dimensional models and holography in curved spacetimes, and  Little Higgs theories.
Fortunately excellent reviews exist on these subjects, such as for example 
the Les Houches lectures by T. Gherghetta
 on holography~[\refcite{Gherghetta:LesHouches2005}] and the review 
by M. Schmaltz and D. Tucker-Smith on Little Higgs models~[\refcite{Schmaltz:LHreview}].
The  lectures by R. Sundrum~[\refcite{Sundrum:TASI2004}]  at TASI 2004 
and the review [\refcite{Serone:2009kf}] by M. Serone
partly overlap with Section~\ref{sec:holoHiggs} and contain
interesting complementary topics and discussions.
General introductions to flat and warped extra dimensions are given for example in the 
parallel TASI lectures  by H.~C.~Cheng~[\refcite{Cheng:TASI2009}] and T. Gherghetta, the
TASI lectures by C. Csaki~[\refcite{Csaki:TASI2002}], G. Kribs~[\refcite{Kribs:TASI2004}], 
and the Cargese lectures by R. Rattazzi~[\refcite{Rattazzi:Cargese2003}].
Extra dimensional models as theories of electroweak symmetry breaking are for example 
discussed in the TASI lectures by C. Csaki, J~Hubisz and P.~Meade~[\refcite{Csaki:TASI2004}].
More detailed references are given throughout the text. They are meant to introduce the reader to the vast literature
on the subject and form a necessarily incomplete and partial list.
I apologize in advance for  the omissions.

%%%%%
%% Section 1

\section{Two paradigms for Electroweak Symmetry Breaking}
\label{sec:2paradigms}

\begin{figure}[t]
\centering
\includegraphics[height=0.3\linewidth]{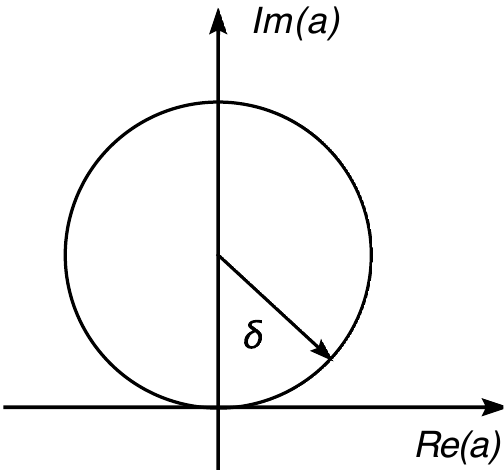} 
\caption{\label{fig:argand}
Unitary circle:  at energies below (above) the inelastic threshold the amplitude $a_l$ is  constrained to  lie on (inside) the circle. 
}
\end{figure}

The vast amount of  data collected so far in high-energy experiments
can be explained and compactly summarized by the Lagrangian 
\begin{equation}
\label{eq:sofar}
\begin{split}
& {\cal L} =  {\cal L}_0 + {\cal L}_{mass} \\[0.3cm]
& {\cal L}_0 =  -\frac{1}{4} W^a_{\mu\nu} W^{a\, \mu\nu} -\frac{1}{4} B_{\mu\nu} B^{\mu\nu} - \frac{1}{4} G_{\mu\nu} G^{\mu\nu} 
 + \sum_{j=1}^3 \,  \bar\Psi^{(j)} i\Dslash \Psi^{(j)} \\[0.1cm]
 & {\cal L}_{mass} = M_W^2\, W^+_{\mu} W^{-\, \mu} + \frac{1}{2} M_Z^2\, Z^\mu Z_\mu \\  
 & \qquad\qquad - \sum_{i,j} \Big( \bar u^{(i)}_L M^u_{ij} u_R^{(j)} + \bar d^{(i)}_L M^d_{ij} d_R^{(j)}  
     + \bar e^{(i)}_L M^e_{ij} e_R^{(j)} + \bar \nu^{(i)}_L M^\nu_{ij} \nu_R^{(j)} \Big) \\
 & \qquad\qquad + h.c. \, ,
\end{split}
\end{equation}
where $\Psi =\{ q^i_L , u^i_R , d^i_R , l^i_L , e^i_R , \nu^i_R \}$ is a collective index for the Standard Model fermions
and $i,j$ are generation indices. A remarkable property of  ${\cal L}$
is that while all the fundamental interactions among the particles (determined by ${\cal L}_0$) are symmetric under  local
$SU(2)_L\times U(1)_Y$ transformations, the observed mass spectrum (determined by ${\cal  L}_{mass}$) is not.
In other words,  the electroweak symmetry is hidden,  \textit{i.e.} spontaneously broken by the vacuum.
Although successful at the energies explored so far, the above mathematical formulation leads to an 
inconsistency if extrapolated to arbitrarily high energies: when used in a perturbative expansion,
it predicts scattering amplitudes that grow with the energy and violate the unitarity bound. The latter
prescribes that the elastic scattering amplitude $a_l$ of each $l$-th partial wave must satisfy
\begin{equation}
\text{Im}(a_l) = |a_l|^2 + |a_l^{in}|^2 \, ,
\end{equation}
where $a_l^{in}$ is the inelastic scattering amplitude.
This means that at energies below the inelastic threshold $a_l$ is  constrained to  lie on  the unitary circle 
$\text{Re}^2(a_l) + (\text{Im}(a_l)-1/2)^2=1/4$, while at higher energies it is bounded to be inside it, see Fig.~\ref{fig:argand}. 
%
%%%%%%%%%%%%%%%%%%%%%%%%%%%%%%%%%%%%%%%%%%%
%%%%%%%%%%%%%%%%%%%%%%%%%%%%%%%%%%%%%%%%%%%
%
Since at tree level the amplitude is real and an imaginary part only arises at the 1-loop level, perturbativity is
lost when the imaginary and real part are of the same order, that is when the scattering phase is large, $\delta\approx \pi$.

It turns out that the violation of perturbative unitarity occurs in processes that involve longitudinally polarized 
vector bosons as external states. 
For example, at tree level the amplitude for the elastic scattering of two longitudinally polarized $W$'s 
grows as $E^2$ at energies $E\gg m_W$:
\\[0.1cm]
\begin{center}
\includegraphics[height=19mm]{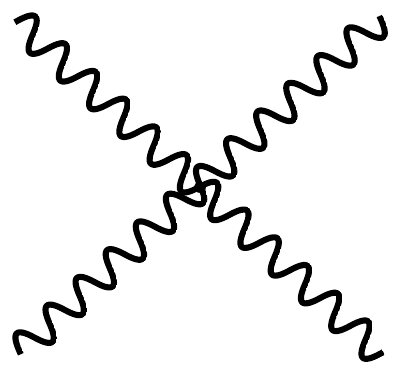} 
\hspace*{1.2cm}
\includegraphics[height=19mm]{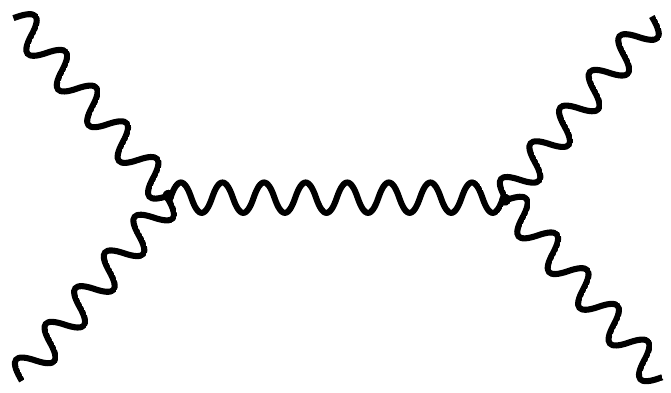} 
\hspace*{1.2cm}
 \begin{minipage}{0.15\linewidth}
   \vspace*{-1.7cm}
 \includegraphics[height=28mm]{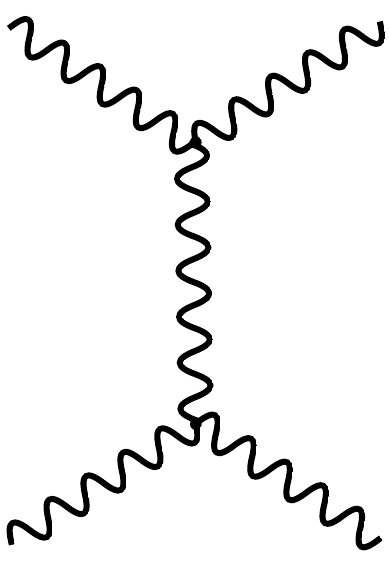} 
 \end{minipage} 
\end{center}
\begin{equation}
{\cal A}(W_L^+W_L^- \to W_L^+ W_L^-) \simeq \frac{g^2}{4 m_W^2} \left( s+t \right)\, .
\end{equation}
Here $s$, $t$ are the Mandelstam kinematic variables, and terms subleading in $m_W/E$ have been dropped.
Each longitudinal polarization brings one factor of $E$, since at large energies $\epsilon_L^\mu(p) = p^\mu/m_W + O(m_W/E)$,
so that each diagram naively grows as $E^4$.   When all the diagrams are summed, however, the 
leading $E^4$ term cancels out, and the amplitude grows as $E^2$.   
We will see shortly  that this cancellation can be easily understood
by performing the calculation in a renormalizable gauge.
By projecting on partial wave amplitudes,
\begin{equation}
a_l = \frac{1}{32\pi} \int_{-1}^{+1} \!\!d\cos\theta\;  {\cal A}(s,\theta) P_l(\cos\theta), \, 
\end{equation}
where $P_l(x)$ are the Legendre polynomials ($P_0(x) = 1$, $P_1(x) = x$, $P_2(x) = 3x^2/2 -1/2$, etc.), 
one finds the following expression for the $s$-wave amplitude ($l=0$):
\begin{equation}
a_0(W_L^+W_L^- \to W_L^+ W_L^-) \simeq \frac{1}{32\pi} \, \frac{s}{v^2}\, .
\end{equation}
The loss of perturbative unitarity in the $s$-wave scattering thus occurs for~\footnote{A slightly stronger bound, 
$\sqrt{s}\lesssim 2\sqrt{2} \pi v = 2.2\,$TeV, 
is obtained by  including the effect of the channel $W^+W^-\to ZZ$, see Ref.~[\refcite{Lee:1977eg}].
Notice that sometimes the bound $\text{Re}(a_l) \leq 1/2$ or $|a_l| \leq 1$ is imposed, instead of $\delta \leq \pi$. 
All are in fact acceptable as an estimates  of the energy where perturbative unitarity is lost. The difference in the values of the cutoff
$\Lambda$ thus obtained can be interpreted as the theoretical uncertainty of the estimate.
}
\begin{equation}
\pi \approx \delta \simeq 2 \text{Re}(a_0)\, , \qquad \text{\textit{i.e.} for:}  \qquad
\sqrt{s} \approx \Lambda = 4 \pi v \simeq 3 \,\text{TeV}\, .
\end{equation}

The role of the longitudinally polarized vector bosons
suggests that the inconsistency of the Lagrangian (\ref{eq:sofar})  is in the sector that breaks spontaneously the electroweak
symmetry and gives mass to the vector bosons. The connection can be made  explicit 
by introducing, as propagating degrees of freedom,  the Nambu-Goldstone bosons $\chi^a$ that 
correspond to the longitudinal polarizations of the $W$ and $Z$ bosons:
\begin{equation}
\Sigma(x) = \exp(i\sigma^a \chi^a(x)/v),  \qquad D_\mu \Sigma = \partial_\mu \Sigma 
 -i g \frac{\sigma^a}{2} W^a_\mu \Sigma + i g^\prime \Sigma \frac{\sigma^3}{2} B_\mu \, .
\end{equation}
In terms of the chiral field $\Sigma$, the mass terms can be rewritten as follows:~\footnote{For simplicity, 
from here on I will omit the lepton terms and concentrate on the quark sector.}
\begin{equation}
\label{eq:chiralLmass}
 {\cal L}_{mass} = \frac{v^2}{4} \text{Tr}\left[ \left( D_\mu \Sigma \right)^\dagger
 \left( D^\mu \Sigma \right) \right] - \frac{v}{\sqrt{2}} \sum_{i,j} \left( \bar u_L^{(i)} d_L^{(i)} \right) 
 \Sigma \begin{pmatrix} \lambda_{ij}^u\,  u_R^{(j)} \\[0.1cm] \lambda_{ij}^d\,  d_R^{(j)} \end{pmatrix} + h.c.
\end{equation}
The local $SU(2)_L\times U(1)_Y$ invariance is now manifest, since  $\Sigma$ transforms~as
\begin{equation}
\begin{gathered}
\Sigma \to U_L(x) \, \Sigma\, U_Y^\dagger(x)\, ,  \\[0.3cm]
  U_L(x) = \exp \big(i\alpha_L^a(x) \sigma^a/2 \big)  \qquad
  U_Y(x) = \exp \big(i\alpha_Y(x) \sigma^3/2 \big)\, ,
\end{gathered} 
\end{equation}
although it is non-linearly realized on the $\chi^a$ fields:
\begin{equation}
\chi^a(x) \to \chi^a(x) + \frac{v}{2}\, \alpha_L^a(x) - \frac{v}{2}\,\delta^{a3}\, \alpha_Y(x) \, .
\end{equation}

In the unitary gauge, $\langle \Sigma \rangle = 1$, 
the chiral Lagrangian (\ref{eq:chiralLmass}) reproduces
the mass term of eq.(\ref{eq:sofar}) with
\begin{equation}
\label{eq:rho}
\rho \equiv \frac{M_W^2}{M_Z^2 \cos^2\theta_W} = 1 \, .
\end{equation}
This relation is consistent with the experimentally measured value to quite good accuracy.
It follows as the consequence of a larger approximate invariance of (\ref{eq:chiralLmass})
under  $SU(2)_L\times SU(2)_R$ global transformations,
\begin{equation}
\Sigma \to U_L \, \Sigma\, U_R^\dagger \, ,
\end{equation}
which is spontaneously broken to the diagonal subgroup  $SU(2)_c$ by $\langle \Sigma \rangle = 1$, and
explicitly broken by $g^\prime\not =0$ and $\lambda^u_{ij} \not = \lambda^d_{ij}$.
In the limit of vanishing $g^\prime$ the fields $\chi^a$ transform as a triplet under the
``custodial'' $SU(2)_c$, so that $M_W = M_Z$.
This equality is  replaced by Eq.(\ref{eq:rho}) at tree level for arbitrary values of $g^\prime$.
Further corrections proportional to $g^\prime$ and $(\lambda^u - \lambda^d)$ arise at the one-loop level 
and are small. In fact, the success of the tree-level prediction $\rho =1$  a posteriori justifies 
the omission in the chiral Lagrangian~(\ref{eq:chiralLmass}) of the additional term 
\begin{equation}
 v^2 \, \text{Tr}\left[ \Sigma^\dagger D_\mu \Sigma \, \sigma^3 \right]^2 
\end{equation}
that is invariant under  the local $SU(2)_L\times U(1)_Y$ but explicitly breaks the global $SU(2)_L\times SU(2)_R$.
In other words, the coefficient  of such additional operator is experimentally constrained to be very small.

The chiral Lagrangian  (\ref{eq:chiralLmass}) makes  the origin of the violation of perturbative unitarity
most transparent. Let us work in a renormalizable $\xi$-gauge, with a gauge-fixing term
\begin{equation}
\begin{split}
{\cal L}_{GF} = & -\frac{1}{2\xi} \left(\partial_\mu W_\mu^3 + \xi \frac{gv}{2} \chi^3 \right)^2 - \frac{1}{2\xi}
 \left(\partial_\mu B_\mu + \xi  \frac{g^\prime v}{2} \chi^3 \right)^2 \\
 & - \frac{1}{2\xi}\left|\partial_\mu W_\mu^+ + \xi \frac{g^\prime v}{2} \chi^+ \right|^2  \, .
\end{split}
\end{equation}
The Equivalence  Theorem~[\refcite{equivth},\refcite{Lee:1977eg}] states that at large energies the amplitude for the emission or absorption of 
a Goldstone field $\chi$ becomes equal  to the amplitude for the emission or absorption of a longitudinally-polarized vector  boson:
\\[0.6cm]
\hspace*{2.2cm}
\includegraphics[height=30mm]{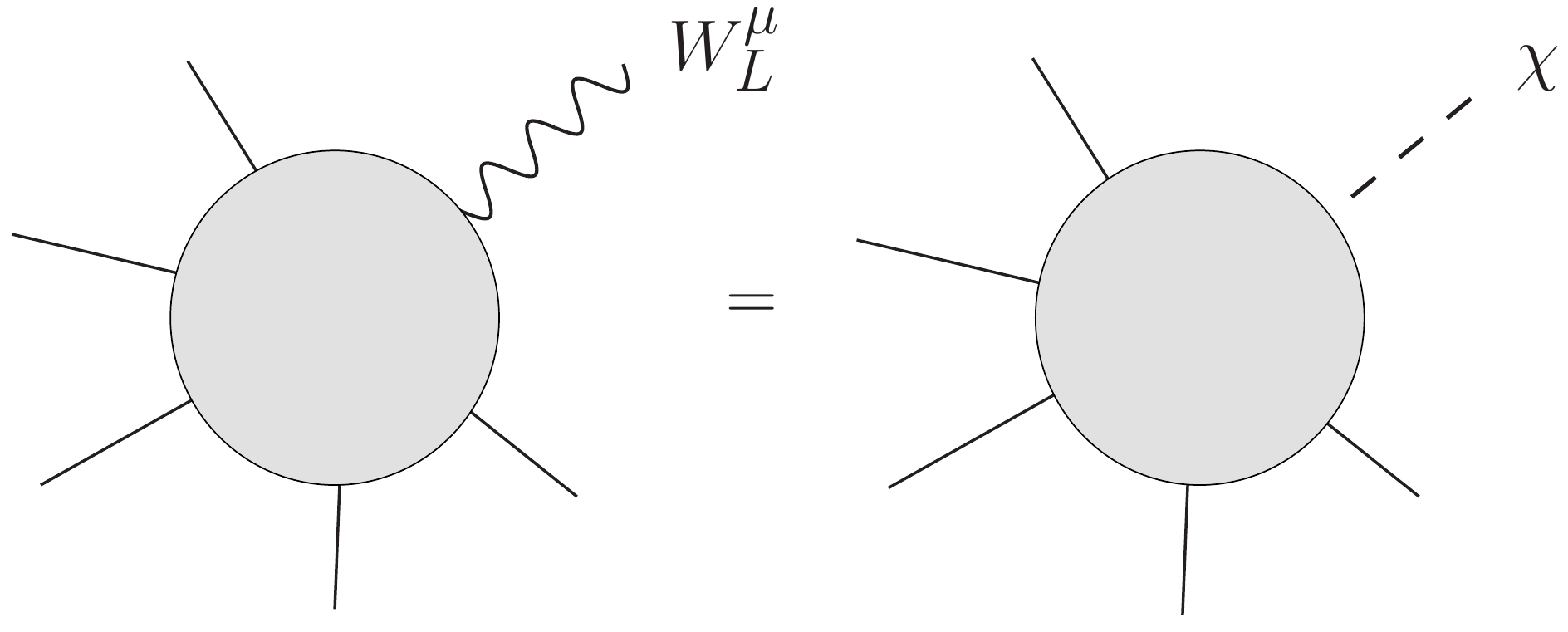} 
\hspace*{-0.2cm}
\begin{minipage}{0.1\linewidth}
  \vspace*{-3.15cm}
\begin{equation*}
\times \ \ \Bigg( 1 + O\left( \frac{m_W^2}{E^2}\right) \Bigg)\, .
\end{equation*}
\end{minipage}
\\[0.2cm]
In particular, the amplitude for the  scattering of two longitudinal $W$'s
becomes equal, at energies $E\gg m_W$, to the amplitude for the scattering of two Goldstone bosons.
For the latter process there is only one diagram which contributes at leading order in $E/m_W$:
\\[0.4cm]
\hspace*{2.5cm}
\includegraphics[height=19mm]{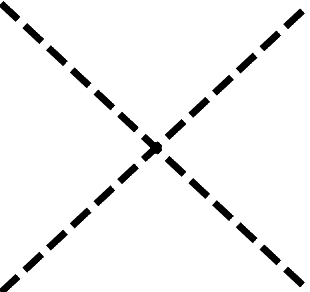} 
\begin{minipage}{0.7\linewidth}
  \vspace*{-1.8cm}
\begin{equation}
{\cal A}(\chi^+ \chi^- \to \chi^+ \chi^-) =  \frac{1}{v^2} (s+t)\, . 
\end{equation}
\end{minipage}
\\[0.3cm]
The growth of the amplitude with $E^2$ thus originates from the derivative interaction among four Goldstones
contained in the  kinetic term of $\Sigma$ in Eq.(\ref{eq:chiralLmass}).
Ultimately, the violation of  perturbative unitarity can be traced back to the non-renormalizability of the Lagrangian~(\ref{eq:chiralLmass}).
The merit of the chiral formulation is that of isolating the problem to the sector of the 
Lagrangian which leads to the  mass terms for the vector bosons and the fermions. 

There are thus two possibilities: \textit{i)} either  new particles associated to new dynamics come in to restore unitarity before
perturbativity is lost, or \textit{ii)} the $\chi\chi$ scattering grows strong until the interaction among four $\chi$'s becomes
non-perturbative.  This latter possibility must also be seen as the emergence of new physics, 
as the description of the theory changes, at the strong scale, in terms of 
new, more fundamental,  degrees of freedom.
These two paradigms for the electroweak symmetry breaking are well exemplified by the two theories 
that we will discuss in the next sections:
the Higgs model, and  Technicolor theories.
Whatever mechanism Nature has chosen, it is generally true that

\vspace{0.3cm}
\textit{There has to be some new  symmetry-breaking dynamics acting as an ultraviolet 
completion of the electroweak chiral Lagrangian~(\ref{eq:chiralLmass}). }
\vspace{0.3cm}

\noindent As required by the experimental evidence, such new dynamics must be (approximately) custodially symmetric,
so as to prevent large corrections to the $\rho$ parameter.
The most important question then is the following: 
% then arises: 
is this dynamics weak or strong~?

\subsection{The Higgs model}
\label{sec:Higgsmodel}

The most economical example of new  
custodially-invariant dynamics is that of just one new scalar field $h(x)$,
singlet under  $SU(2)_L\times SU(2)_R$.
Assuming that $h$ is coupled to the SM gauge fields and
fermions only via weak gauging and (proto)-Yukawa couplings, the most general EWSB Lagrangian
has three free parameters $a$, $b$, $c$~\footnote{In general $c$
can be a matrix in flavor space. We will assume it is proportional to unity, so that no
flavor-changing neutral current effects originate from the tree-level exchange of $h$.} 
at the quadratic order in $h$~[\refcite{stronghh}]:
\begin{equation}
\label{eq:CHLag}
\begin{split}
{\cal L}_{H} =& \frac{1}{2} \left(\partial_\mu h\right)^2 + V(h) + 
 \frac{v^2}{4} \text{Tr}\left[ \left( D_\mu \Sigma \right)^\dagger \left( D_\mu \Sigma \right) \right] 
 \left( 1 + 2 a\, \frac{h}{v} + b\, \frac{h^2}{v^2} + \dots \right) \\[0.1cm]
 & - \frac{v}{\sqrt{2}} \sum_{i,j} \left( \bar u_L^{(i)} d_L^{(i)} \right)  \Sigma \left( 1+ c\, \frac{h}{v} + \cdots\right) 
 \begin{pmatrix} \lambda_{ij}^u\,  u_R^{(j)} \\[0.1cm] \lambda_{ij}^d\,  d_R^{(j)} \end{pmatrix} + h.c.
\end{split}
\end{equation}
Here $V(h)$ denotes the potential, including a mass term, for $h$.
Each of these parameters controls the unitarization of a different sector of the theory.
For $a=1$ the exchange of the scalar unitarizes the $\chi\chi\to\chi\chi$ scattering
\footnote{In the diagrams showed in present section, dashed and solid lines denote respectively the fields $\chi$ and $h$, whereas 
solid lines with an arrow denote fermions.}
\\[0.2cm]
\begin{center}
\includegraphics[height=17mm]{chichiscattering.pdf} 
\hspace*{1cm}
\includegraphics[height=17mm]{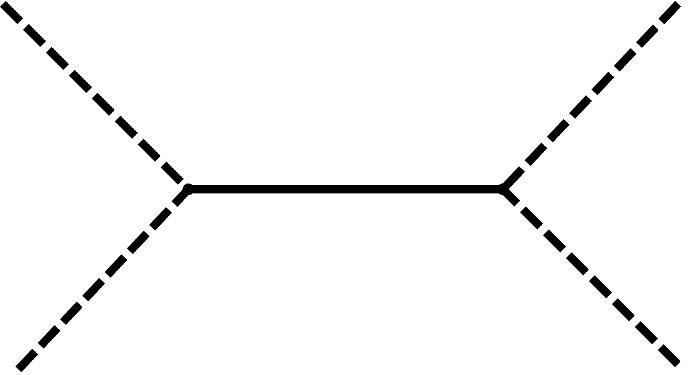} 
\hspace*{1cm}
\begin{minipage}{0.18\linewidth}
  \vspace*{-1.5cm}
\includegraphics[height=25mm]{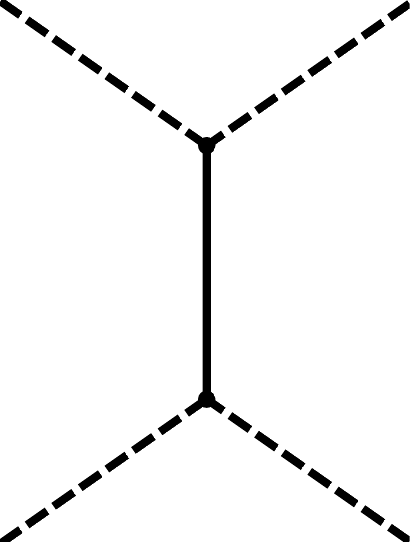} 
\end{minipage} 
\end{center}
\begin{equation*}
\begin{split}
  {\cal A}(\chi^+\chi^- \to \chi^+\chi^-) 
 & = \frac{1}{v^2} \left[ s - a^2  \frac{s^2}{s- m_h^2} + (s \leftrightarrow t) \right] \\
 & = \frac{s+t}{v^2}  \left(1-a^2\right) + O\left( \frac{m_h^2}{E^2} \right) \, .
\end{split}
\end{equation*}
\\
Since  we have introduced a new particle in the theory, we have to check that also the inelastic channels
involving $h$ are unitarized.   The $\chi\chi\to hh$ scattering (equivalent to $W_L W_L\to hh$ at high energy), 
is perturbatively unitarized  for $b=a^2$:
\\
\begin{center}
\includegraphics[height=19mm]{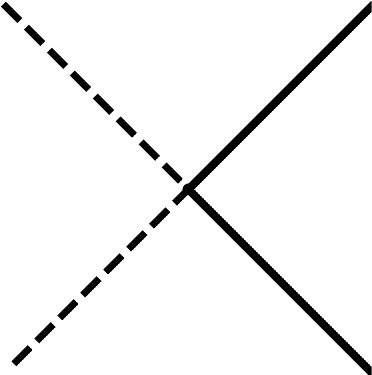} 
\hspace*{1.5cm}
 \begin{minipage}{0.15\linewidth}
   \vspace*{-1.5cm}
 \includegraphics[height=25mm]{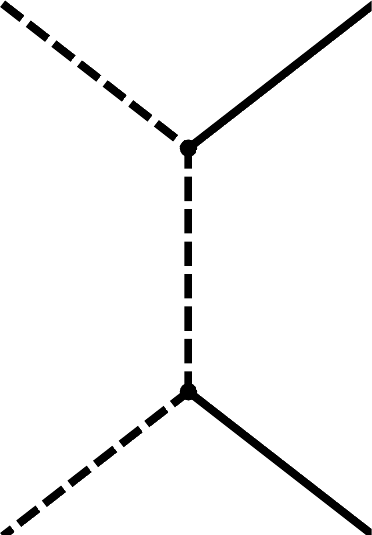} 
 \end{minipage} 
\end{center}
\begin{equation*}
 {\cal A}(\chi^+\chi^-\to hh) = \frac{s}{v^2} \left(b-a^2\right) + O\left( \frac{m_h^2}{E^2} \right) \, .
 \end{equation*} 
\\[0.1cm]
Finally, the $\chi\chi\to\psi\bar\psi$ scattering (equivalent to $W_L W_L \to\psi\bar\psi$ at high energy)  is unitarized  
for $ac=1$  \\
\begin{center}
\includegraphics[height=19mm]{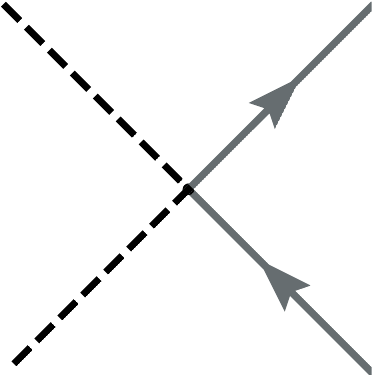} 
\hspace{1cm}
\includegraphics[height=19mm]{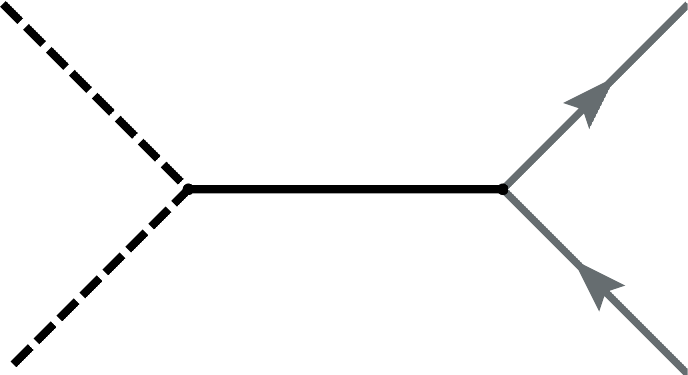} \hspace{1cm}
\end{center}
\begin{equation*}
{\cal A}(\chi^+\chi^-\to \psi\bar\psi) =  \frac{m_\psi \sqrt{s}}{v^2} \left(1-a c \right) + O\left( \frac{m_h^2}{E^2} \right)  \, .
\end{equation*} 
\\
Only for $a=b=c=1$  the EWSB sector is  weakly interacting (provided the scalar $h$ is light), 
as for example $a\not =1$ implies  a strong $WW\to WW$ scattering with violation of perturbative unitarity at energies 
$\sqrt{s} \approx 4\pi v/\sqrt{1-a^2}$, and similarly for the other channels.

The point $a=b=c=1$ in fact defines what I will call the ``Higgs model'': ${\cal L}_{H}$ (with vanishing higher-order terms in $h$) 
can be rewritten in terms of the $SU(2)_L$ doublet
\begin{equation}
H(x) = \frac{1}{\sqrt{2}}\, e^{i \sigma^a \chi^a(x)/v} \begin{pmatrix} 0 \\ v + h(x) \end{pmatrix}
\end{equation}
and gets the usual form of the  Standard Model Higgs Lagrangian.
In other words, $\chi^a$ and $h$ together form a linear representation of  $SU(2)_L\times SU(2)_R$.
The unitarity of the model can be thus traced back to its renormalizability.
In terms of the Higgs doublet $H$, the custodial invariance of the Lagrangian
appears like an accidental symmetry:  at the renormalizable
level, all the  ($SU(2)_L\times U(1)_Y$)-invariant operators are functions of $H^\dagger H = \sum_i \omega_i^2$,
where $\omega_i$ are the four real components parametrizing the complex doublet $H$.
This implies that the theory is invariant under an $SO(4)\sim SU(2)_L\times SU(2)_R$ invariance,
broken down to $SO(3)\sim SU(2)_c$ in the vacuum $\langle H^\dagger H\rangle = v^2$, 
under which the $\omega_i$ components are rotated.

The weakly-interacting Higgs model has two main virtues: it is theoretically attractive because of its
calculability, and it is insofar phenomenologically successful, as it satisfies the LEP and SLD 
electroweak precision tests~[\refcite{LEPEWWG}].
Both calculability  (which stems from  perturbativity) and the success in passing the precision tests follow
from the Higgs boson being light. 
It is however well known that an elementary light scalar, such as $h$, is  unstable under
radiative corrections: its mass receives quadratically divergent corrections, which makes a light Higgs scalar
highly unnatural in absence of some symmetry protection.
In this sense, the Higgs model should perhaps be regarded as a \textit{parametrization} rather than a dynamical explanation
of the electroweak symmetry breaking.

\subsection{Technicolor models}
\label{sec:TCmodels}

The Higgs model is an extremely economical way to perturbatively  unitarize the theory and parametrize the 
symmetry breaking, but we know that it is not the solution that Nature has chosen in another similar physical
system: QCD.  At low energy the condensation of the color force dynamically breaks the $SU(2)_L\times SU(2)_R$ 
chiral symmetry to its vectorial subgroup $SU(2)_V$,  the three pions $\pi^a$ being the associated Nambu-Goldstone bosons.
Their dynamics is described by a non-linear sigma model analogous to the chiral Lagrangian (\ref{eq:chiralLmass}) for the $\chi$ fields
\begin{equation}
\label{eq:Lpion}
 {\cal L}_{\pi} = \frac{f_\pi^2}{4}\, \text{Tr}\!\left[ \left( \partial_\mu \Sigma \right)^\dagger
 \left( \partial^\mu \Sigma \right) \right] \, , \qquad \Sigma(x) = \exp(i\sigma^a \pi^a(x)/f_\pi) \, ,
\end{equation}
where $f_\pi = 92\,$MeV is the pion decay constant.
Consequently, the pion-pion scattering is affected by the same unitarity problems  encountered in the $WW$ scattering.
In this case however, we know from experiment that there is no  light scalar resonance playing the role of the Higgs boson $h$.
Rather,   a whole tower of heavier resonances is exchanged in pion-pion scattering at high energies, which eventually
enforces unitarity. Experimentally, the most important contribution comes from the lightest vector resonance, the $\rho$ meson
($J=1$, $I=1$)
\\
\begin{center}
\hspace*{0.3cm}
\includegraphics[height=28mm]{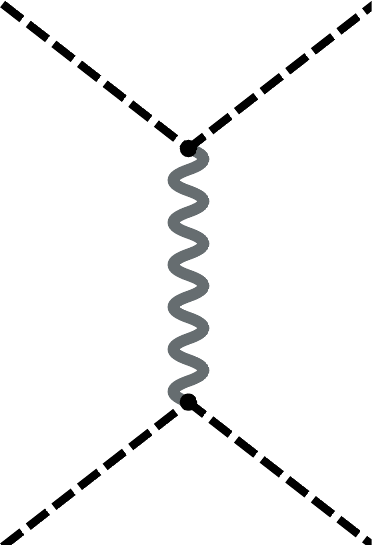} 
\hspace*{1.6cm}
 \begin{minipage}{0.32\linewidth}
   \vspace*{-2.6cm}
 \includegraphics[height=19mm]{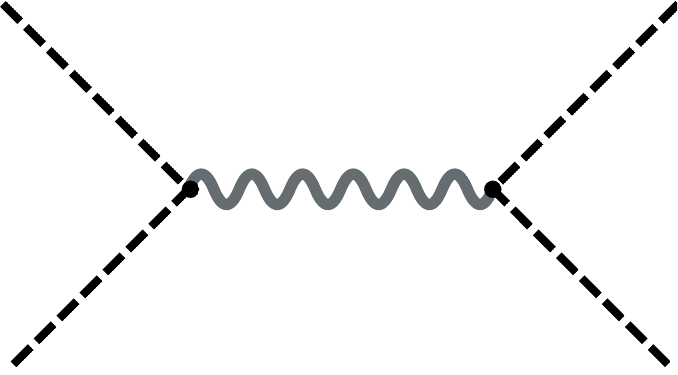} 
 \end{minipage} 
\end{center}
The  new symmetry breaking dynamics is thus strongly interacting  in this case, and its
dual description in terms of more fundamental degrees of freedom is the quark Lagrangian.
QCD can be considered as the prototype for strong symmetry breaking, and the study of its properties
can shed light on the UV completion of the electroweak Lagrangian (\ref{eq:chiralLmass}).

It is interesting, for example, to discuss what happens to QCD and to the pions when one turns on the weak interactions.
~\footnote{In the following we consider for simplicity QCD with two quark flavors, the generalization
to the six-flavor case is trivial.}
In the limit of vanishing quark masses (chiral limit), and before turning on the weak interactions,
the pions are exact NG bosons associated to the global  symmetry breaking 
$SU(2)_L \times SU(2)_R \times U(1)_B\to SU(2)_V \times U(1)_B$,  and are thus massless.
\begin{figure}[t]
\centering
\includegraphics[height=40mm]{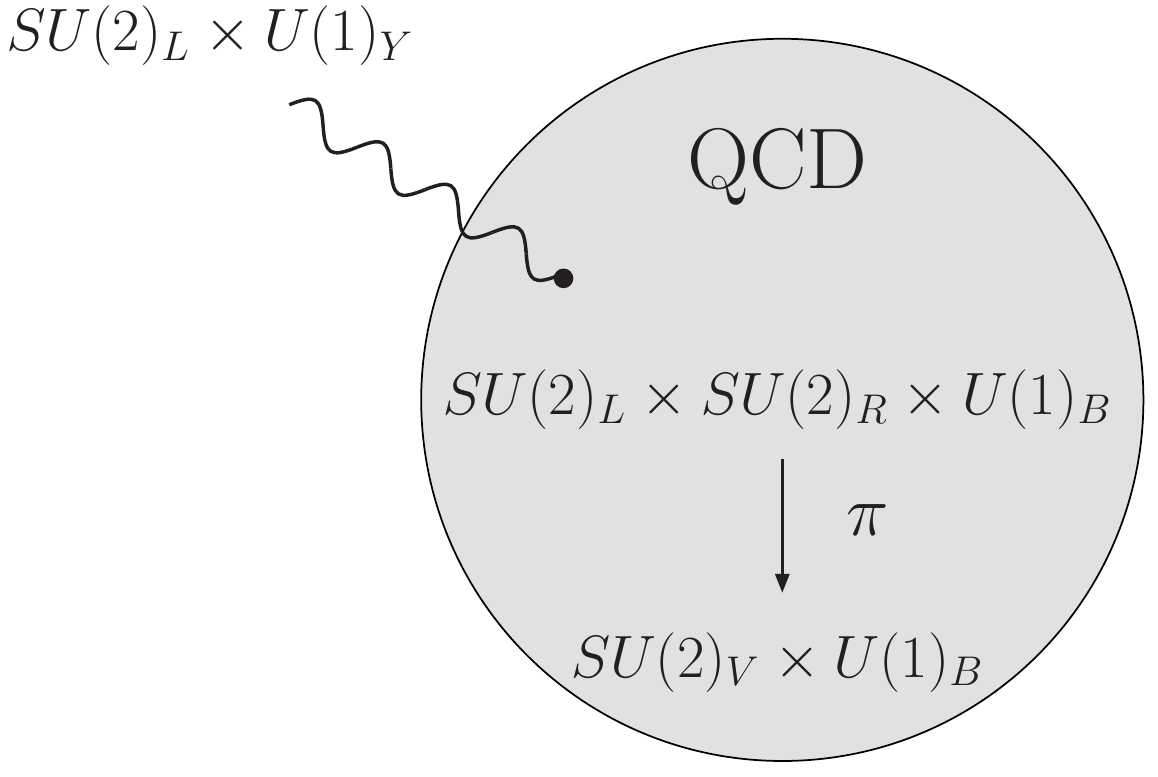} 
\caption{\label{fig:QCDpicture}
Cartoon of QCD  with part of its chiral symmetry gauged by the weak interactions. 
}
\end{figure}
The $SU(2)_L \times U(1)_Y$ interactions gauge only part of the full global symmetry, 
and in this way they introduce an explicit breaking of the symmetry, see Fig.~\ref{fig:QCDpicture}.
In other words, the QCD vacuum breaks the electroweak
invariance and the pions are eaten to give mass to the $W$ and the $Z$. The surviving unbroken group is exactly the electromagnetic
$U(1)_{em}$, with the electric charge given by $Q=T_{L}^3 + T_R^3 + B/2$.
To see how the weak bosons get mass, let us consider for example the $W$ propagator in the Landau gauge $\xi =0$.
As the result of the coupling of the $W$ to the conserved weak current $J^{\mu\, \pm} = \bar q_L \gamma^\mu T^{\pm} q_L$ 
($T^{\pm} = T_1 \pm i T^2$),  its propagator gets corrected from the QCD dynamics:
\vspace{0.2cm}
\begin{center}
\includegraphics[height=22mm]{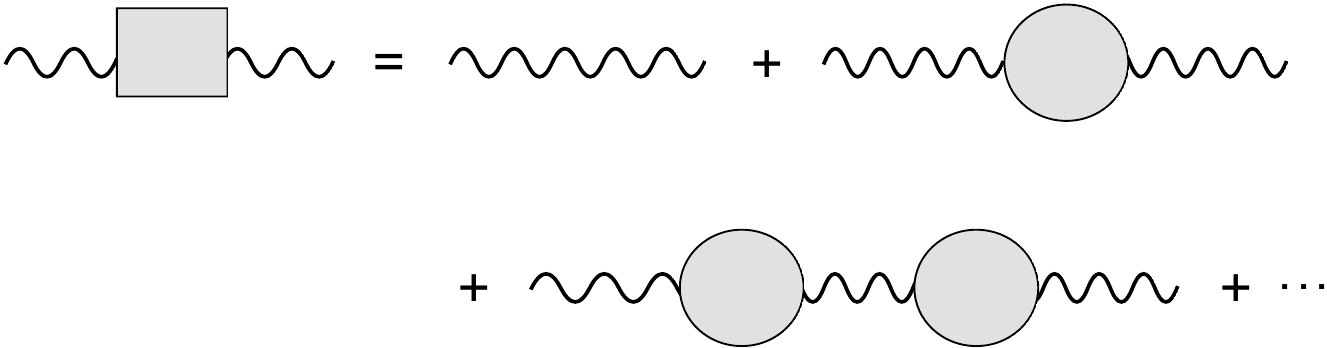} 
\end{center}
\vspace{0.2cm}
\begin{equation}
\begin{gathered}
\label{eq:Wprop}
G_{\mu\nu}(q) = \frac{-i}{q^2-g^2 \Pi(q^2)/2} \,(P_T)_{\mu\nu} \,  , \qquad
(P_T)_{\mu\nu} \equiv \eta_{\mu\nu} - \frac{q_\mu q_\nu}{q^2} \, ,
\end{gathered}
\end{equation}
where
\begin{equation}
\label{eq:vacuumpol}
\begin{split}
i \Pi_{\mu\nu}(q) =& - \int \!d^4 x \; e^{-iq\cdot x} \langle 0|T\left(J^+_\mu(x) J^-_{\nu}(0) \right) |0\rangle \\[0.1cm]
\Pi_{\mu\nu}(q) =&  \left( \eta_{\mu\nu}  - \frac{q_\mu q_\nu}{q^2} \right) \Pi(q^2)\, .
\end{split}
\end{equation}
Then, a mass for the $W$ arises if $\Pi_{\mu\nu}(q^2)$ has a pole at $q^2=0$. 
The pole in fact exists as a result of the symmetry breaking, due to the exchange of the pion:
\begin{equation}
\langle 0| J^+_\mu | \pi^-(p) \rangle = i\frac{f_\pi}{\sqrt{2}} p_\mu 
\end{equation}
\vspace{0.02cm}
\begin{center}
\includegraphics[width=43mm]{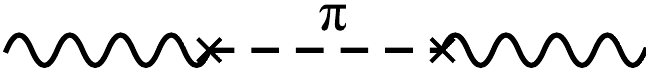} 
\hspace{0.7cm}  
\begin{minipage}[t]{150pt}
  \vspace{-0.65cm}
  \centering
  {$\Longrightarrow$ \qquad $\displaystyle \Pi(q^2) = \frac{f_\pi^2}{2}$.}
\end{minipage} 
\end{center}
\vspace{0.15cm}
This implies that the $W$ acquires a mass 
\begin{equation*}
m_W = \frac{gf_\pi}{2} \simeq 29\,\text{MeV} \, .
\end{equation*}
Although this number is far from the experimental value, the above discussion shows that QCD is, at the qualitative level, a good example
of electroweak symmetry breaking sector. This is even more true considering that the unbroken $SU(2)_V$ isospin invariance acts as a custodial
symmetry so that $\rho=1$ at tree level in the QCD vacuum.

\begin{figure}[t]
\centering
\includegraphics[height=45mm]{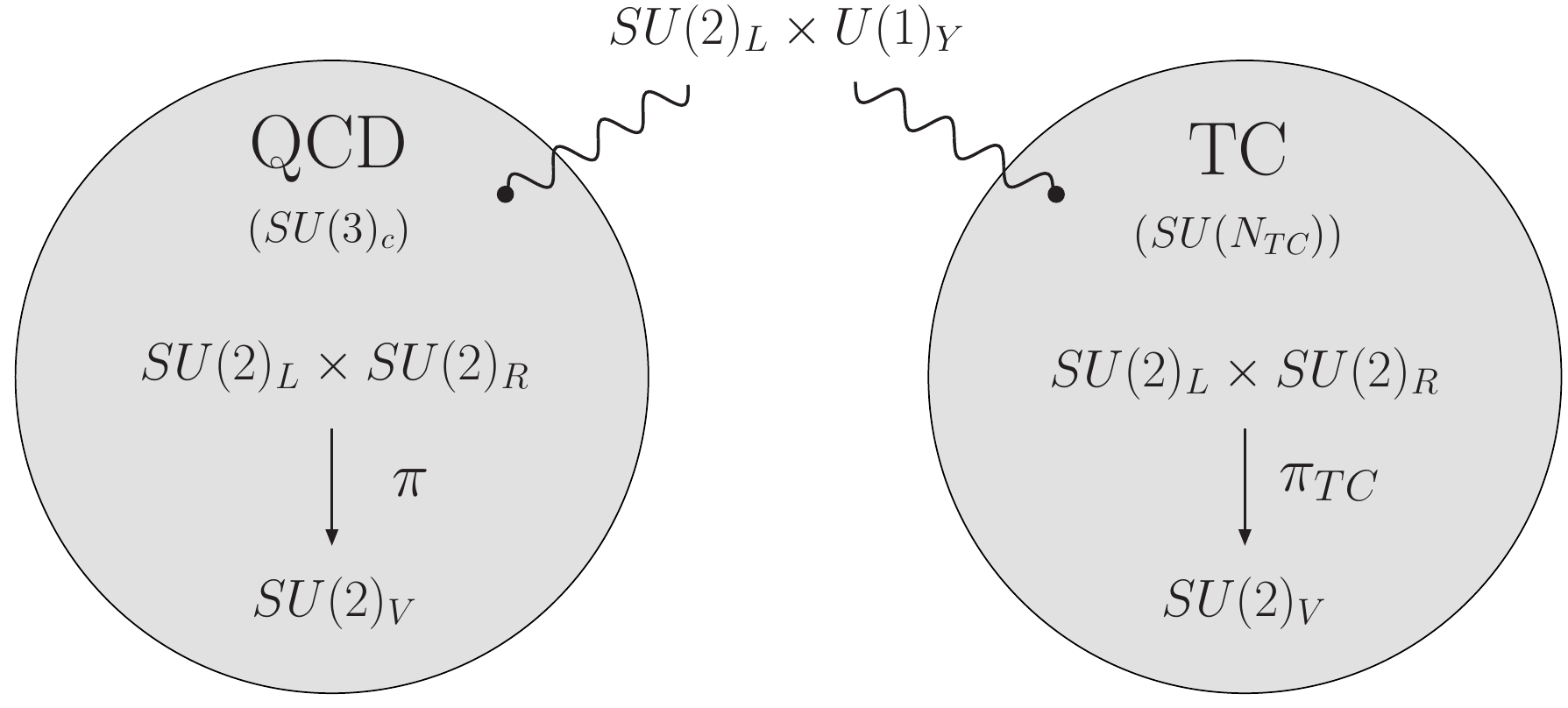} 
\caption{\label{fig:TCpicture}
Cartoon of a new Technicolor sector and QCD with part of their global symmetries gauged by the weak interactions.
}
\end{figure}

This suggests that the actual EWSB dynamics could  be just a scaled-up version of QCD,  with
\begin{equation}
f_\pi \quad\longrightarrow\quad F_\pi \simeq v = 246\,\text{GeV}\, .
\end{equation}
In general, one can think of an $SU(N_{TC})$ ``Technicolor'' gauge group with a  global $SU(2)_L\times SU(2)_R$ invariance 
broken down to $SU(2)_V$ at low energy due to confinement~[\refcite{TC}], see Fig.~\ref{fig:TCpicture}.
A linear combination of the QCD pions and the new set of `technipions' is thus eaten to form the longitudinal polarizations
of the $W$ and the $Z$
\begin{equation}
\begin{gathered}
|V_L \rangle = \sin\alpha \,| \pi_{QCD} \rangle + \cos\alpha \,| \pi_{TC} \rangle \\[0.1cm]
v^2 = f_\pi^2 + F_\pi^2 \qquad \tan \alpha = \frac{f_\pi}{F_\pi}\, ,
\end{gathered}
\end{equation}
while the orthogonal combination can be identified with the physical pion.
Since $f_\pi \ll F_\pi \simeq v$, the longitudinal polarizations of the $W$ and the $Z$ will mostly come from
the technipions,  although it is true that they will have a small component of  the QCD pions. 
In fact, this is true in general for any EWSB sector, Technicolor being just a specific example of symmetry breaking dynamics
where the role of the $\chi$ fields in the chiral Lagrangian (\ref{eq:chiralLmass}) is played by the technipions.

In order to derive the properties of the Technicolor sector, the large-$N$ formalism developed in 
Refs.~[\refcite{largeN-tHooft,largeN-Witten}] is extremely useful.  Here we just summarize the results that we will use and refer to 
Ref.~[\refcite{largeN-Witten}] for the proof.
Let us consider an $SU(N)$ gauge theory, with a large number $N$ of `colors'.  We know that QCD at $N=3$ is a confining theory and
we will assume that this behavior persists for $N\gg 1$.
Under this assumption, the large-$N$ theory has the following properties:
\begin{enumerate}
\item At leading order in $N$ the two-point function of a quark local bilinear operator $J(x)$, like the scalar $\bar q q$
or the current $\bar q\gamma^\mu q$, is given by an infinite exchange of one-meson states:
\begin{equation*}
\langle J(q) J(-q)\rangle = \sum_n \frac{f_n^2}{q^2 - m_n^2} \, ,
\end{equation*}
where $m_n$ is the mass of the $n$-th meson and $f_n = \langle 0|J|n\rangle$ the amplitude for $J$ to create it from the vacuum.
This in turn implies that:
\item  For large $N$ the mesons  are free, stable and non-interacting.  Their  number  is infinite and their masses 
have a smooth large-$N$ limit. The mass of the lowest lying modes is of the order
\begin{equation} \label{eq:mrhoNDA}
m_\rho  \sim g_{\rho} f_\pi \, ,
\end{equation}
where  $g_\rho$  denotes the coupling among mesons. 
\item Since the two-point function $\langle JJ\rangle$ is of order $N/16\pi^2$, it follows that $f_n$ scales like
\begin{equation} \label{eq:fNDA}
f_n \sim \frac{\sqrt{N}}{4\pi} \, .
\end{equation}
From the behavior of the $n$-point Green functions of $J$ it follows that a local vertex with $n$ mesons scales like $\sim g_\rho^{n-2}$,
where
\begin{equation} \label{eq:grhoNDA}
g_\rho \sim \frac{4\pi}{\sqrt{N}}\, .
\end{equation}
\end{enumerate}
Using the above results,  the vectorial and axial conserved currents of the Technicolor sector can be written, for large $N_{TC}$, in terms
of an infinite sum over vectorial ($\rho_n$) and axial ($a_n$) resonances:
\begin{align}
\langle T\left\{ J_V^\mu(q) J_V^\nu(-q) \right\}\rangle =&
 \left(q^2 \eta^{\mu\nu} - q^\mu q^\nu \right) \sum_n \frac{f_{\rho_n}^2}{q^2- m_{\rho_n}^2} \\[0.1cm]
\langle T\left\{ J_A^\mu(q) J_A^\nu(-q) \right\}\rangle =&
 \left(q^2 \eta^{\mu\nu} - q^\mu q^\nu \right) \left[ \sum_n \frac{f_{a_n}^2}{q^2- m_{a_n}^2} + \frac{F_\pi^2}{q^2} \right]\, .
\end{align}
Here we have defined the amplitude for $J_V$ to create a vectorial resonance with momentum $q$ and polarization $\epsilon_r$ 
to be $\langle 0|J_V^\mu|\rho (q,\epsilon_r)\rangle = \epsilon_r^\mu m_\rho f_\rho$, and similarly for the axial current.
Notice that the latter have the quantum numbers to create, in addition to spin-1 axial mesons, also the technipion.
As a consequence, the two-point function $\langle J_A J_A \rangle$ has a pole at $q^2=0$.
At large $N_{TC}$ the vector, axial and technipion decay constants scale like $f_{\rho}, f_{a}, F_\pi \sim \sqrt{N_{TC}}/4\pi$, while
all the masses are constants.
In particular, the mass of the lowest-lying vectorial resonance of the Technicolor sector, the `technirho', is
expected to be of order
\begin{equation}
m_{\rho_{TC}} \sim \sqrt{\frac{3}{N_{TC}}} \,\frac{F_\pi}{f_\pi} \,m_\rho \, ,
\end{equation}
where $m_\rho = 770\,$MeV is the mass of the QCD rho meson, $f_\pi = 92\,$MeV is the QCD pion decay constant and
$m_{\rho_{TC}}$, $F_\pi$ are the corresponding Technicolor quantities.
For example, for $N_{TC} =4$ one has $m_{\rho_{TC}} \sim 1.8\,$TeV.

The most attractive feature of  Technicolor theories, and in general of theories with strong electroweak symmetry breaking,
is that the hierarchy problem of the Higgs model is solved by dimensional transmutation:
the electroweak scale $v$ is generated dynamically as the scale at which the Technicolor  coupling $g_{TC}$ 
grows strong in the infrared ($\beta_0 < 0$):
\begin{equation}
\mu \frac{d}{d\mu} \frac{1}{g_{TC}^2}(\mu) = -\frac{\beta_0}{8\pi^2}
\hspace{0.3cm} \Longrightarrow \hspace{0.3cm} 
v = M_{Pl} \,\exp\left( -\frac{8\pi^2}{g_{TC}^2(M_{Pl}) (-\beta_0)} \right) \, .
\end{equation}
This is in complete analogy with the dynamical generation of the QCD scale from the Planck scale  $M_{Pl}$.
On the other hand, the simplest Technicolor constructions, like the naive scaled-up version of QCD, lead to predictions
in conflict with the experimental data. The two most serious problems are a parametrically  too large correction to the Peskin-Takeuchi
$S$ parameter, and too fast flavor-changing neutral-current  processes. Let us review both in turn.

The Peskin-Takeuchi $S$ parameter is defined as~[\refcite{PesTak}]
\begin{equation} \label{eq:defS}
S \equiv -16\pi \frac{\partial}{\partial q^2} \Pi_{3B}(q^2) \big|_{q^2=0}
\end{equation}
where the vacuum polarization of a $W_{3L}^\mu$ and an hypercharge boson $B^\mu$, $\Pi_{3B}(q^2)$, is defined according to
eq.(\ref{eq:vacuumpol}). The leading contribution to $S$ from new heavy states can be parametrized in terms of the dimension-6 operator
(see Refs.~[\refcite{Georgi:1991ci,Barbieri:2004qk}])
\begin{equation} \label{eq:chiralS}
\frac{S}{16\pi} \text{Tr} \left[ T^{aL} W_{\mu\nu}^{aL} \, \Sigma \, B^{\mu\nu} T^{3R} \, \Sigma^\dagger \right]
\end{equation}
where $T^{aL} = \sigma^a/2 = T^{aR}$ are the generators of $SU(2)_L \times SU(2)_R$.
Since the Technicolor sector is strongly coupled, a perturbative calculation of the $S$ parameter is not possible. However,  one can
estimate its size using Naive Dimensional Analysis (NDA)~[\refcite{NDA}]: it will arise at the 1-loop level, thus carrying a factor $N_{TC}/16\pi^2$,
and it will be proportional to the number of technidoublets~$N_{D}$:
\begin{center}
 \includegraphics[width=50mm]{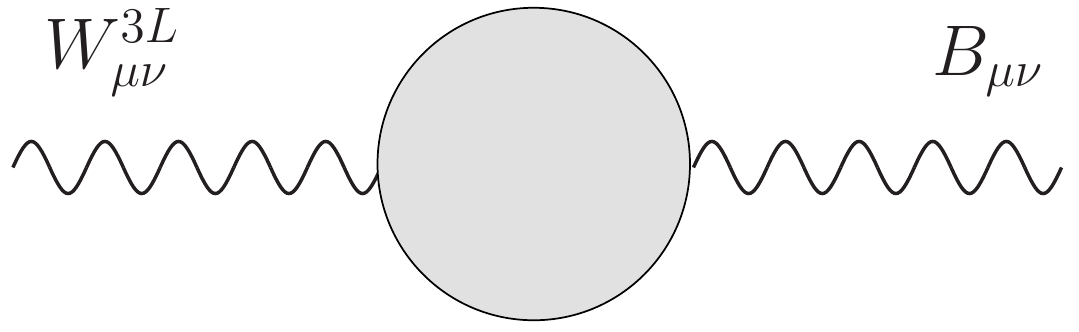}
\end{center}
\begin{equation} \label{eq:SNDA}
\frac{S}{16\pi} \sim \frac{N_{TC} N_D}{16 \pi^2} \qquad \Longrightarrow  \qquad S \sim \frac{N_{TC} N_D}{\pi}\, .
\end{equation}
A more sophisticated calculation that makes use of QCD data rescaled up to the EW scale gives a similar result~[\refcite{PesTak}].
From the estimate above one deduces that even minimal models (with $N_{TC}$ and $N_{D}$ small) tend to predict $S\sim 1$.
Such values are ruled out by the LEP data, which bound (assuming an optimal contribution to the $T$ parameter)~[\refcite{LEPEWWG}]
\begin{equation}
S \lesssim 0.3 \quad  @ \; 99\% \; \text{CL}\, .
\end{equation}
Conversely, $S < 0.3$ requires $m_{\rho_{TC}} \gtrsim 3\,$TeV$\sqrt{N_D}$, which is difficult to accommodate in Technicolor models
given that the mass of the first vectorial resonance is tied to the EW scale $v$.

The second important difficulty with  simplest Technicolor models is the way in which quark masses are generated.
So far we have not discussed how the quark sector feels the electroweak symmetry breaking.  For this to occur, some interaction
must exist between quarks and techniquarks.  A simple solution is to assume that both the color group $SU(3)_c$ and the
Technicolor $SU(N_{TC})$ are embedded in a larger Extended Technicolor (ETC) group,
\begin{equation*}
SU(N_{ETC}) \supset SU(3)_c \times SU(N_{TC})  \, ,
\end{equation*}
which is assumed to be spontaneously broken at the scale $\Lambda_{ETC}$~[\refcite{ETC}].
The exchange of the broken ETC gauge bosons connects quarks with techniquarks and
generates, at the scale $\Lambda_{ETC}$, four-fermion operators with two SM quarks and two technifermions:
\begin{center}
 \includegraphics[height=35mm]{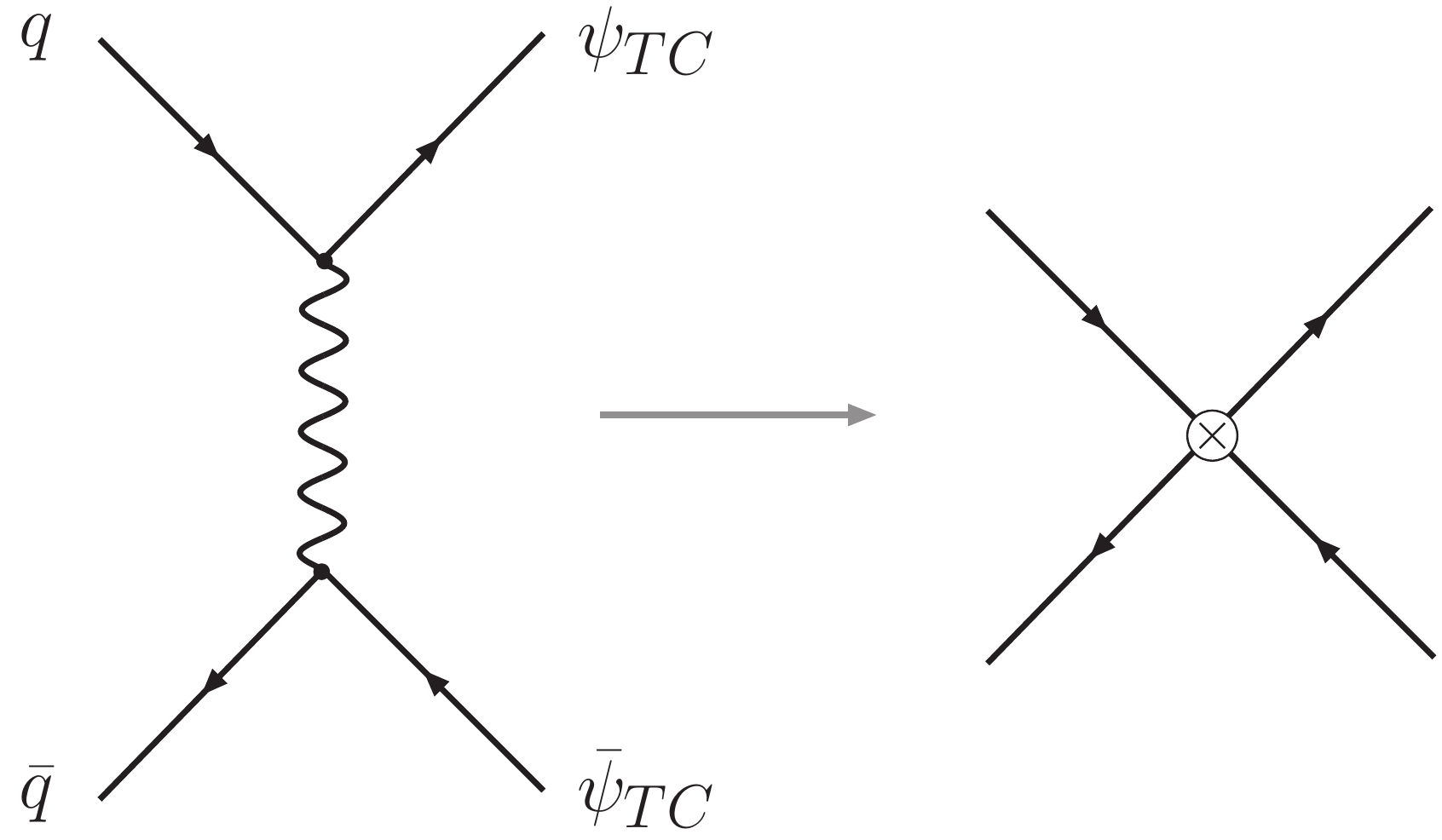} 
\end{center}
 \begin{equation}
\label{eq:4fermion}
{\cal L}_{int} = \frac{g_{ETC}^2}{\Lambda_{ETC}^2} \left(\bar q q\right)\!\left( \bar\psi_{TC} \psi_{TC} \right)\, .
\end{equation}
At the lower scale $\Lambda_{TC}\approx F_\pi \simeq v$ the  $SU(N_{TC})$ group condenses giving rise to the
quark masses
\begin{equation}
\label{eq:quarkmass}
m_q = \frac{g_{ETC}^2}{\Lambda_{ETC}^2} \,\langle \bar\psi_{TC} \psi_{TC} \rangle \sim \Lambda_{TC} \left( \frac{\Lambda_{TC}}{\Lambda_{ETC}} \right)^2\, .
\end{equation}
In order to explain the large hierarchies in the quark  masses it is thus clear that the generation of the four-fermion interactions (\ref{eq:4fermion})
for different flavors cannot happen just at one single scale $\Lambda_{ETC}$. Rather, one has to assume that different SM quark families are
embedded into a single ETC multiplet and that $SU(N_{ETC})$ undergoes a cascade of breakings, thus generating several different scales.
The ETC breaking scale relevant for any given quark flavor cannot be too large, otherwise the  corresponding quark mass 
that follows from eq.(\ref{eq:quarkmass}) becomes too small. For example,  if $\Lambda_{TC} \simeq v$
one needs $\Lambda_{ETC} \approx 10\,\text{TeV}$ in order to reproduce the $s$ quark mass.

The same exchange of ETC gauge fields that leads to  the four-fermion operator (\ref{eq:4fermion}), however, also generates operators
with four SM fermions, $(\bar q q)^2/\Lambda_{ETC}^2$. Quite generically, these operators  violate flavor and CP, since different
SM flavors have to be embedded into the same ETC multiplet, and thus give rise to various FCNC processes.
The bounds from $K\bar K$ and $B\bar B$ mixing and rare meson decays, for example,
put very stringent limits on the scale at which such operators
can be generated: $\Lambda_{ETC} \gtrsim 10^5\,$TeV ($10^3-10^4\,$TeV)  from CP-violating  (-conserving) processes~[\refcite{UTfit}].
Thus, there is a tension between generating large enough quark masses and avoiding too fast FCNC processes.

\begin{figure}[t]
\centering
\includegraphics[height=40mm]{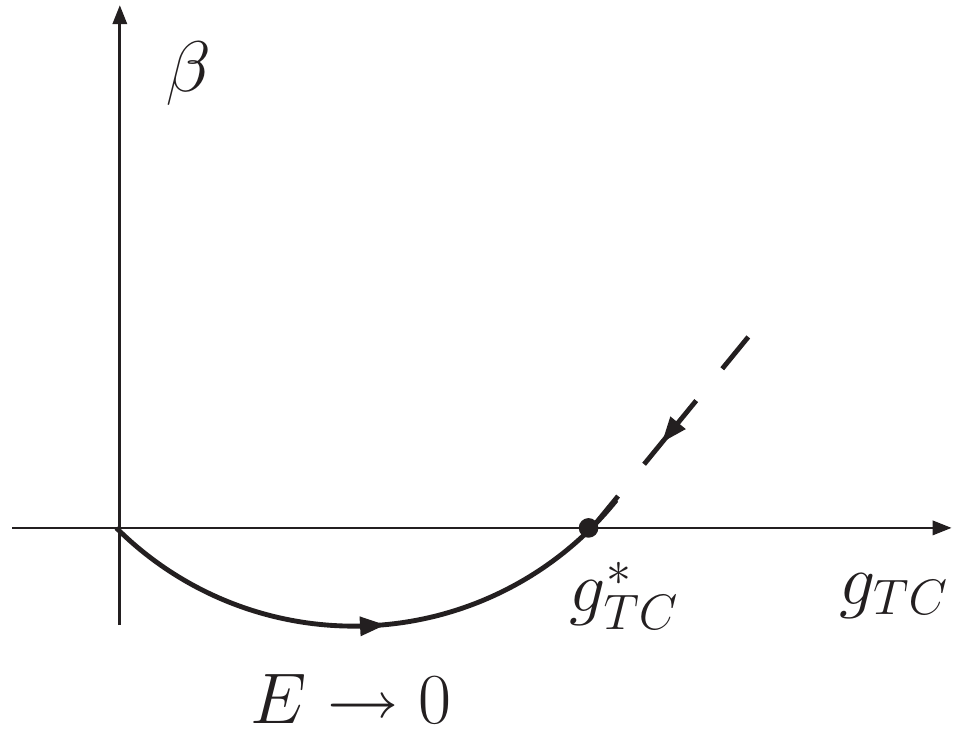} 
\caption{\label{fig:IRfixedpoint}
Flow to an IR non-perturbative fixed point.
}
\end{figure}

One mechanism that has been proposed to resolve this tension is that of Walking Technicolor~[\refcite{walkingTC}].
It is based on the following general observation:  if a term  $\Delta {\cal L} = \lambda\, O(x)$ is generated  in the Lagrangian at the scale $\Lambda$
with dimension $[O(x)] =d$, its contribution to physical amplitudes at the low-energy scale $E$ goes like $\sim \lambda\, (E/\Lambda)^{d-4}$.
The energy factor is due to the classical running of the coupling $\lambda$ for $d\not =4$, so that the higher the dimension of the operator $O$,
the more suppressed its contribution  at low energy. 
In writing the formula (\ref{eq:quarkmass}) for the quark masses  we have implicitly assumed that the dimension of the four-fermion
operator (\ref{eq:4fermion}) is equal to its classical  value $[(\bar q q)(\bar\psi_{TC} \psi_{TC})] =6$, although quantum corrections due to the 
Technicolor interactions can change it. In general, the anomalous dimension $\gamma$ is small if the $SU(N_{TC})$ theory is asymptotically 
free above the scale $\Lambda_{TC}$,
so that the coupling $g_{TC}(\mu)$ quickly runs to small values for $\mu > \Lambda_{TC}$:
\begin{equation}
[(\bar q q)(\bar\psi_{TC} \psi_{TC})] = 6 + \gamma \qquad \gamma(\mu) \sim O(\alpha_{TC})\, .
\end{equation}
However, it is possible that starting from high energies, $\Lambda_{ETC} \gtrsim  E\gg \Lambda_{TC}$, and flowing down to lower scales, 
the $SU(N_{TC})$ dynamics  reaches a non-perturbative infrared fixed point, see Fig.~\ref{fig:IRfixedpoint}.
In that case  the theory behaves like a conformal field theory down to energies $E\sim \Lambda_{TC}$, at which it 
condenses  and the conformal behavior is lost.
In the conformal regime, any operator $O$ is characterized by its scaling  dimension $d_*$ at the fixed point:
\begin{equation}
[O] = d_* \, , \quad \qquad \langle O(x) O(0) \rangle \propto \frac{1}{x^{2d_*}} \, .
\end{equation}
Notice that $d_*$ can significantly differ from its perturbative (classical) value, since the Technicolor coupling at the fixed point is large 
and its evolution above $\Lambda_{TC}$ is slow: it `walks' towards the fixed-point value $g^*_{TC}$.
Once applied to the operator $O=(\bar q q)(\bar\psi_{TC} \psi_{TC})$, the above argument shows that the formula for the quark masses 
generalizes to
\begin{equation}
\label{eq:quarkmassWT}
m_q \sim \Lambda_{TC} \left( \frac{\Lambda_{TC}}{\Lambda_{ETC}} \right)^{2+\gamma}\, .
\end{equation}
Hence, if the anomalous dimension $\gamma$ is sizable and negative, the suppression in the quark masses can be reduced or, equivalently,
$\Lambda_{ETC}$ can be larger.  This ameliorates the FCNC problem, since no large anomalous dimension is generated by
the SM color and weak interactions, so that the suppressing factor in front of the flavor-violating four-quark operators is still  $1/\Lambda_{ETC}^2$.
However, naive arguments suggest that the smallest consistent value for the anomalous dimension is $\gamma =-1$, so that
the suppression in the quark masses can be ameliorated but not completely avoided.
For example, a simple way to deduce the lower bound $\gamma > -1$ is the following~[\refcite{Luty:2004ye}]:
Neglecting the contribution  coming from SM interactions, $\gamma$ entirely arises from the anomalous dimension of the quark
bilinear $H = (\bar\psi_{TC} \psi_{TC})$, which plays the role of the Higgs field in acquiring a vacuum expectation value and giving mass
to the SM quarks. 
The unitarity bound on primary scalar operators of a 
conformal field theory requires $\gamma$ to be larger than  $-2$, a value at which
the dimension of $H$ becomes equal to that of the corresponding free field.
In the limit of large $N_{TC}$ or $\gamma\to -2$ the dimension of the SM scalar singlet $H^\dagger H$ is well approximated
by twice the dimension of $H$: $[H^\dagger H] \simeq 2 [H] = 6+ 2\gamma$.  Then, for $\gamma < -1$ the operator $H^\dagger H$ becomes relevant
and it will reintroduce the problem of  UV instability that plagues the Higgs model.
In particular,  for $\gamma = -2$ the operator $H^\dagger H$ has dimension 2 and its radiative correction will be quadratically divergent.
The possibility is still open for a Walking Technicolor theory at small $N_{TC}$ where $[H]$ is not too much above 1 (in order not to suppress
the quark masses), while $[H^\dagger H]$ stays close to 4 (so that no hierarchy problem is present)~[\refcite{Luty:2004ye}], although
strong constraints have been derived on this scenario based  on  general properties of conformal field 
theories~[\refcite{Rattazzi:2008pe,Rychkov:2009ij}].

%%%%%
%% Section 2
\section{The Higgs as a composite Nambu-Goldstone boson}
\label{sec:CHM}

There is an interesting variation of the strong symmetry breaking paradigm that interpolates between 
simple Technicolor theories and the Higgs model:  a light Higgs boson could emerge as the bound state
of a strongly interacting sector, rather than being an elementary field.
A composite Higgs would solve the hierarchy problem of the Standard Model, as its mass is not
sensitive to virtual effects above the compositeness scale, in the same way  as the mass of the QCD pion does not 
receive Planckian corrections.
Having a light Higgs in the spectrum, on the other hand, would allow the theory to satisfy 
the LEP electroweak precision tests more easily  than in the case of simple Technicolor constructions.

\begin{figure}[t]
\begin{minipage}{0.54\linewidth}
\centering
\includegraphics[height=40mm]{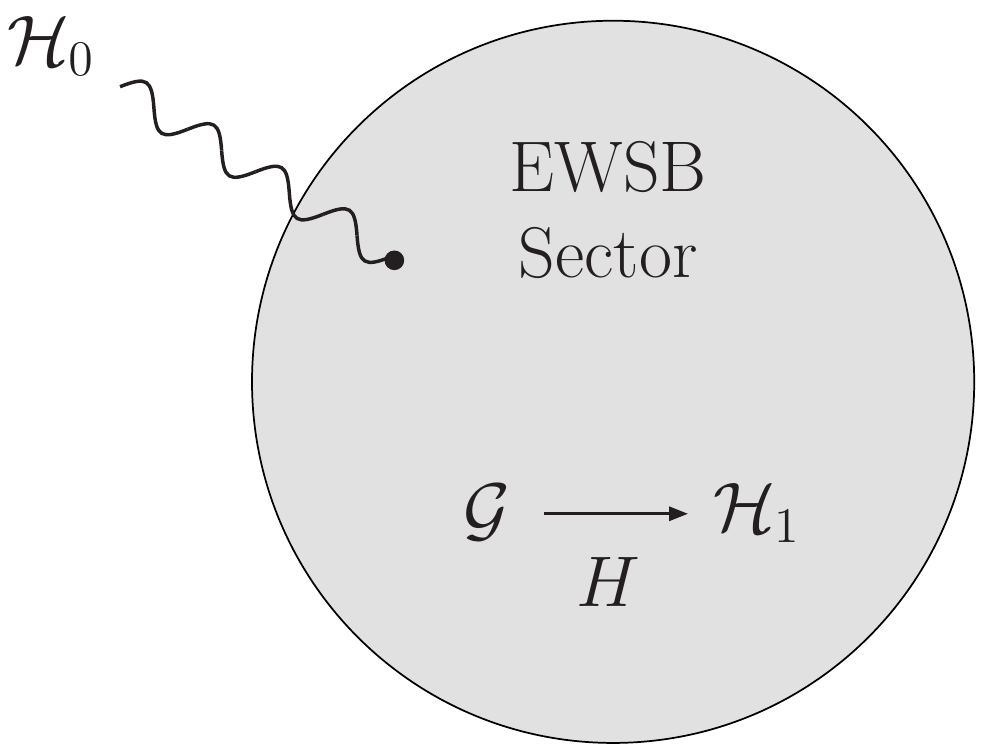} 
\caption{Cartoon of a strongly interacting EWSB sector with global symmetry ${\cal G}$ 
broken down to ${\cal H}_1$ at low energy. The  subgroup ${\cal H}_0 \subset {\cal G}$ is gauged by  external vector bosons.}
\label{fig:PGBcartoon}
\end{minipage}
\hspace{0.5cm}
\begin{minipage}{0.40\linewidth}
\centering
\includegraphics[height=45mm]{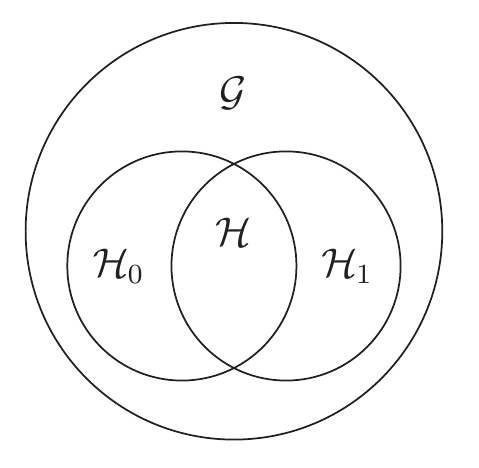} 
\caption{The pattern of symmetry breaking.}
\label{fig:PGBdiagram}
\end{minipage}
\end{figure}

As first pointed out by Georgi and Kaplan in the eighties in a series of seminal papers, the composite Higgs
boson can  be naturally lighter than the other resonances if it 
emerges as the pseudo Nambu-Goldstone boson of an enlarged global symmetry of the strong dynamics
[\refcite{GK1,GK2,Banks:1984gj,GK3,GK4,GK5}], see also [\refcite{Dimopoulos:1981xc}].
Consider for example the general case in which the strongly interacting sector has a global symmetry~${\cal G}$
dynamically broken to ${\cal H}_1$ at the scale $f$ (the analog of the pion decay constant $f_\pi)$,  
and the subgroup ${\cal H}_0 \subset {\cal G}$ is gauged by  external vector bosons, see Fig.~\ref{fig:PGBcartoon}. 
The global symmetry breaking ${\cal G}\to {\cal H}_1$ implies $n=\text{dim}({\cal G})-\text{dim}({\cal H}_1)$ Goldstone bosons, 
$n_0 = \text{dim}({\cal H}_0)-\text{dim}({\cal H})$ of which are eaten to give mass to as many vector bosons, 
so that ${\cal H} = {\cal H}_1 \cap {\cal H}_0$ is the unbroken gauge group, see Fig.~\ref{fig:PGBdiagram}.
The remaining $n-n_0$   are pseudo Nambu-Goldstone bosons.
In this picture the SM fields, both gauge bosons and fermions,  are assumed to  be external to the strong sector,
and in this sense we will refer to them as `elementary',  as opposed to the composite nature of the resonances 
of the strong dynamics. The SM gauge fields, in particular, are among the vector bosons associated to gauge group ${\cal H}_0$.  
For simplicity, in the following  we will identify ${\cal H}_0$ with the
SM  electroweak group, ${\cal H}_0 = G_{SM} \equiv SU(2)_L \times U(1)_Y$, so that the SM vectors are the only
elementary gauge fields coupled to the strong sector.

In order to have a composite pNG Higgs boson one has to require two conditions:
\begin{enumerate}
\item The SM  electroweak group $G_{SM}$ must be embeddable in the unbroken subgroup ${\cal H}_1$:
\begin{equation*}
{\cal G} \to {\cal H}_1 \supset G_{SM}
\end{equation*}
\item ${\cal G}/{\cal H}_1$ contains at least one $SU(2)_L$ doublet, to be identified with the Higgs doublet.
\end{enumerate}
If the above two conditions are realized,  at tree level $G_{SM}$ is unbroken and the Higgs doublet is one of the pNG bosons
living on the coset ${\cal G}/{\cal H}_1$.
Its potential vanishes at tree level as a consequence of the non-linear Goldstone symmetry acting on it.
On the other hand, the global symmetry ${\cal G}$ is explicitly broken by 
the couplings of the SM fields to the strong sector, as they will be invariant under $G_{SM}$ but
not in general under ${\cal G}$.
Loops of SM fermions and gauge bosons thus  generate a  Higgs potential, which in 
turn can  break the electroweak symmetry.
In this context the electroweak scale $v$ is dynamically determined and can be smaller than
the sigma-model scale $f$, differently from Technicolor theories where no separation of scale exists. 
The ratio $\xi =(v/f)^2$ is determined by the orientation of  $G_{SM}$ with respect to ${\cal H}$ in the true vacuum (degree of misalignment),
and sets the size of the parametric suppression in all corrections to the precision observables.
By naive dimensional analysis, indeed, the  mass scale of the  resonances of the strong sector
is $m_\rho \sim g_\rho f$, with $1 \lesssim g_{\rho} \lesssim 4\pi$. The Higgs instead gets a much lighter mass at one-loop,
$m_h \sim g_{SM} v$  where $g_{SM}\lesssim 1$ is a generic SM coupling.
The limit $f\to\infty$ ($\xi\to 0$) with fixed $v$
is thus a decoupling limit where the Higgs stays light and all the other resonances become
infinitely heavy.  

Let us explain in detail all the above points by considering an explicit example.

\subsection{An $SO(5)/SO(4)$ example}
\label{sec:SO5example}

Let us consider the case in which the strongly interacting sector has  a global symmetry ${\cal G} = SO(5)\times U(1)_X$
broken down to ${\cal H}_1 = SO(4)\times U(1)_X$~[\refcite{Agashe:2004rs,Contino:2006qr}].~\footnote{For an analysis of the less 
minimal coset $SO(6)/SO(5)$ see Ref.[\refcite{Gripaios:2009pe}].}
In section~\ref{sec:holoHiggs}  we will provide an explicit  example of dynamics that leads to this  pattern of  global symmetries.
As shown in the Appendix,  $SO(4)$ is isomorphic to (that is: it has the same Lie algebra of)  $SU(2)_L\times SU(2)_R$.
The SM electroweak  group $SU(2)_L \times U(1)_Y$ can  be thus embedded into 
$SO(4)\times U(1)_X \sim SU(2)_L\times SU(2)_R \times U(1)_X$, so that  hypercharge is realized as $Y = T^{3R} + X$.
The coset $SO(5)/SO(4)$ implies four real NG bosons transforming as a fundamental of $SO(4)$, 
or equivalently as a complex doublet $H$ of $SU(2)_L$.
The doublet $H$ is the composite Higgs. Under an $SU(2)_R$  rotation it mixes 
with its conjugate $H^c = i\sigma^2 H^*$, so that
$(H,H^c)$ transforms as a bidoublet $(2,2)$ representation of $SU(2)_L\times SU(2)_R$.

Let us derive  the effective action that describes the composite Higgs and the SM elementary fields.
As our final goal is to compute the Higgs potential generated at one-loop by the virtual exchange of SM fields,
we will integrate out the strong dynamics encoding its effects into form factors and  keep terms up to quadratic order in the SM fields.
The four NG bosons living on the coset $SO(5)/SO(4)$ can be parametrized in terms of the linear field $\Sigma$,
\begin{equation}
\Sigma(x) = \Sigma_0 e^{\Pi(x)/f}  \qquad 
\begin{array}{ll}
& \Sigma_0 = (0,0,0,0,1) \\[0.15cm]
& \Pi(x) = -i T^{\hat a} h^{\hat a}(x) \sqrt{2}\, ,
\end{array} 
\end{equation}
where $T^{\hat a}$ are the $SO(5)/SO(4)$ generators.  Using the basis of $SO(5)$ generators given in the Appendix,
one can easily compute the explicit expression of $\Sigma$ in terms of its four real components~$h^{\hat a}$:
\begin{equation}
\label{eq:sigma}
\Sigma = \frac{\sin(h/f)}{h}\, \big(  h^1,  h^2,  h^3,  h^4,  h \cot(h/f) \big) \, , \qquad h \equiv \sqrt{(h^{\hat a})^2}\, .
\end{equation}
The most general effective action for  the SM gauge fields in the background of $\Sigma$ can be derived just 
based on symmetries by using a trick: let us assume that the full $SO(5) \times U(1)_X$ global symmetry
of the strong sector is gauged, so that the external gauge fields form a complete adjoint representation
of $SO(5) \times U(1)_X$.   Then, at the quadratic level and in momentum space, the most general ($SO(5) \times U(1)_X$)-invariant action 
has the form:
\begin{equation}
\label{eq:effSO(5)L}
{\cal L} = \frac{1}{2} (P_T)^{\mu\nu} \left[ \Pi_0^X(q^2)\, X_\mu X_\nu + \Pi_0(q^2)\, \text{Tr}(A_\mu A_\nu) +
 \Pi_1(q^2)\, \Sigma A_\mu A_\nu \Sigma^t \right] \, .
\end{equation}
Here $X_\mu$ and $A_\mu = A_\mu^{a} T^a+A_\mu^{\hat a} T^{\hat a}$ are the $U(1)_X$ and $SO(5)$ gauge bosons~\footnote{Here and in the following 
$T^a$ and $T^{\hat a}$ denote respectively the unbroken ($SO(4)$) and broken 
($SO(5)/SO(4)$) generators. Among the $SO(4)$ generators, those of $SU(2)_L$ ($SU(2)_R$) will be denoted as $T^{a_L}$ ($T^{a_R}$).},
and $P_T$ is the transverse projector defined by eq.(\ref{eq:Wprop}).
Since we want to derive only the Higgs potential and not its derivative interactions,
the field $\Sigma$ has been treated as a classical background, with vanishing momentum. 
The form factors $\Pi^X_0$, $\Pi_{0,1}$ encode the dynamics of the strong sector, including the 
effect of the   fluctuations around the background $\Sigma$ (\textit{i.e.} the NG fields).
A few useful properties of the form factors can be derived as follows.

By expanding around the $SO(4)$-preserving vacuum $\Sigma = \Sigma_0$, the effective action (\ref{eq:effSO(5)L}) can be rewritten as
\begin{equation}
{\cal L} = \frac{1}{2} (P_T)^{\mu\nu} \left[ \Pi_0^X(q^2)\, X_\mu X_\nu + \Pi_a(q^2)\, A^a_\mu A^a_\nu+
\Pi_{\hat a}(q^2)  \, A^{\hat a}_\mu A^{\hat a}_\nu \right] \, ,
\end{equation}
where
\begin{equation}
\label{eq:ff}
\Pi_a = \Pi_0 \, , \qquad\quad \Pi_{\hat a} = \Pi_0 + \frac{\Pi_1}{2} 
\end{equation}
are the form factors associated respectively to the unbroken and broken generators.
In the limit of large number of `colors'  $N$ of the strong sector,   
they can be written in terms of an infinite sum of narrow resonances using the large-$N$ results of section~\ref{sec:TCmodels}:
\begin{align}
(P_T)^{\mu\nu} \Pi_a(q^2) = \langle  J_a^\mu J_a^\nu \rangle =&
 \left(q^2 \eta^{\mu\nu} - q^\mu q^\nu \right) \sum_n \frac{f_{\rho_n}^2}{q^2- m_{\rho_n}^2} \\[0.1cm]
(P_T)^{\mu\nu} \Pi_{\hat a}(q^2) = \langle  J_{\hat a}^\mu J_{\hat a}^\nu \rangle =&
 \left(q^2 \eta^{\mu\nu} - q^\mu q^\nu \right) \left[ \sum_n \frac{f_{a_n}^2}{q^2- m_{a_n}^2} + \frac{1}{q^2} \frac{f^2}{2} \right]\, .
\end{align}
We have used the fact that the current $ J_{\hat a}^\mu$ has the correct quantum numbers to excite the NG bosons $h^{\hat a}$ from
the vacuum. Thus, we deduce that at zero momentum $\Pi_0$ must vanish (and similarly $\Pi_0^X$), while $\Pi_1$ does not:
\begin{equation}
\Pi_0(0) = 0 = \Pi_0^X(0) \, , \qquad \quad \Pi_1(0) = f^2\, .
\end{equation}

At this point we  turn back to the original action (\ref{eq:effSO(5)L}) and  switch off the unphysical gauge fields 
keeping only those of $SU(2)_L\times U(1)_Y$. By using eq.(\ref{eq:sigma}) we obtain:
\begin{equation} \label{eq:holoL}
\begin{split}
{\cal L} = \frac{1}{2} (P_T)^{\mu\nu} \bigg[ 
 &\left( \Pi_0^X(q^2)+\Pi_0(q^2) + \frac{\sin^2(h/f)}{4}\, \Pi_1(q^2) \right) \, B_\mu B_\nu  \\
 & + \left( \Pi_0(q^2) + \frac{\sin^2(h/f)}{4}\, \Pi_1(q^2) \right)  \, A^{a_L}_\mu A^{a_L}_\nu \\
 & + 2 \sin^2(h/f)\, \Pi_1(q^2)  \;
     \hat H^\dagger T^{a_L} Y \hat H\, A_\mu^{a_L} B_\nu \bigg] \, ,
\end{split}
\end{equation}
where $B_\mu$ is the hypercharge field and we defined
\begin{equation} \label{eq:Hhatdef}
\hat H \equiv \frac{1}{h}\, H = \frac{1}{h}  \begin{pmatrix} h^1 - i h^2 \\ h^3 - i h^4 \end{pmatrix}\, .
\end{equation}
This is the  effective action for the SM gauge fields in the background of $\Sigma$ that we were looking for.
By expanding the form factors at momenta small compared to the mass scale of the strong resonances, $q^2\ll m_\rho^2$, one obtains
an effective Lagrangian in terms of local operators. 
Without loss of generality,  one can always perform an $SO(4)$ rotation and 
align the Higgs vev along the $h^3$ direction, so that $(h^1,h^2,h^3,h^4) = (0,0,1,0)$ and $\hat H^t = (0,1)$. 
Hence, at order $q^2$ one has
\begin{equation}
\begin{split}
{\cal L} =  (P_T)^{\mu\nu} \bigg\{
   & \frac{1}{2} \left( \frac{f^2 \sin^2(\langle h \rangle/f)}{4} \right) \left( B_\mu B_\nu + W^3_\mu W^3_\nu -2 W^3_\mu B_\nu \right) \\
   & + \left( \frac{f^2 \sin^2(\langle h \rangle/f)}{4} \right)  W^+_\mu W^-_\nu  \\
   & + \frac{ q^2}{2} \Big[ \Pi_0^\prime(0)\, W^{a_L}_\mu W^{a_L}_\nu + \left(  \Pi_0^\prime(0) +  \Pi_0^{X\, \prime}(0) \right) B_\mu B_\nu \Big]
 + \dots \bigg\}
\end{split}
\end{equation}
where $\Pi^\prime$ denotes the first derivative of $\Pi$ with respect to $q^2$.
From the above Lagrangian we can thus identify 
\begin{equation}
\label{eq:gfromstrong}
\frac{1}{g^2} = -\Pi_0^\prime(0) \, , \qquad \ \frac{1}{g^{\prime 2}} = -\left( \Pi_0^\prime(0) + \Pi_0^{X\, \prime}(0)  \right)
\end{equation}
and
\begin{equation} \label{eq:defxi}
v = f \,\sin \frac{\langle h\rangle}{f}\, , \qquad \text{so that} \quad \ \xi \equiv \frac{v^2}{f^2} = \sin^2 \frac{\langle h\rangle}{f}\, .
\end{equation}
Notice that the formulas in eq.(\ref{eq:gfromstrong}) show  the  contribution to the low-energy gauge couplings from the strong dynamics only.
If one adds to the effective action (\ref{eq:effSO(5)L}) bare kinetic terms for the external $SU(2)_L\times U(1)_X$ fields, the expressions for $g$ 
and $g^\prime$ will be modified to
\begin{equation}
\frac{1}{g^2} = -\Pi_0^\prime(0) +\frac{1}{g_0^2} \, , \qquad \
 \frac{1}{g^{\prime 2}} = -\left( \Pi_0^\prime(0) + \Pi_0^{X\, \prime}(0)  \right) + \frac{1}{g_0^{\prime 2}} \, .
\end{equation}

Starting from eq.(\ref{eq:holoL}) it is simple to derive the couplings of the physical Higgs boson to the gauge fields. By 
expanding around the vev $\langle h \rangle$,
\begin{equation}
h^{\hat a} = \begin{pmatrix} 0 \\ 0 \\ \langle h \rangle + h \\ 0 \end{pmatrix}\, ,
\end{equation}
one has
\begin{equation}
\begin{split}
f^2 \sin^2 \frac{h}{f} = & f^2 \bigg[ \sin^2\frac{\langle h\rangle}{f} + 2\sin\frac{\langle h\rangle}{f}\cos\frac{\langle h\rangle}{f} 
 \left( \frac{h}{f} \right) \\
 & \phantom{f^2 \bigg[} + \left(  1- 2\sin^2\frac{\langle h\rangle}{f} \right) \left( \frac{h}{f} \right)^2 + \dots \bigg] \\[0.2cm]
 = & v^2 + 2 v \sqrt{1-\xi} \, h + (1-2\xi) \, h^2 + \dots
\end{split}
\end{equation}
where, with a slight abuse of notation, $h$ stands for $\sqrt{h^{\hat a} h^{\hat a}}$ on the left hand side, while it denotes the physical
Higgs boson on the right hand side. Compared to their SM prediction, the couplings of the composite Higgs to the  gauge bosons 
$V=W,Z$ are thus modified  as follows:
\begin{equation} \label{eq:modcoupl}
g_{VVh} = g_{VVh}^{SM} \sqrt{1-\xi} \, , \quad \qquad g_{VVhh} = g_{VVhh}^{SM}  (1-2\xi)\, .
\end{equation}
If one compares with the effective Lagrangian for a generic scalar eq.(\ref{eq:CHLag}), one finds that the $SO(5)/SO(4)$ theory
predicts  
\begin{equation}
\label{eq:ab}
a=\sqrt{1-\xi}\, , \qquad b=1-2\xi\, .
\end{equation}
Using the results of section~\ref{sec:Higgsmodel}  on the $WW$ scattering, we  deduce that both the
$WW\to WW$ and $WW\to hh$  scattering amplitudes grow as $\sim(E/v)^2\xi $ at large energies, violating perturbative unitarity 
at a scale $\Lambda \approx 4\pi v/\sqrt{\xi}$. This is a factor $\sqrt{\xi}$ larger than what we found for a theory with no Higgs.

We see that  the composite Higgs  only \textit{partly} unitarizes the scattering amplitudes, 
simply postponing the loss of perturbative unitarity to larger scales. In the limit $\xi\to 0$ (with $v$ fixed) one recovers the 
standard Higgs model: the resonances of the strong sector become infinitely heavy and decouple, while the Higgs boson fully unitarizes 
the theory.  For $\xi \to 1$, on the other hand,  the Higgs contribution vanishes and unitarity in $WW\to WW$ scattering is  enforced 
solely by the strong resonances. Furthermore,  $f=v$ and there is no gap of scales in theory: in this limit the strong dynamics behaves
quite similarly to a minimal Technicolor theory, although a light scalar exists in the spectrum.
In the general case, for $\xi$ small enough the strong resonances can be made  sufficiently heavy  and their correction to the
electroweak observables sufficiently small to pass the LEP precision tests.
We will illustrate this point in detail later on, in section~\ref{sec:CHandEWPT}, as we are now ready to derive the 
Coleman-Weinberg  potential for the composite Higgs.

\begin{figure}[t]
\centering
%\fbox{ 
\begin{minipage}[t]{0.16\linewidth}
  \vspace{0pt}
  \centering
  \includegraphics[scale=0.65]{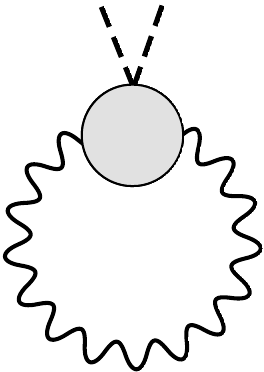} 
\end{minipage} 
%}
\hspace{0.03cm} 
\begin{minipage}[t]{7pt}
  \vspace{1.6cm}
  \centering
  {$\mathbf{+}$}
\end{minipage} 
\hspace{0.03cm} 
\begin{minipage}[t]{0.16\linewidth}
  \vspace{0pt}
  \centering
  \includegraphics[scale=0.65]{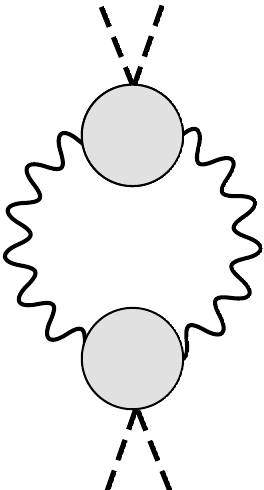} 
\end{minipage} 
\hspace{0.03cm} 
\begin{minipage}[t]{7pt}
  \vspace{1.6cm}
  \centering
  {$\mathbf{+}$}
\end{minipage} 
\hspace{0.03cm} 
\begin{minipage}[t]{0.3\linewidth}
  \vspace{0pt}
  \centering
  \includegraphics[scale=0.65]{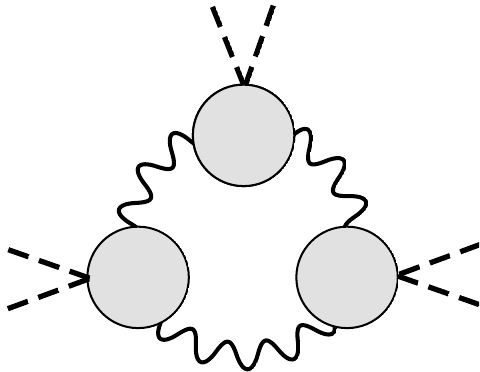} 
\end{minipage} 
\hspace{0.03cm} 
\begin{minipage}[t]{12pt}
  \vspace{1.6cm}
  \centering
  {$\mathbf{ + \; \cdots}$}
\end{minipage} 
\caption{1-loop  contribution of the SM gauge fields to the Higgs potential. A grey blob represents the strong dynamics encoded by the 
form factor $\Pi_1$.
}
\label{fig:Hpotdiags}
\end{figure}

We will concentrate on the  contribution from the $SU(2)_L$ gauge fields, neglecting the smaller correction from hypercharge.
The contribution from fermions will be derived in section~\ref{sec:fermionpot}.
The 1-loop Coleman-Weinberg potential resums the class of diagrams in Fig.~\ref{fig:Hpotdiags}.
From the effective action~(\ref{eq:holoL}), after the addition of the gauge-fixing term 
\begin{equation}
{\cal L}_{GF} = - \frac{1}{2g^2 \zeta} \left(\partial^\mu A_\mu^{a_L} \right)^2\, ,
\end{equation}
it is easy to derive the  Feynman rules for the gauge propagator and vertex:
\\[0.4cm]
\begin{minipage}{0.45\linewidth}
\centering
\includegraphics[width=28mm]{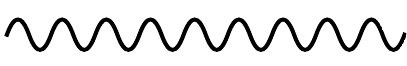}
\end{minipage}
\hspace{0.3cm}
$\displaystyle G_{\mu\nu} =  \frac{i}{\Pi_0(q^2)} (P_T)_{\mu\nu} - \zeta \frac{ig^2}{q^2} (P_L)_{\mu\nu}$ 
\\[0.8cm]
\begin{minipage}{0.45\linewidth}
\centering
\vspace{-0.6cm}
\includegraphics[width=28mm]{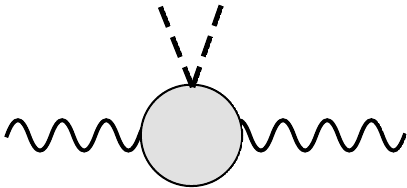}
\end{minipage}
\hspace{0.3cm}
$\displaystyle  i\Gamma_{\mu\nu} =  \frac{i\Pi_1(q^2)}{4}  \sin^2(h/f) (P_T)_{\mu\nu}$ 
\\[0.5cm]
where $(P_L)_{\mu\nu} = q_\mu q_\nu/q^2$ is the longitudinal projector.
Resumming  the series of 1-loop diagrams of Fig.~\ref{fig:Hpotdiags} then leads to the potential:
\begin{equation}
\label{eq:gaugepotential}
V(h) = \frac{9}{2} \int\! \frac{d^4Q}{(2\pi)^4} \, \log\left( 1+ \frac{1}{4} \frac{\Pi_1(Q^2)}{\Pi_0(Q^2)} \, \sin^2(h/f) \right)
\end{equation}
where $Q^2=-q^2$ is the Euclidean momentum. The factor $9$ originates from the sum over three Lorentz polarizations
and three $SU(2)_L$ flavors. 

Let us argue on the behavior of the form factors at large Euclidean momentum and on the convergence of the integral.
We have seen that  $\Pi_0$ is related to the product of two $SO(4)$ currents
\begin{equation}
\langle J_\mu^a(q) J^a_\nu (-q)\rangle = \Pi_0(q^2) (P_T)_{\mu\nu}
\end{equation}
where, we recall, the notation $\langle O_1 O_2 \rangle$ denotes the vacuum expectation of the time-ordered product
of the operators $O_1$ and $O_2$. The form factor $\Pi_1$, on the other hand, is given by the difference (see eq.(\ref{eq:ff}))
\begin{equation}
\langle J_\mu^a(q) J^a_\nu (-q) \rangle - \langle J_\mu^{\hat a}(q) J^{\hat a}_\nu (-q) \rangle = -\frac{1}{2} \Pi_1(q^2) (P_T)_{\mu\nu}\, .
\end{equation}
At energies much above the scale of symmetry breaking, the $SO(5)$ invariance is restored, and the difference of two-point
functions along broken and unbroken directions is expected to vanish.
In this sense $\Pi_1$ is an \textit{order parameter}: it is sensitive
to the  symmetry-breaking IR dynamics, and it vanishes at large momenta.
If $\Pi_1$ goes to zero fast enough, the integral in eq.(\ref{eq:gaugepotential}) will be convergent and the Higgs potential finite.
This agrees with the intuition that if the Higgs is a bound state of the strong dynamics, then its mass cannot receive
corrections larger than the compositeness scale. To support this intuition with a more rigorous argument, let us consider
the Operator Product Expansion (OPE) of two currents.

Following Wilson, the time-ordered product of two  operators $A(x_1)$, $B(x_2)$ can be expressed as an infinite sum
of local operators of increasing dimension multiplied by  coefficients that depend on the separation $(x_1-x_2)$:
\begin{equation}
T\left\{ A(x) B(0) \right\} = \sum_n C_{12}^{(n)}(x) O_n(0)\, .
\end{equation}
The equality is at the level of operators, thus implying the equality  of any Green function made of them.
The sum extends over all  operators with the same global symmetries of the product $A B$.
In particular, the OPE of the product of two conserved currents $J_\mu$ reads, in momentum space:
\begin{equation} \label{eq:JJOPE}
i\int\! d^4x \, e^{iq\cdot x}\, T\left\{ J_\mu(x) J_\nu(0) \right\} = (q^2 \eta_{\mu\nu} - q_\mu q_\nu) \sum_n C^{(n)}(q^2) O_n(0) \, .
\end{equation}
By dimensional analysis, the larger is the dimension of the operator $O_n$, the more suppressed is its coefficient at large 
Euclidean momenta $Q^2 = -q^2$:
\begin{equation}
C^{(n)}(Q^2) \sim \frac{1}{Q^{[O_n]}} \qquad \  \text{for } Q \text{ large.} 
\end{equation}
The convergence of the integral in the Higgs potential then requires that the first operator to contribute to the 
difference of the product of $SO(4)$ and $SO(5)/SO(4)$ currents must have dimension 5 or greater:
\begin{equation}
\langle J_\mu^a(q) J^a_\nu (-q) \rangle - \langle J_\mu^{\hat a}(q) J^{\hat a}_\nu (-q) \rangle = (q^2 \eta_{\mu\nu} - q_\mu q_\nu) 
 \left[ C^{(5)}(q^2) \langle O_5 \rangle + \dots \right]
\end{equation}
so that
\begin{equation}
\label{eq:convcond}
\Pi_1(Q^2) \sim \frac{1}{Q^{n-2}}  \qquad n\geq 5 \qquad \text{for } Q^2\to \infty\, .  
\end{equation}
This makes use of the fact that $\Pi_0$ grows at least as $Q^2$ at large momenta, see eq.(\ref{eq:gfromstrong}).

Clearly, without knowing the details of the strong dynamics we cannot say more about the behavior of the form factors,
nor  can we prove that the condition (\ref{eq:convcond}) is satisfied in general.  There is however a similar physical situation
where we have enough experimental and theoretical information to  reconstruct the OPE and deduce the convergence of the
integral: this is the case of the electromagnetic correction to the pion mass.

\subsection{Comparing with QCD: the pion potential}

Let us consider QCD in the chiral limit, so that the pion is an exact NG boson at tree level, and turn on the electromagnetic interaction.
Differently from the cartoon of Fig.~\ref{fig:QCDpicture}, where the full $SU(2)_L\times U(1)_Y$ symmetry was gauged,  
in this case the external  $U(1)_{em}$ group can be embedded into the unbroken subgroup $SU(2)_V$. 
This means that the pion is not eaten to form a massive photon, but remains in the spectrum as a pseudo NG boson.
At the radiative level,  diagrams with loops of the elementary photon will generate a potential and a mass term for the charged pion, 
while the neutral pion remains massless. This is in complete analogy to the composite Higgs theory considered in the previous section,
although in the case of QCD we dispose of much more detailed information on the strong dynamics. Following the same steps as we did
for the case of the composite Higgs, we can derive the effective action for the pion  and compute its potential.
\footnote{Much of the material of this section is reviewed, for example, in Ref.[\refcite{deRafael:1997ea}]. See also 
the original papers [\refcite{SVZ}] and [\refcite{Knecht:1997ts}].}

In order to write down the effective action that describes the photon and the pion we use the same trick of section~\ref{sec:SO5example}
and assume that the whole $SU(2)_L \times SU(2)_R$ chiral invariance of QCD is gauged by external fields.
Treating the pion field $\Sigma$ (see eq.(\ref{eq:Lpion})) as a constant classical background, the most general
$(SU(2)_L \times SU(2)_R)$-invariant action, in momentum space and at the quadratic order in the gauge fields, is
\begin{equation} \label{eq:pionaction}
\begin{split}
{\cal L} = \frac{1}{2}  (P_T)_{\mu\nu} \Big[  & \Pi_L(q^2)\, \text{Tr}\left\{ L_\mu L_\nu \right\} +  \Pi_R(q^2)\, \text{Tr}\left\{ R_\mu R_\nu \right\} \\
 & - \Pi_{LR}(q^2)\,  \text{Tr}\left\{ \Sigma^\dagger L_\mu \Sigma R_\nu \right\}  \Big]\, .
\end{split}
\end{equation}
Here $L_\mu$, $R_\mu$ are the external gauge fields associated respectively to $SU(2)_L$ and $SU(2)_R$ transformations.
Since in the vacuum $\langle \Sigma\rangle =1$ the chiral $SU(2)_L \times SU(2)_R$ symmetry is broken down to $SU(2)_V$,
it is useful to rewrite the left and right gauge fields in term of vectorial and axial ones:
\begin{equation}
V_\mu = \frac{1}{\sqrt{2}} \left( R_\mu + L_\mu \right)\, ,  \quad A_\mu = \frac{1}{\sqrt{2}} \left( R_\mu - L_\mu \right)\, .
\end{equation}
In the $\langle \Sigma\rangle =1$ vacuum the effective action thus reads
\begin{equation}
\begin{split}
{\cal L} = \frac{1}{2}  (P_T)_{\mu\nu} \Big[ & \Pi_{VV}(q^2)\, \text{Tr}\left\{ V_\mu V_\nu \right\} +  \Pi_{AA}(q^2)\, \text{Tr}\left\{ A_\mu A_\nu \right\} \\
 & + \Pi_{VA}(q^2)\, \text{Tr}\left\{ V_\mu A_\nu + A_\mu V_\nu  \right\}  \Big] \, .
\end{split}
\end{equation}
where we have defined
\begin{equation}
\begin{split}
\Pi_{VV} =& \frac{1}{2} \left( \Pi_L + \Pi_R - \Pi_{LR} \right) \\[0.1cm]
\Pi_{AA} =& \frac{1}{2} \left( \Pi_L + \Pi_R + \Pi_{LR} \right) \\[0.1cm]
 \Pi_{VA} =& \frac{1}{2} \left( \Pi_R - \Pi_L \right)\, .
\end{split}
\end{equation}
We know that only the axial current has the right quantum numbers to excite a pion from the vacuum, so we expect that only
the $\langle A_\mu A_\nu \rangle$ correlator has a pole at $q^2=0$ from the pion exchange.
This implies that at zero momentum the form factors $\Pi_{VV}$ and $\Pi_{VA}$ vanish, whereas $\Pi_{AA}(0)=f_\pi^2$.
Equivalently,
\begin{equation}
\Pi_{LR}(0) = 2\, \Pi_L(0) = 2\, \Pi_R(0) = f_\pi^2 \, .
\end{equation}

At this point we turn back to the effective action (\ref{eq:pionaction}) and switch off all the external gauge fields but the photon: we set
$L_\mu(x) = T^3 v_\mu(x) = R_\mu(x)$.
Using
\begin{equation}
\begin{gathered}
\Sigma = \exp\left( i \sigma^a \pi^a/f_\pi\right) = 1 \cos\left(\pi/f_\pi \right) + i\, \hat\pi^a \sigma^a \sin\left(\pi/f_\pi \right) \\[0.1cm]
\pi \equiv \sqrt{(\pi^a)^2}\, , \qquad \hat\pi^a = \frac{\pi^a}{\pi}\, ,
\end{gathered}
\end{equation}
we obtain
\begin{equation}
{\cal L} = \frac{1}{2}  (P_T)_{\mu\nu} v^\mu v^\nu \left[ \frac{1}{2} \left(\Pi_L(q^2) + \Pi_R(q^2) \right) -
 \Pi_{LR}(q^2)\,  \text{Tr}\left\{ \Sigma^\dagger T^3 \Sigma T^3 \right\}  \right] \, .
\end{equation}
After a bit of algebra one finds ($\pi^+ \pi^- \equiv  (\pi_1)^2 + (\pi_2)^2$)
\begin{equation}
\text{Tr}\left\{ \Sigma^\dagger T^3 \Sigma T^3 \right\} = \frac{1}{2} - \frac{\sin^2(\pi/f_\pi)}{\pi^2} (\pi^+\pi^-)\, ,
\end{equation}
hence
\begin{equation}
{\cal L} = \frac{1}{2}  (P_T)_{\mu\nu} v^\mu v^\nu \left[ \Pi_{VV}(q^2)  +
 \Pi_{LR}(q^2)\,  \frac{\sin^2(\pi/f_\pi)}{\pi^2} (\pi^+\pi^-) \right] \, .
\end{equation}
As expected the neutral pion does not couple to the photon at the quartic level, although there can be interactions involving
both the neutral and the charged pion. The 1-loop diagrams associated to the Coleman-Weinberg potential are these same
as those in Fig.~\ref{fig:Hpotdiags}.  Their resummation gives
\begin{equation}
V(\pi) = \frac{3}{16\pi^2} \int_0^\infty \!\! dQ^2\,  Q^2 
 \log\left( 1 + \frac{1}{2} \, \frac{\Pi_{LR}(Q^2)}{\Pi_{VV}(Q^2)} \, \frac{\sin^2(\pi/f_\pi)}{\pi^2} (\pi^+\pi^-) 
\right)\, .
\end{equation}
The convergence of the integral thus depends on the behavior of the form factors $\Pi_{LR}(Q^2)$ and $\Pi_{VV}(Q^2)$ at large Euclidean
momenta $Q^2$. To infer such behavior we can use the information that comes from the OPE of the product of two vector and axial currents,
see eq.(\ref{eq:JJOPE}).  The  color-singlet, scalar~\footnote{Operators of spin $1/2$ and higher do not contribute to the  vacuum expectation value
 $\langle J_\mu J_\nu \rangle$ and are thus irrelevant to the following argument.} operators of dimension 6 or less are:
\begin{center}
\begin{tabular}{lc}
$1$ \ (identity operator) \  \ \ \ & (d=0) \\[0.2cm]
$O_m = \bar\psi m_q \psi$ & (d=4) \\[0.2cm]
$O_G = G_{\mu\nu}^a G^{a\, \mu\nu}$ & (d=4) \\[0.2cm]
$O_\sigma = \bar\psi \sigma^{\mu\nu} t^a m_q \psi G^a_{\mu\nu}$  & (d=6) \\[0.2cm]
$O_\Gamma = \left(\bar\psi \Gamma_1 \psi\right)\left(\bar\psi \Gamma_2 \psi\right)$ & (d=6) \\[0.2cm]
$O_f = f^{abc} G^{a\, \mu}_\nu G^{b\, \nu}_\rho G^{c\, \rho}_\mu$ & (d=6) 
\end{tabular}
\end{center}
where $a,b,c$ are color indices and $\Gamma_{1,2}$ are matrices in flavor, color and Lorentz space.
Notice that the operators $O_m$ and $O_\sigma$ break explicitly the chiral symmetry and must be thus proportional
to the quark mass matrix $m_q$.  As such they vanish in the chiral limit.
On the other hand $O_\Gamma$ is the only chiral-invariant operator among those listed above whose vacuum expectation value
can violate the chiral  symmetry and thus distinguish between the axial and vector currents.
In other words,  $O_\Gamma$ is the  operator with lowest dimension to contribute to the form factor $\Pi_{LR}$:
\begin{equation} \label{eq:PiLRbehavior}
\Pi_{LR}(Q^2) = Q^2\,  C_{O_\Gamma}(Q^2) \langle O_\Gamma \rangle + \dots = Q^2 \left( \frac{\delta}{Q^6} + O\left(\frac{1}{Q^8}\right) \right) \, ,
\end{equation}
where $\delta$ is a numerical coefficient.
\footnote{The coefficient $\delta$ can be computed perturbatively expanding in powers of $\alpha_s$ and $1/N_c$. 
In the large $N_c$ limit, the matrix element $\langle O_\Gamma \rangle$ factorizes into $(\langle \bar\psi \psi\rangle)^2$, and one finds:
$\delta = 8\pi^2 \left( \alpha_s/\pi + O(\alpha_s^2) \right)(\langle \bar\psi \psi\rangle)^2$~[\refcite{SVZ,Knecht:1997ts}].}
Since the form factor $\Pi_{VV}$ grows as $Q^2$ at large Euclidean momenta (the leading term in its expansion
corresponds to the kinetic term of the photon), we deduce that the integral in the pion potential is convergent.
A reasonable approximation to the full potential is obtained by setting $\Pi_{VV}(Q^2)\simeq Q^2/e^2$ and expanding the logarithm
at first order:
\begin{equation} \label{eq:pionpotential}
V(\pi) \simeq \frac{3}{8\pi^2} \alpha_{em} \frac{\sin^2(\pi/f_\pi)}{\pi^2} (\pi^+\pi^-)  \int_0^\infty \!\! dQ^2\,  \Pi_{LR}(Q^2) \, .
\end{equation}

The information of the OPE  on the asymptotic behavior of $\Pi_{LR}$ allows us to proceed further and compute the integral explicitly 
provided we make two approximations: the large $N_c$ limit and vector meson dominance.
At leading order in $1/N_c$ the product of two vector or axial currents can be written in terms of an infinite sum of resonances poles, so that
\begin{equation}
\begin{split}
\Pi_{VV}(Q^2) =&  Q^2 \sum_n \frac{f_{\rho_n}^2}{Q^2+ m_{\rho_n}^2} \\[0.1cm]
\Pi_{AA}(Q^2) =& Q^2  \left[ \sum_n \frac{f_{a_n}^2}{Q^2+ m_{a_n}^2} + \frac{f_\pi^2}{Q^2}\right]\, .
\end{split}
\end{equation}
Given that
\begin{equation}
\Pi_{LR}(q^2) = \Pi_{AA}(q^2) - \Pi_{VV}(q^2)\, ,
\end{equation}
the large-$Q$ behavior that follows from eq.(\ref{eq:PiLRbehavior}),
\begin{equation}
\Pi_{LR}(Q^2) \propto \frac{1}{Q^4} + O\left( \frac{1}{Q^6}\right) \quad \Longrightarrow  \quad
\begin{cases}  
\displaystyle \lim_{\substack{Q^2\to\infty}} \Pi_{LR}(Q^2) = 0 \\[0.4cm]
\displaystyle \lim_{Q^2\to\infty} Q^2\, \Pi_{LR}(Q^2) = 0\, ,
\end{cases}
\end{equation}
implies two \textit{sum rules} on the spectrum of masses and decay constants of the strong resonances:
\begin{align}
\label{eq:1stWSR}
& \sum_n \Big[ f_{\rho_n}^2 - f_{a_n}^2 \Big] = f_\pi^2 \\[0.15cm]
& \sum_n \Big[ f_{\rho_n}^2 m_{\rho_n}^2 - f_{a_n}^2 m_{a_n}^2 \Big] = 0 \, .
\end{align}
These relations where first derived by Weinberg~[\refcite{Weinberg:1967kj}], and are known  respectively as his first and second sum rules.

The vector meson dominance approximation then consists in assuming that the dominant contribution to these relations,
as well as to other observables,  comes from the first vector and axial resonances (the $\rho$ and the $a_1$).
By neglecting the higher resonances and saturating the two Weinberg sum rules with the $\rho$ and the $a_1$ we then obtain
\begin{align}
f_{\rho}^2 & = f_\pi^2\, \frac{m_{a_1}^2}{m_{a_1}^2 - m_{\rho}^2} \\[0.1cm]
f_{a_1}^2 & = f_\pi^2\, \frac{m_{\rho}^2}{m_{a_1}^2 - m_{\rho}^2} \, ,
\end{align}
and the $\Pi_{LR}$ form factor can be written as
\begin{equation} \label{eq:PiLRanalytic}
\Pi_{LR}(Q^2) \simeq f_\pi^2 \, \frac{m_{a_1}^2 m_\rho^2}{(Q^2 + m_{a_1}^2)(Q^2 + m_{\rho}^2)}\, .
\end{equation}
Using the above expression of $\Pi_{LR}$, the integral appearing in the pion potential gives
\begin{equation} \label{eq:int}
\int_0^\infty \!\! dQ^2\,  \Pi_{LR}(Q^2)  = f_\pi^2\, \frac{m_\rho^2 m_{a_1}^2}{m_{a_1}^2-m_\rho^2}\,  \log\left(\frac{m_{a_1}^2}{m_\rho^2}\right)\, .
\end{equation}
For any value of the masses, the above expression is always positive (reflecting the positivity of $\Pi_{LR}$ in eq.(\ref{eq:PiLRanalytic})). 
This means that the  pion potential is minimized  for 
\begin{equation} \label{eq:pionvacuum}
\langle \pi^1 \rangle = \langle \pi^2 \rangle = 0\, .
\end{equation}
In other words, the radiative corrections align the vacuum along the $U(1)$-preserving  direction, and the photon remains massless.
It turns out that the positivity of the integral (\ref{eq:int}) and the above conclusion on the alignment of the vacuum 
are much more general that our approximate result. 
Witten~[\refcite{Witten:1983ut}] has shown that in a generic vector-like confining gauge theory  one has
\begin{equation}
\Pi_{LR}(Q^2) \geq 0  \qquad \text{for} \qquad 0 \leq Q^2 \leq \infty\, ,
\end{equation}
so that the radiative contribution from gauge fields always tends to align the vacuum in the direction that preserves the gauge symmetry.

The effect of the  one-loop potential (\ref{eq:pionpotential}) is  that of lifting the degeneracy of vacua and 
give a (positive) mass to the charged pion, while leaving the neutral one massless.  
Notice indeed that the potential vanishes in the vacuum (\ref{eq:pionvacuum}), so that there is still a flat direction along $\pi^0$. 
All the results derived above are valid in the chiral limit, that is for vanishing quark masses. When  the quark masses is turned on, 
both the charged and neutral pion get a mass, as a consequence of the explicit breaking of the chiral symmetry.
The \textit{difference} of the charged and neutral pion mass, however, is still dominantly accounted for by the electromagnetic
correction that we have derived. Thus, we can compare our prediction with the experimentally measured value and check the
accuracy of our approximations.
From eqs.(\ref{eq:pionpotential}) and (\ref{eq:int}) one gets
\begin{equation} \label{eq:dmpi}
m_{\pi^\pm}^2 - m_{\pi_0}^2 \simeq \frac{3\, \alpha_{em} }{4\pi} \,
\frac{m_\rho^2 m_{a_1}^2}{m_{a_1}^2-m_\rho^2}\,  \log\left(\frac{m_{a_1}^2}{m_\rho^2}\right)\, .
\end{equation}
This result was first derived in 1967 by Das et al. using current algebra techniques~[\refcite{Das:1967it}].
Inserting the experimental values $m_\rho=770\,$MeV and $m_{a_1}=1260\,$MeV into eq.(\ref{eq:dmpi}) one obtains the theoretical prediction
\begin{equation}
(m_{\pi^\pm} - m_{\pi_0})|_\text{TH} \simeq 5.8\,\text{MeV}\, ,
\end{equation}
to be compared with the experimentally measured value
\begin{equation}
(m_{\pi^\pm} - m_{\pi_0})|_\text{EXP} \simeq 4.6\,\text{MeV}\, .
\end{equation}
Considering that corrections to the  large-$N_c$ approximation are expected to be of order $\sim 30\%$, we
conclude that the agreement of our theoretical prediction with the experimental value is fully satisfactory.

As an exercise useful for the following, we also derive the prediction for the chiral coefficient $L_{10}$ under the same
assumptions that led to eq.(\ref{eq:dmpi}).   $L_{10}$ is defined in terms of the difference of the derivative of the axial and
vector form factors at zero momentum:
\begin{equation}
- 4 L_{10} \equiv \Pi_{AA}^\prime(0) - \Pi_{VV}^\prime(0) = \Pi_{LR}^\prime(0) \simeq \frac{f_\rho^2}{m_\rho^2} - \frac{f_{a_1}^2}{m_{a_1}^2}\, .
\end{equation}
Under the assumption of vector meson dominance, the first Weinberg sum rule, eq.(\ref{eq:1stWSR}), requires $f_{a_1} < f_\rho$, 
and we know  experimentally that $m_{a_1} > m_\rho$.
This implies that the sign of $L_{10}$ is fixed to be negative. Using both the Weinberg sum rules one obtains
\begin{equation} \label{eq:L10}
- 4 L_{10} \simeq \frac{f_\pi^2}{m_\rho^2} \left( 1+ \frac{m_\rho^2}{m_{a_1}^2} \right)\, .
\end{equation}

\subsection{Electroweak precision tests and flavor constraints in composite Higgs models}
\label{sec:CHandEWPT}

Having discussed the QCD example  in detail, we now turn back to the case of the composite Higgs.
We will assume that the form factor $\Pi_1(Q^2)$ goes to zero fast enough for $Q^2\to\infty$, so that
the integral in eq.(\ref{eq:gaugepotential}) is convergent.
As for the pion potential, we can expand the logarithm at first order and approximate $\Pi_0(Q^2)\simeq Q^2/g^2$ to obtain 
\begin{equation} 
V(h) = \frac{9}{8}\frac{g^2}{16\pi^2}  \sin^2(h/f)  \int_0^\infty \!\! dQ^2\,  \Pi_{1}(Q^2) \, .
\end{equation}
According to Witten's argument on vector-like gauge theories~[\refcite{Witten:1983ut}], we expect that the above integral is positive and that the 1-loop
gauge contribution to the potential aligns the vacuum in an $SU(2)_L$-preserving direction: $v/f = \sin\langle h \rangle/f =0$.
This is indeed verified  in explicit models, see for example Refs.[\refcite{Agashe:2004rs,Contino:2006qr}].

In the context of the original composite Higgs models, various  solutions have been proposed   to solve this problem.
For example, in their first paper Georgi and Kaplan make use of an additional elementary scalar that mixes with the composite 
Higgs~[\refcite{GK1}], while in the model of Ref.~[\refcite{GK2}] the vacuum is misaligned by the explicit breaking of the global symmetry 
mediated at a higher scale by the exchange of  (extended ultracolor) heavy vectors.
A more attractive mechanism, which does not rely on the existence of elementary scalars or
any hard  breaking of the global symmetry, has been proposed by  Banks~[\refcite{Banks:1984gj}] and subsequently implemented
in Refs.[\refcite{GK3,GK4,GK5}]. The idea is that of 
enlarging the external gauge group to include an additional axial $U(1)_A$
and designing the pattern of global symmetry breaking $\mathcal{G} \to \mathcal{H}_1$ such that
while the electroweak $SU(2)_L\times U(1)_Y$ can be embedded in the unbroken subgroup $\mathcal{H}_1$, 
the full $SU(2)_L\times U(1)_Y\times U(1)_A$ cannot.
The impossibility of preserving the full gauge group implies that  the 1-loop contribution from the $U(1)_A$ vector boson to the potential 
necessarily destabilizes the ($SU(2)_L\times U(1)_Y$)-symmetric vacuum, leading to  $(v/f)^2  =\xi \not = 0$.
In such models, the degree of vacuum misalignment $\xi$ depends on the ratio of the $U(1)_A$ and $SU(2)_L$ gauge couplings, $g_A/g$.

More recently, it has been shown that 
the SM top quark contribution can also misalign the vacuum 
and break the electroweak symmetry in a natural way~[\refcite{Agashe:2004rs}]. 
This will be discussed in detail in section~\ref{sec:fermionpot}, where we compute the contribution
of the SM fermions to the Higgs potential.
Here we want to discuss how  composite Higgs theories face the electroweak
precision tests of LEP and the constraints from FCNC processes.
We will thus assume that  some other contribution to the Higgs potential exists, 
for example (though not necessarily) coming from the SM top quark, 
which triggers the EWSB and gives $\xi \not = 0$.  
As before we will analyze the $SO(5)/SO(4)$ model, although the results
that we will derive are generic.

Let us consider first the correction to the Peskin-Takeuchi $S$ parameter.
According to its definition (\ref{eq:defS}), $S$
is given by the following term in the
expansion of the effective action (\ref{eq:holoL}) at small momenta:
\begin{equation}
{\cal L} \supset \frac{1}{2} \Pi_1^\prime(0) \sin^2\left( h/f \right) 
 W^{a_L}_{\mu\nu} B^{\mu\nu}\, \hat H^\dagger T^{a_L} Y \hat H \ \to\  -\frac{\xi}{8}  \Pi_1^\prime(0)\, W_{\mu\nu}^3 B^{\mu\nu} \, . 
\end{equation}
In analogy with the previous section, in the limit in which the number of colors $N$ of the strong sector is large we can use the
results of section \ref{sec:TCmodels} and write the form factor $\Pi_1$ in terms of a sum over resonances. 
This gives
\begin{equation} \label{eq:SlargeN}
S = 2\pi \xi\, \Pi_1^\prime(0) = 4 \pi \xi \sum_n \left( \frac{f_{\rho_n}^2}{m_{\rho_n}^2} - \frac{f_{a_n}^2}{m_{a_n}^2} \right)\, ,
\end{equation}
where  $m_{\rho_n}$ and $f_{\rho_n}$ ($m_{a_n}$ and $f_{a_n}$)  denote respectively the mass and decay constant of the $SO(4)$ ($SO(5)/SO(4)$)
spin-1 resonances.
Equation (\ref{eq:SlargeN}) represents the large-$N$ leading contribution to $S$  from the strong dynamics, 
interpreted as due to the tree-level exchange of spin-1 resonances.
It is clear at this point the strict analogy with the  chiral coefficient $L_{10}$ that we computed in the previous section
for the case of QCD. Indeed,  by using the Weinberg sum rules and assuming vector meson dominance we obtain 
\begin{equation}
S = 4 \pi \xi \, \frac{f^2}{m_\rho^2}  \left( 1 + \frac{m_\rho^2}{m_a^2} \right)\, ,
\end{equation}
which is in complete analogy with the expression of $L_{10}$ in eq.(\ref{eq:L10}). In particular,  the sign of $S$ is fixed to be positive.
By means of the  large-$N$ relation $f/m_\rho \sim g_\rho \sim 4\pi/\sqrt{N}$ (where $g_\rho$ is the coupling among three composite states,
see eqs.(\ref{eq:mrhoNDA})-(\ref{eq:grhoNDA})), 
it is also easy to see that its size is parametrically suppressed by a factor $\xi$ compared to the Technicolor estimate of eq.(\ref{eq:SNDA}).
This was in fact expected, considering that  $\xi \to 0$ with fixed $v$ is a  limit in which all the
resonances of the strong sector except the Higgs become infinitely heavy and decouple.
Hence, for $\xi$ small enough the LEP constraints can be satisfied.
If we use  eq.(\ref{eq:defxi}) and the relation $m_\rho/m_a \simeq 3/5$ valid in the 5-dimensional $SO(5)/SO(4)$ models 
of Refs.~[\refcite{Agashe:2004rs,Contino:2006qr}],  we get
\begin{equation} \label{eq:boundfromS}
S = 4 \pi \left( 1.36 \right)  \left( \frac{v}{m_\rho} \right)^2\, ,
\end{equation}
which leads to  a constraint on the mass of the lightest spin-1 resonance $m_\rho$.~\footnote{\label{ftnt:5DS}
Notice that resumming the effect of the whole tower  of resonances, without assuming vector meson dominance,
will in general make the bound stronger. For example, the  calculation of $S$ in the  5-dimensional models 
of Refs.~[\refcite{Agashe:2004rs,Contino:2006qr}]
leads to a formula analogous to eq.(\ref{eq:boundfromS}) where the coefficient $1.36$ is replaced by $2.08$, see [\refcite{Agashe:2004rs}].}

Concerning $\Delta\rho$ (or equivalently the Peskin-Takeuchi $T$ parameter), the 
 tree-level correction  due to the exchange of heavy spin-1 resonances identically vanishes in the $SO(5)/SO(4)$ model
as a consequence of the  custodial symmetry of the strong sector. 
In fact, the absence of this otherwise large correction to
$\Delta\rho$ is the main reason to consider this symmetry breaking pattern
instead of more minimal ones (like for example $SU(3) \to SU(2)\times U(1)$, see Ref.[\refcite{Contino:2003ve}]),
where no custodial symmetry is present.
Non-vanishing corrections  to $\Delta\rho$ will  follow in general from loops of  heavy fermions and vectors.
We do not discuss these effects here, referring to the  literature~[\refcite{Carena:2006bn,Barbieri:2007bh,loopsdrho}] for more details.

\begin{figure}[t]
\centering
\includegraphics[height=30mm]{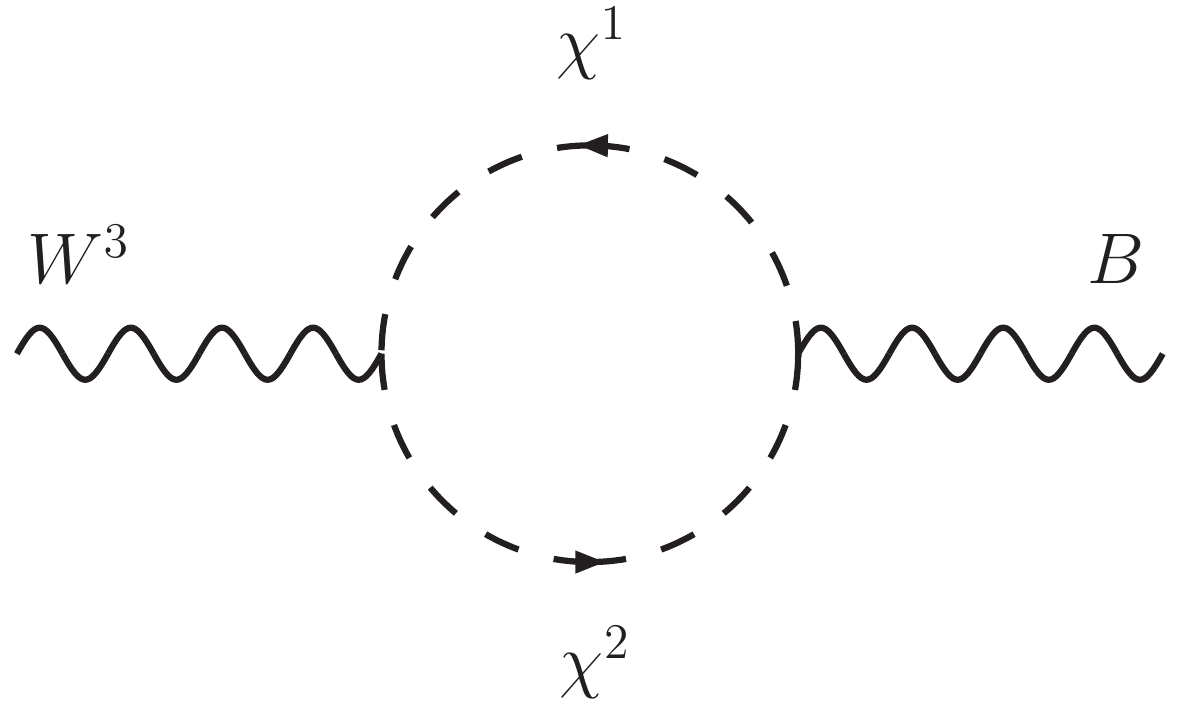} \hspace{1cm} 
\includegraphics[height=30mm]{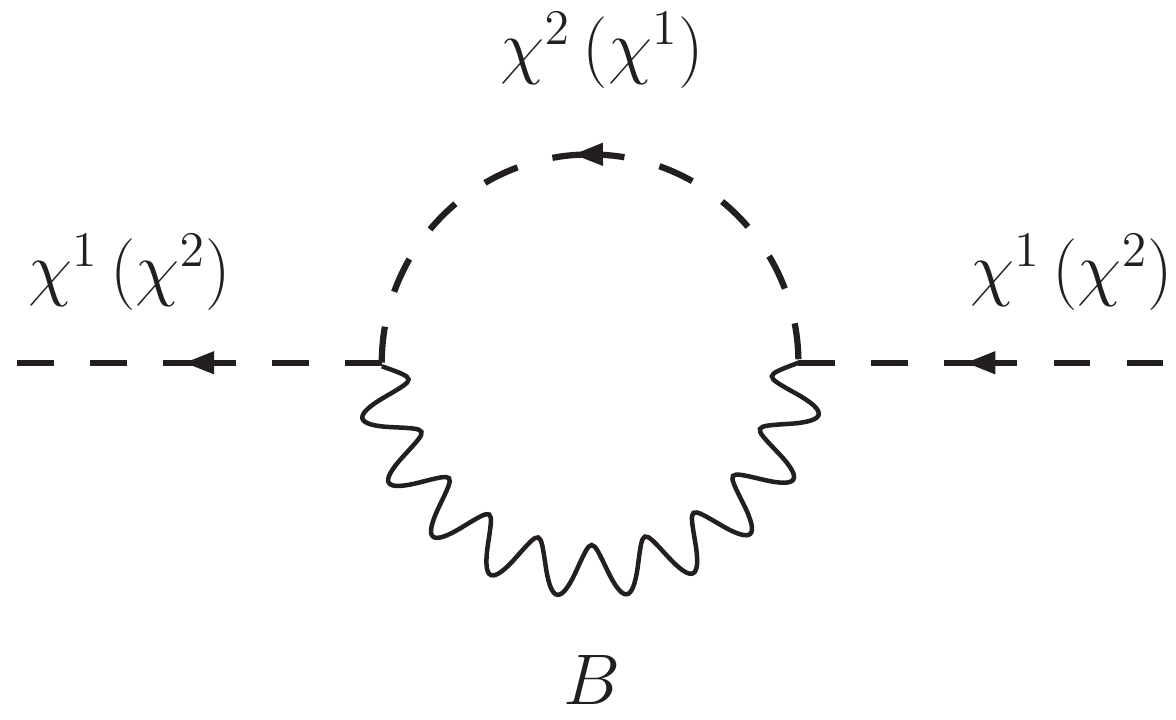} 
\\[0.5cm]
\includegraphics[height=30mm]{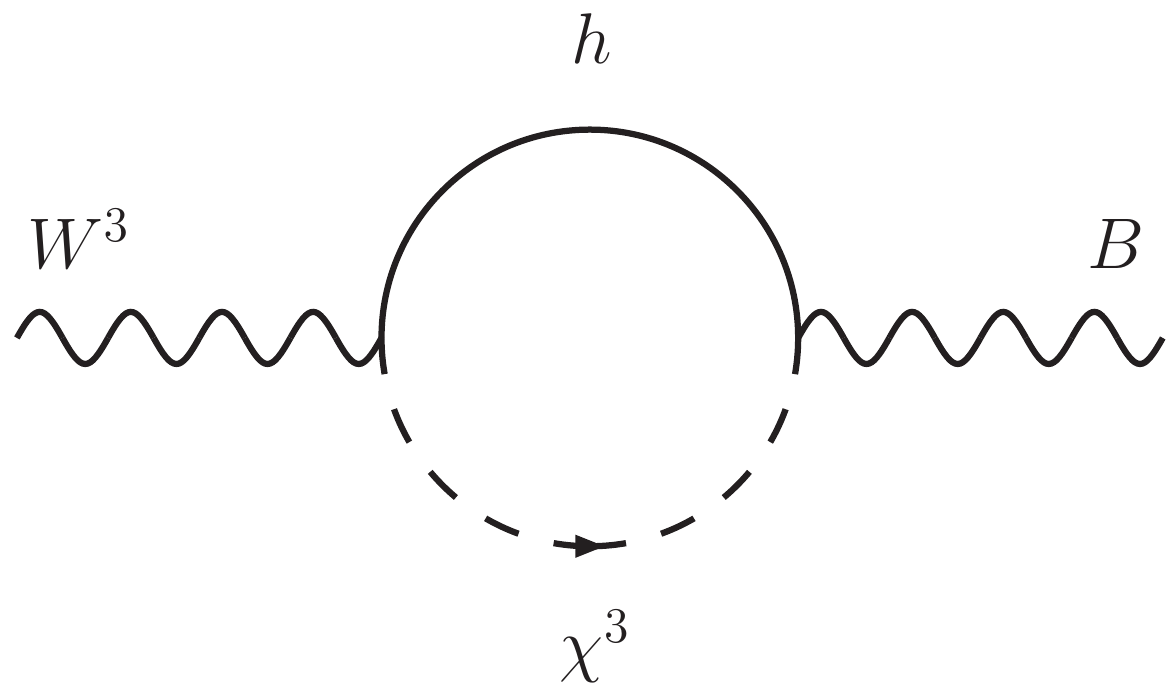} \hspace{1cm} 
\includegraphics[height=30mm]{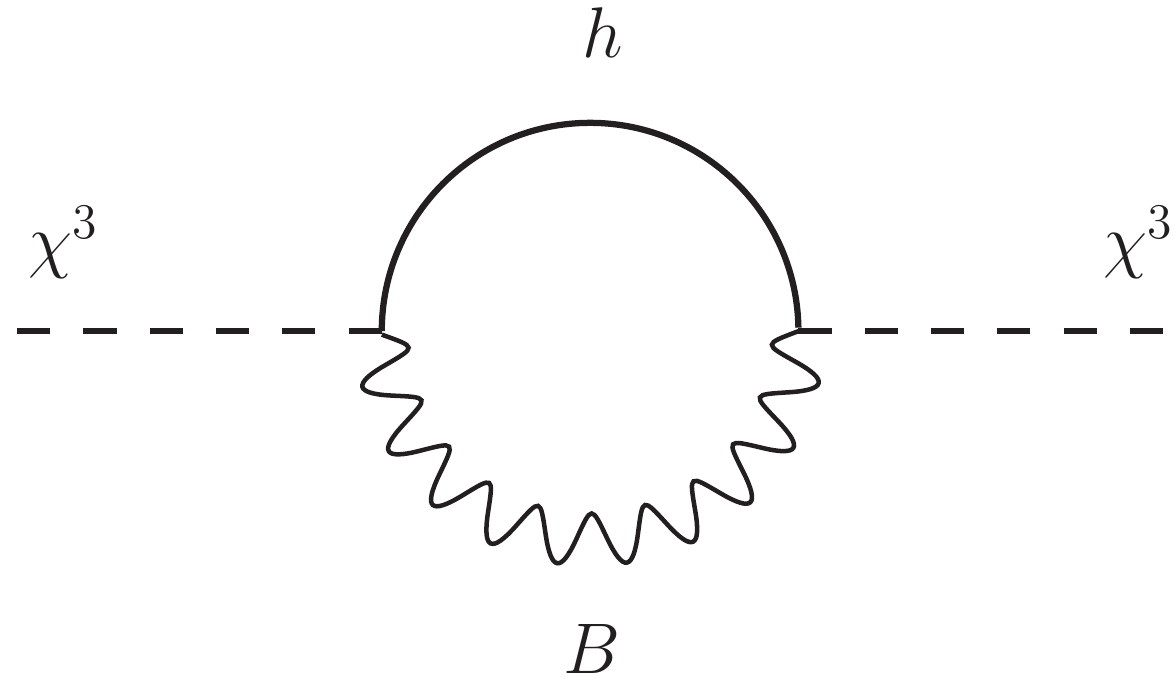} 
\caption{ \label{fig:IRST}
Logarithmically divergent contributions to $S$ (left diagrams) and $T$ (right diagrams) from
loops of  would-be NG Goldstones $\chi$'s (upper row) and of the Higgs boson (lower row).
In the SM the Higgs divergent contribution exactly matches that from the $\chi$'s to give a finite result.
At scales below $m_h$, the upper left diagram contributes to the running of the coefficient of the operator 
$\text{Tr}\left[ \Sigma^\dagger W_{\mu\nu} \Sigma B^{\mu\nu} \right]$, see eq.(\ref{eq:chiralS}).
Similarly, the upper right diagram contributes to the running of the coefficient of 
$\left( \text{Tr}\left[ T^3 \Sigma^\dagger D_\mu \Sigma \right] \right)^2$. See Ref.~[\refcite{Georgi:1991ci}].
}
\end{figure}

There is another important correction to both the $S$ and $T$ parameters,
calculable within the low-energy effective theory, that follows from the modified couplings of the composite
Higgs to the SM gauge bosons, see eq.(\ref{eq:modcoupl}).
In the Standard Model the 1-loop  contribution of the Higgs boson to the vector self energy exactly cancels the logarithmic
divergence arising from loops of would-be NG bosons $\chi^a$ (see for example Ref.~[\refcite{Georgi:1991ci}]).
The relevant diagrams are shown in  Fig.~\ref{fig:IRST}.
The cancellation follows from the fact that the Standard Model is a renormalizable theory, and there are no additional divergences that cannot
be reabsorbed by a renormalization of the gauge couplings and the Higgs wave function.
In other words, there is no counterterm 
% at the level of renormalizable operators 
that can cancel a possible divergence in $S$ and $T$ (at the level of renormalizable operators).
Thus, in the full theory the contribution to $S$ and $T$  must be finite when expressed in terms of the
renormalized parameters. On the other hand,  working in a renormalizable $\xi$ gauge one finds that 
loops of NG bosons $\chi^a$ give a logarithmically divergent contribution to $S$ and $T$.~\footnote{Of course the same result
is obtained with any choice of gauge fixing. The renormalizable $\xi$ gauge is convenient because it shows that the log
divergences solely arise from the EWSB sector, \textit{i.e.} from the contribution of the NG fields $\chi^a$, and not
from the transverse gauge bosons.}
This must be then exactly matched by the Higgs boson contribution at 1-loop to give a finite result.

\begin{figure}[t]
\centering
\includegraphics[height=70mm]{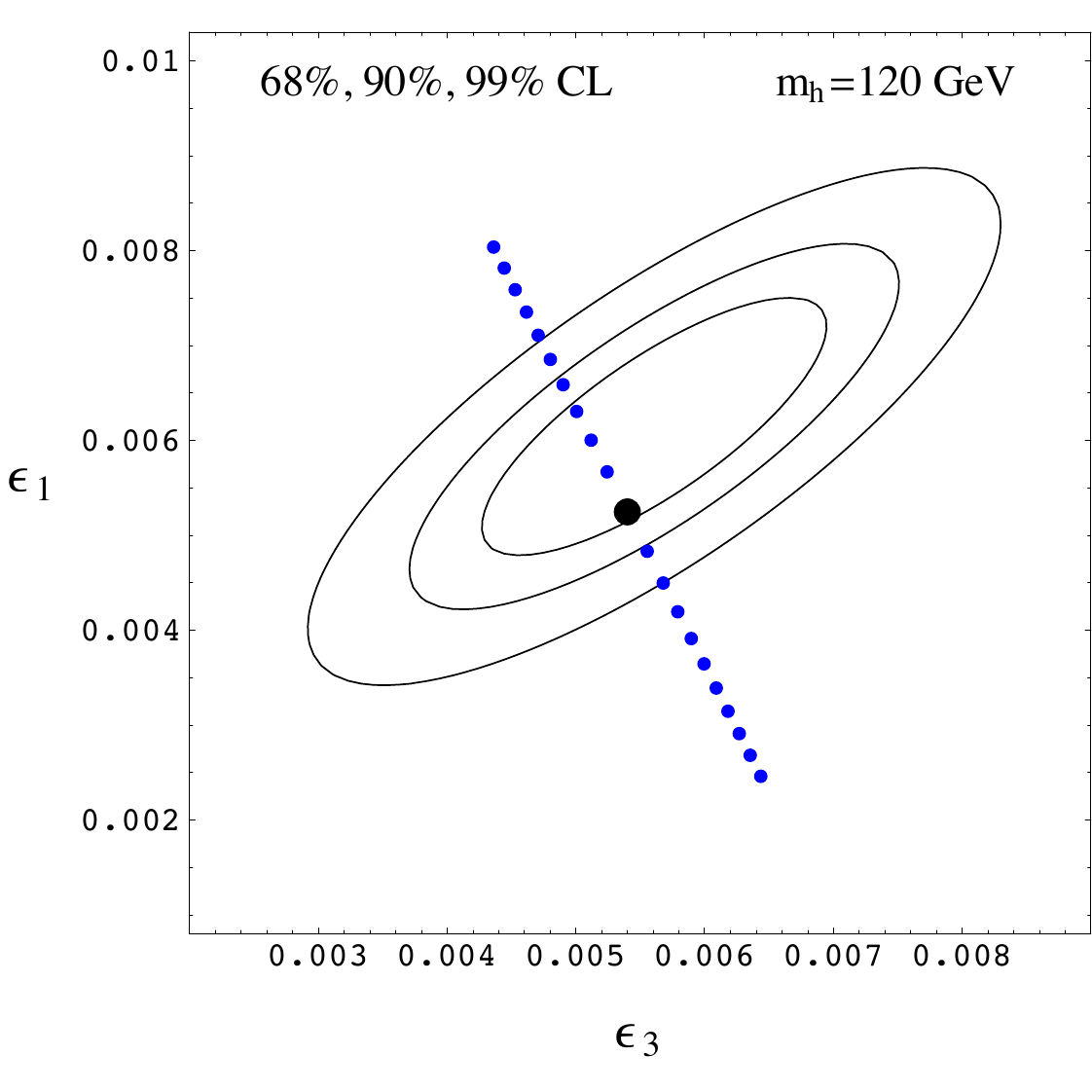} 
\caption{
$\chi^2$ fit to the parameters $\eps_{1,3}$ of Altarelli and Barbieri [\refcite{Altarelli:1990zd,Altarelli:1991fk}] obtained from LEP data~[\refcite{EWfit}]. 
The relation to the $S$ and $T$ parameters is as follows: $\epsilon_1 = \epsilon_1^{SM} + \alpha\, T$, 
$\epsilon_3 = \epsilon_3^{SM} + \alpha/(4 \sin^2\!\theta_W)\, S$. 
The solid curves represent the 68\%, 90\% and 99\% CL contours obtained by setting  $\eps_2$, $\eps_b$ to their SM value with 
$m_h=120\,$GeV and $m_t = 171.3\,$GeV.
The black fat dot shows the SM prediction for $m_h=120\,$GeV ($a=1$).
The blue smaller dots show how $\eps_1$ and $\eps_3$ are modified by varying $a^2$ from 0 to 2 in steps of 0.1 
(for $\Lambda = 1.2\,\text{TeV}/\sqrt{|1-a^2|}$ and $m_h=120\,$GeV). No additional correction to $S$ and $T$ has been included other than 
that of eqs.(\ref{eq:IRdS}),(\ref{eq:IRdT}).
}
\label{fig:eps13}
\end{figure}

In a non-renormalizable composite Higgs theory the above argument on the finiteness of the $S$ and $T$ parameters does not
hold anymore.
In particular, as noticed by the authors of Ref.~[\refcite{Barbieri:2007bh}], the modified Higgs couplings  to the SM gauge bosons 
imply that the   contribution of the composite Higgs to the self-energy does not  exactly cancel  the infrared log divergence arising
from the $\chi$'s. This mismatch leads to a correction to $S$ and $T$ given by ($\Delta T = \Delta\rho/\alpha$)
\begin{align}
\label{eq:IRdS}
\Delta S = & + \frac{1}{12\pi}  (1-a^2) \log \left( \frac{\Lambda^2}{m_h^2} \right)\, ,  \\[0.2cm]
\label{eq:IRdT}
\Delta T =& -\frac{3}{16\pi} \frac{1}{\cos^2\theta_W} (1-a^2) \log \left( \frac{\Lambda^2}{m_h^2} \right)\, ,
\end{align}
where $a$ parametrizes the shift of the coupling of one Higgs boson to two $W$'s, see eq.(\ref{eq:CHLag}), 
and $\Lambda\approx 4\pi f$ is the strong cutoff scale of the theory (\textit{i.e.} the scale at which unitarity is ultimately restored 
in $WW$ scattering). The  LEP precision tests thus imply a constraint on the parameter~$a$.
For example, assuming that the only correction to $S$ and $T$ comes from eqs.(\ref{eq:IRdS}),(\ref{eq:IRdT}) and setting $m_h = 120\,$GeV, 
$\Lambda = 1.2\,\text{TeV}/\sqrt{|1-a^2|}$,  one obtains $0.8 \lesssim a^2 \lesssim 1.6$ at 99$\%$~CL, see Fig.~\ref{fig:eps13}. 
%%%%%%%%%%%%%%%%
%%%%%%%%%%%%%%%%
Larger deviations of $a$ from 1 can of course be accommodated if  the 1-loop contribution to $S$ and $T$ from the heavy resonances 
(partly) compensates the infrared  correction of eqs.(\ref{eq:IRdS}),(\ref{eq:IRdT}).

\begin{figure}[tb]
\centering
\includegraphics[width=0.67\linewidth]{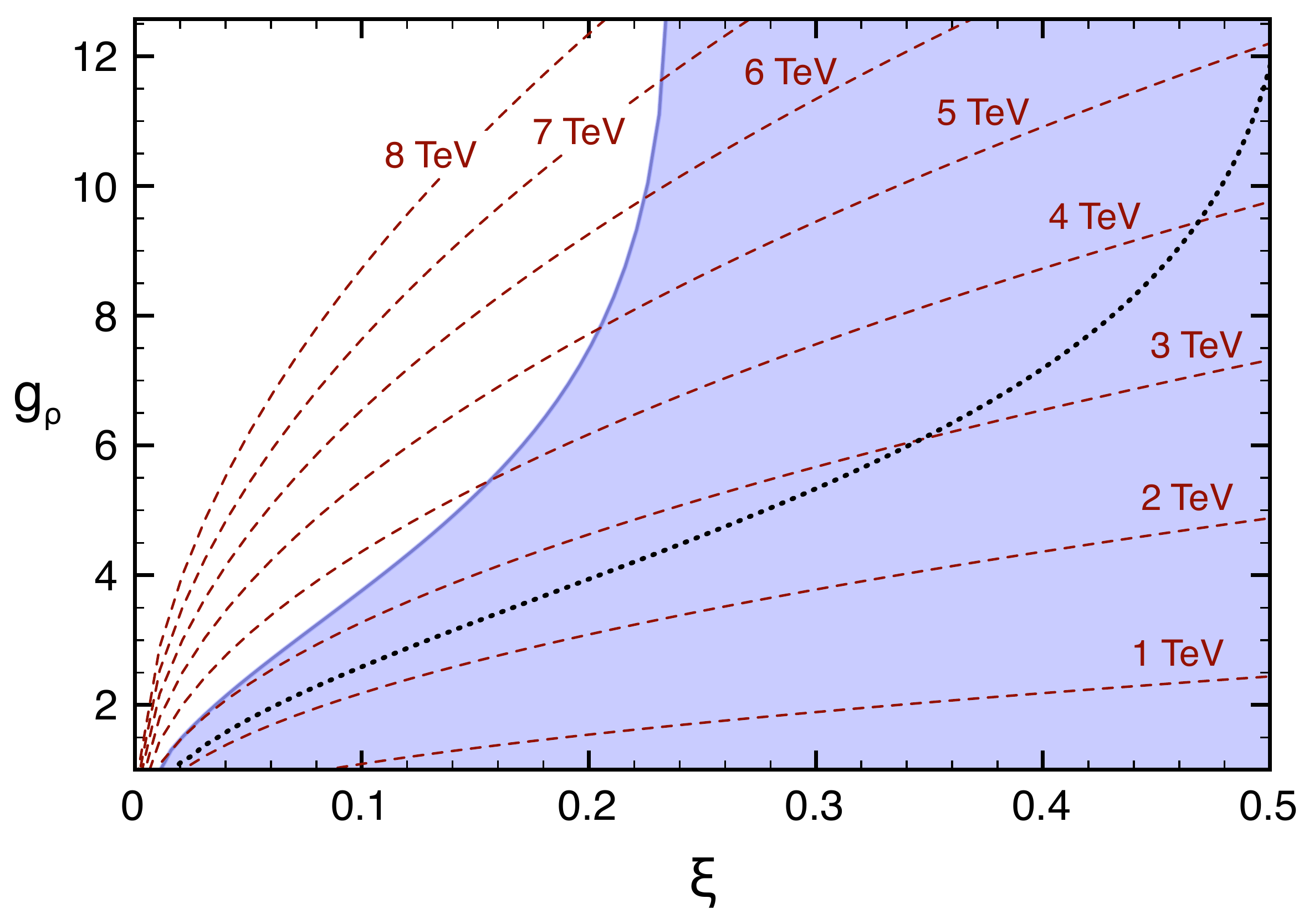} 
\caption{
The blue area denotes the region in the plane $(\xi,g_\rho)$ excluded at 99\% CL by the LEP data 
after including the UV and IR corrections to $S$ and $T$ as predicted in the $SO(5)/SO(4)$ models
of Refs.[\refcite{Agashe:2004rs,Contino:2006qr}], see text.  Superimposed in red are the isocurves of constant $m_\rho$ (dashed lines).
The  dotted black curve  shows how the excluded region is reduced by adding an extra $\Delta T = + 2\times 10^{-3}$.
The Higgs mass has been set to $m_h =120\,$GeV. 
}
\label{fig:xigrhoexcl}
\end{figure}

The corrections to $S$ and $T$ of Eqs. (\ref{eq:SlargeN}), (\ref{eq:IRdS}) and (\ref{eq:IRdT}) together put a strong bound on the value of the 
mass of the lightest vector resonances.  The size of the corrections is controlled by the value of $\xi$ and $g_\rho$, which in turn determine
$a$ and $m_\rho$.  As an illustrative example we consider the constraint that follows in the 5-dimensional $SO(5)/SO(4)$ models of 
Refs.[\refcite{Agashe:2004rs,Contino:2006qr}], where $a=\sqrt{1-\xi}$, $m_\rho=(3\pi/8) g_\rho v/\sqrt{\xi}$ and the UV correction to the $S$ 
parameter is $\Delta S= 4\pi (2.08) (v/m_\rho)^2$ (see Eq.(\ref{eq:boundfromS}) and footnote~\ref{ftnt:5DS}).  Figure~\ref{fig:xigrhoexcl}  shows 
the  region in the plane $(\xi,g_\rho)$ excluded at 99\% CL  (blue area) and the isocurves of constant $m_\rho$.
%%%%%%%%%%%%%%%%
%%%%%%%%%%%%%%%%
The mass of the Higgs has been set to $m_h =120\,$GeV and the argument of the IR logarithm of Eqs.(\ref{eq:IRdS},\ref{eq:IRdT}) 
to $(m_\rho^2/m_h^2)$.  One can see that values of $\xi$ up to $\sim 0.2$ are allowed for large $g_\rho$, while a smaller $\xi$
is required to have more weakly coupled vector resonances. In any case the mass of these latter, $m_\rho$, must be larger than $3\,$TeV, 
which makes their detection at the LHC challenging (see for example~[\refcite{Agashe:2008jb}]).
The bound on $m_\rho$ can be  relaxed if an additional \textit{positive} contribution to $T$ is present: for example the dotted black
curve of Fig.~\ref{fig:xigrhoexcl}  shows how the excluded region is reduced by adding an extra $\Delta T = + 2\times 10^{-3}$.

We have seen that  LEP data constrain the parameter $a$.  It is worth stressing, however, that 
no  bound exists on the coupling of two Higgses to two vector bosons, \textit{i.e.} on the parameter $b$ defined in Eq.(\ref{eq:CHLag}).
Although $a$, $b$ and $c$ are related in specific composite Higgs models (see for example Eqs.(\ref{eq:ab}) and (\ref{eq:SO5c}) for their
prediction  in the $SO(5)/SO(4)$ model),
this shows that a direct  measurement of $b$ from the experiment would be highly desirable.
Unfortunately this seems to be quite difficult at the LHC: while $a$ can be extracted  from both the  Higgs decay branching
fractions and  the analysis of the $WW\to WW$ scattering, the parameter $b$ can be extracted only from the
$WW\to hh$ scattering. The exploratory analysis of Ref.[\refcite{stronghh}] shows that observing this process will be quite challenging at the LHC,
although  it should be possible at its planned luminosity upgrade.

From the above discussion we conclude that a mild gap between $v$ and $f$, such as for example $(v/f)^2 = \xi \lesssim 0.1$, can
make a composite Higgs compatible with the electroweak precision data from LEP. 
The original  models constructed by Georgi and Kaplan, however,  suffer from
a much more severe bound on $f$  from CP-violating and FCNC processes.  
In those theories the breaking of the electroweak symmetry  is transmitted to the quark sector
trough   the same mechanism of Extended Technicolor theories:
at some high scale $\Lambda_{UV}$ the exchange of massive vectors generates four-fermion operators
made of two SM fermions and two technifermions:
\begin{equation} \label{eq:bilinear}
\Delta {\cal L} = \lambda\, \bar q q O \, , \qquad O(x) =  \bar \psi_{TC}(x) \psi_{TC}(x)\, .
\end{equation}
Below the scale $\Lambda$ the strong dynamics condenses and
the composite operator $O$ interpolates a  Higgs field,  $\lambda\, O(x) \approx (\Lambda/\Lambda_{UV})^{[O]-1} H(x)$.
The term of eq.(\ref{eq:bilinear}) thus becomes  a Yukawa coupling between $H$ and the SM quarks, and 
gives quark masses of the order
\vspace{0.1cm}
\begin{center}
\begin{minipage}{0.4\linewidth}
\vspace*{-1.3cm}
$\displaystyle  m_q \sim v\, \frac{4\pi}{\sqrt{N}}\left(\frac{\Lambda}{\Lambda_{UV}}\right)^{[O]-1} $ 
\end{minipage} \hspace{0.7cm}
 \includegraphics[width=45mm]{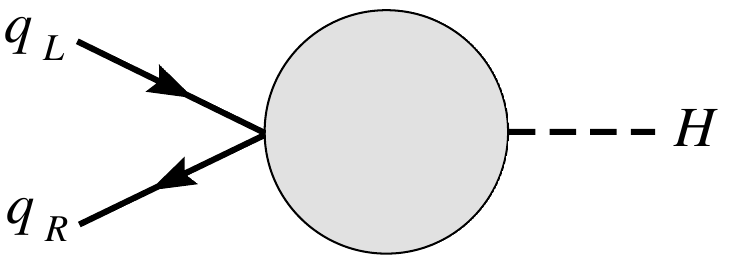}  
\end{center}
\vspace{0.05cm}
Similarly to the case of Technicolor,
one can assume that above $\Lambda$ the strong dynamics stays close to an IR fixed point where its coupling slowly walks,
and the  dimension of the operator $O$ can significantly differ from its classical value.
As discussed in section~\ref{sec:TCmodels},  a naive  argument shows that
 if one does not want to reintroduce  UV instabilities in the theory,
the dimension of the operator $O$ cannot be smaller than 2, thus implying at least a factor $(\Lambda/\Lambda_{UV})$
in the quark masses.~\footnote{Strongly coupled theories at small $N$ can however evade
this conclusion, see the discussion at the end of section~\ref{sec:TCmodels}.}
Besides the term in eq.(\ref{eq:bilinear}), the UV dynamics at the scale $\Lambda_{UV}$  will also generate operators made of four SM fermions.
These are suppressed by $1/\Lambda_{UV}^2$ and are expected to violate flavor and CP,  thus leading  to a strong  
bound on $\Lambda_{UV}$. 
While in Technicolor theories this in turn  implies too small quark masses as a consequence of the suppression factor
$(\Lambda/\Lambda_{UV}) \approx (v/\Lambda_{UV})$,  in composite Higgs models one can
still obtain large enough quark masses by making $\Lambda$ large, $\Lambda \approx f \gg v$.
This can be  achieved, however, only at the price of fine tuning the vacuum alignment parameter to be very small,
$\xi = (v/f)^2 \ll 1$.  Therefore, the FCNC problem of Technicolor can be solved, but the resulting model is highly tuned.
At the same time, a simple explanation of the  hierarchy among quark masses is still missing, 
the only possible mechanism being a
complicated cascade of symmetry breakings as in  Extended Technicolor theories.

There is however a different mechanism that can transmit the EWSB  to the SM fermions and  leads to much milder experimental constraints on $f$.
Suppose that some UV physics at the scale $\Lambda_{UV}$ generates a \textit{linear} coupling between a composite operator $O$ and one SM fermion,
\begin{equation} \label{eq:linear}
\Delta {\cal L} = \lambda\, \bar q  O + h.c. 
\end{equation}
In this case $O$ must be a fermionic composite operator (made for example, but not necessarily, of three technifermions) with the same
$SU(3)_c \times SU(2)_L \times U(1)_Y$ quantum numbers of the SM fermion to which it couples. Hence, there must be at least one composite 
operator  for each $SU(3)_c \times SU(2)_L \times U(1)_Y$ quark multiplet.   At low energy the composite Higgs field is interpolated by pairs of 
fermionic operators $O_R O_L$, and the naive estimate for the quark masses is as follows:
\vspace{0.05cm}
\begin{center}
 \includegraphics[width=50mm]{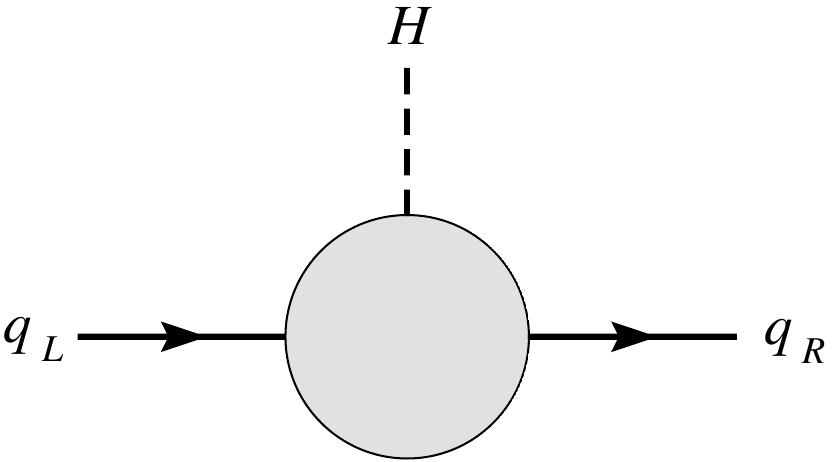}  \hspace{0.8cm}
\begin{minipage}{0.4\linewidth}
\vspace*{-1.3cm}
$\displaystyle  m_q = v\, \frac{\sqrt{N}}{4\pi}\, \lambda_L(\Lambda) \lambda_R(\Lambda)$
\end{minipage} 
\end{center}
\vspace{0.05cm}
The value of the coupling $\lambda$ at low energy  is determined by 
the dimension of the corresponding operator $O$.
Particularly interesting is the case in which from $\Lambda_{UV}$ down to the scale $\Lambda$ the strong dynamics is almost conformal  
and the dimension of $O$ is constant. Then, at large $N$ the RG evolution of $\lambda$  is governed by the following equation:
\begin{equation} \label{eq:RGeq}
\mu \frac{d}{d\mu} \lambda = \gamma \lambda + c \frac{N}{16\pi^2} \lambda^3 + \dots
\end{equation}
Additional terms (indicated by the dots) can be neglected as long as $(\lambda^2/16\pi^2)\ll 1$ and $N$ is large.
The first term in (\ref{eq:RGeq}) corresponds to the classical scaling of $\lambda$ according to the anomalous dimension $\gamma = [O]-5/2$.
The second term, instead, comes from the wave-function renormalization  that (\ref{eq:linear}) induces on the SM fermion, $c$ being a numerical
coefficient of order~1.

For $\gamma > 0$ (so that eq.(\ref{eq:linear}) is an irrelevant term in the Lagrangian), the coupling $\lambda$ becomes smaller at lower energies,
and the second term in the RG equation can be neglected as long as $(\lambda^2 N/16\pi^2)\ll 1$ at the scale $\Lambda_{UV}$.
At low-energy  one has:
\begin{equation}
\lambda(\Lambda) = \lambda(\Lambda_{UV})  \left( \frac{\Lambda}{\Lambda_{UV}} \right)^\gamma\, .
\end{equation}
For $\gamma < 0$ (which means that (\ref{eq:linear}) is a relevant deformation of the Lagrangian),  if the coupling $\lambda$ starts
small at the UV scale $\Lambda_{UV}$, its RG evolution will be initially driven by the first term of (\ref{eq:RGeq}),
so that $\lambda$ increases when evolving to lower energies.  For $c$ negative, $\lambda$ rapidly grows and becomes non-perturbative,
 driving the strong sector away from the fixed point. On the other hand, if $c$ is positive, then the second term has opposite sign 
compared to the first,  and the strong sector is driven to a new fixed point at which
\begin{equation}
\lambda \simeq \lambda_* = \sqrt{\frac{-\gamma}{c}}\,  \frac{4\pi}{\sqrt{N}}\, .
\end{equation}
For $N$ large the value of the coupling at the fixed point is  perturbative and thus our derivation (where we neglected the additional terms
in the RG equation) can be trusted.
Finally, for $\gamma = 0$ the second term in the RG equation leads to a logarithmic evolution of the coupling.

Depending on whether the anomalous dimensions of the operators $O_{L,R}$ are positive or negative,  
the corresponding quark mass can be large or very much suppressed. 
For example, if both $O_{L,R}$ have positive anomalous dimensions, $\gamma_{L,R} > 0$, the naive estimate
for the quark mass reads
\begin{equation}
m_q \sim v \frac{\sqrt{N}}{4\pi} \left( \frac{\Lambda}{\Lambda_{UV}} \right)^{\gamma_L + \gamma_R} \, .
\end{equation}
Although this expression looks similar to that obtained in the case of a bilinear coupling, here the difference is that 
one can have $\gamma_L + \gamma_R$  close or equal to zero without reintroducing any UV instability.
 The unitary bound  on the dimension of a fermionic operator is 3/2 (which corresponds to the dimension of the free field),  implying
$\gamma_{L,R} \geq -1$. Furthermore, for $\gamma \geq 0$ the operator $\bar OO$  (singlet under the SM gauge group)
is  irrelevant at large $N$, and radiative corrections do not reintroduce any UV divergence.
This means that the UV scale $\Lambda_{UV}$ can be arbitrarily large, possibly equal to the Planck scale, without suppressing the quark masses.
On the other hand, a large value for $\Lambda_{UV}$  suppresses any flavor- and CP-violating operator with four SM fermions
generated  at that scale, thus  resolving the problem of  Technicolor theories without any fine-tuning. 
As a bonus, when $(\Lambda/\Lambda_{UV})\ll 1$,  differences of $O(1)$ in the anomalous dimensions can generate large hierarchies in the 
light quark masses. In other words, assuming linear couplings between the SM fermions and the strong sector, and a vast energy range 
over which these can evolve, gives a natural explanation of the hierarchies in the quark masses~[\refcite{Grossman:1999ra,Gherghetta:2000qt}] 
and can  lead to a qualitative explanation  of the pattern of observed flavor mixings~[\refcite{Huber:2000ie,Huber:2003tu,Agashe:2004cp}].
Furthermore, the case with negative anomalous dimensions can be relevant for reproducing the top quark mass.
For example, if both $\gamma_{L,R}$ are negative, and assuming that the strong sector flows to a new IR fixed-point, one has
\begin{equation}
m_q \sim v \, \frac{4\pi}{\sqrt{N}}\, \sqrt{\gamma_L  \gamma_R} \, ,
\end{equation}
which can easily reproduce the experimental top mass in the range $-1 \leq \gamma_{L,R} < 0$ even for moderately large $N$.

To summarize,  linear couplings between the SM fermions and the strong sector represent an
extremely interesting mechanism to communicate the breaking of the electroweak symmetry to the quark sector
and generate the quark masses. The possibility of coupling one SM fermion to three technifermions was first proposed
by D.B. Kaplan~[\refcite{Kaplan:1991dc}], as an alternative mechanism to the more standard Extended Technicolor approach. 
Ref.~[\refcite{Kaplan:1991dc}] however assumed a QCD-like dynamics for the strong sector, and did not exploit the natural generation of
hierarchies in the quark masses that follows from the RG evolution over a vast energy domain.
The importance of linear couplings was re-discovered only later in the context of extra-dimensional warped field 
theories~\footnote{To my knowledge, the connection between the mechanism of Ref.~[\refcite{Kaplan:1991dc}] and extra-dimensional 
warped theories was pointed out for the first time by Ref.~[\refcite{Agashe:2002pr}].},
when their relevance to explain the flavor structure was realized.

\begin{figure}[t]
\begin{center}
 \includegraphics[width=48mm]{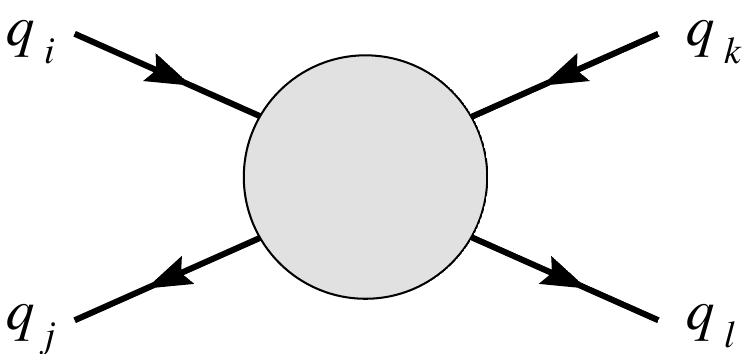}  \hspace{0.4cm}
\begin{minipage}{0.36\linewidth}
\vspace*{-2.0cm}
$\displaystyle  \sim \, \frac{\lambda_i \lambda_j \lambda_k \lambda_l}{\Lambda^2} \, \sim\,  \frac{\sqrt{y_i y_j y_k y_l}}{\Lambda^2}$
\end{minipage} 
\end{center}
\caption{\label{fig:FCNC}
Flavor-Changing four fermion operators generated by the exchange of composite states at the scale $\Lambda$ are suppressed
by four powers of the  couplings $\lambda_i$. For $(\lambda_L )_i\simeq (\lambda_R)_i$ this suppression corresponds to the square root
of the product of the Yukawa couplings $y_i$ of the external quarks.
}
\end{figure}

Although flavor-violating local interactions generated at the scale $\Lambda_{UV}$ can be safely suppressed in the case of linear couplings,
there are still important flavor-violating effects that can be mediated by the strong sector at the lower scale $\Lambda$.
In particular, four-SM fermion operators can be generated by the exchange of the composite resonances of the strong sector, see Fig.\ref{fig:FCNC}.
An interesting possibility is that the strong sector is flavor anarchic, and the flavor structure of the SM Yukawa couplings
entirely  arises from the RG evolution of the composite-elementary couplings $\lambda_i$.
In this case, four-fermion operators
involving light external quarks will be suppressed by their small couplings to the strong sector at low energy,
providing a sort of GIM protection against large FCNC~[\refcite{Gherghetta:2000qt,Huber:2000ie,Huber:2003tu,Agashe:2004ay,Agashe:2004cp}].
Important effect can still arise, however, from the sizable coupling of the third generation quarks to the strong sector.
In particular, it has been shown that important constraints on the scale $\Lambda$ arise from 
CP violation in the $K\bar K$ system~[\refcite{Csaki:2008zd,Blanke:2008zb,Agashe:2008uz,Gedalia:2009ws}],
$b\to s\gamma$~[\refcite{Agashe:2004ay,Agashe:2004cp,Agashe:2008uz,Gedalia:2009ws}] 
and lepton-violating processes such as $\mu\to e \gamma$ and 
$\mu\to 3e$~[\refcite{Huber:2003tu,Agashe:2006iy,Agashe:2009tu}].
Here we will not discuss these constraints, referring to the vast literature on the subject for more details.

There is an extremely interesting phenomenological consequence of linear couplings which was already noticed 
in Ref.~[\refcite{Kaplan:1991dc}]:  similarly to QCD, where a  current made of quarks has the quantum numbers to excite a heavy 
spin-1 resonance from the vacuum,  at energies below the scale $\Lambda$, at which the strong dynamics is assumed to condense,
 a composite operator $O$ can excite a  heavy fermionic resonance. More exactly, there will be a full tower of composite fermions
of increasing mass that can be excited by the operator $O$.~\footnote{We are assuming that the operator $O$ is vector-like, 
so that the excited composite fermions are massive Dirac states.}
The linear coupling (\ref{eq:linear}) thus becomes a mass mixing term at low energy between the elementary fermion $\psi$ and the 
tower of composite fermions $\chi_n$:
\begin{equation} \label{eq:fmix}
{\cal L}_{mix} = \sum_n \, \Delta_n \left( \bar\psi \chi_n + h.c. \right)\, , \qquad  \langle 0 | O | \chi_n \rangle = \Delta_n\, .
\end{equation}
Similarly, and in  complete analogy with QCD, a conserved current $J_\mu$ associated with the global symmetry ${\cal G}$
of the strong sector will excite a tower of spin-1 resonances $\rho_n$ which will mix with the elementary gauge fields $A_\mu$:
\begin{equation} \label{eq:gmix}
{\cal L}_{mix} = \sum_n \, m_{\rho_n} f_{\rho_n} A_\mu \rho_n^\mu \, , \qquad \langle 0 | J_\mu | \rho_n (\epsilon_r) \rangle = \epsilon_\mu^r  m_{\rho_n} f_{\rho_n}\, .
\end{equation}
The corresponding phenomenon is known as $\rho$-photon mixing in the QCD literature.

As a consequence of the mass mixings (\ref{eq:fmix}) and (\ref{eq:gmix}), the physical fermion and vector eigenstates 
(to be identified with the SM fields) will be
admixtures of elementary and composite states. In this case one speaks of \textit{partial compositeness} of the SM 
particles~[\refcite{Kaplan:1991dc,Contino:2006nn}].
A qualitative and simple understanding of the phenomenology of such scenarios can be obtained by
considering the simplifying limit in which one includes only the first  resonance of each tower in the low-energy theory, and
neglects the other heavy states~[\refcite{Contino:2006nn}].
For example, the effective Lagrangian describing one elementary chiral field $\psi_L$ and one composite heavy fermion $\chi$ is 
\begin{equation} \label{eq:Llin}
{\cal L} = \bar\psi_L  \, i\!\not\!\partial \, \psi_L + \bar\chi \left( i \!\not\!\partial - m \right) \chi + \Delta_L \bar \psi_L \chi_R + h.c.
\end{equation}
Notice that, as a result of the RG evolution above $\Lambda$, the  mass mixing parameter  $\Delta_L$  
can be naturally much smaller than the mass $m_*$ of the composite fermion.
The Lagrangian (\ref{eq:Llin}) can be easily diagonalized by rotating the left-handed fields:
\begin{equation}
\begin{pmatrix} \psi_L \\ \chi_L \end{pmatrix} \to \begin{pmatrix} \cos\varphi_L & \sin\varphi_L \\ -\sin\varphi_L  & \cos\varphi_L \end{pmatrix}
\begin{pmatrix} \psi_L \\ \chi_L \end{pmatrix}  \, , \qquad \tan\varphi_L = \frac{\Delta_L}{m_*}\, .
\end{equation}
The mass eigenstate fields, a light left-handed fermion (to be identified with the SM field), and a heavy Dirac fermion of mass 
$m = \sqrt{m_*^2 + \Delta_L^2}$,  are superpositions of elementary and composite states:
\begin{equation}
\begin{split}
| \text{light} \rangle  =& \cos\varphi_L | \psi \rangle + \sin\varphi_L | \chi \rangle \\[0.2cm]
| \text{heavy} \rangle  =& -\sin\varphi_L | \psi \rangle + \cos\varphi_L | \chi \rangle \, .
\end{split}
\end{equation}
The angle $\varphi_L$ thus parametrizes the degree of partial compositeness of the corresponding SM field.
Similar formulas can be derived in the case of the mixing of a right-handed elementary field in terms of a right handed angle $\varphi_R$.
Since the origin of the breaking of the electroweak symmetry  resides, by assumption, in the composite sector,
the mass acquired by a SM fermion  $\psi$ entirely stems from  the composite components of $\psi_L$ and $\psi_R$,
\begin{equation}
y = Y_* \sin\varphi_L \sin\varphi_R \, , 
\end{equation}
where $Y_*$ is a Yukawa coupling among composites.
Thus, heavier SM fields must have larger degree of compositeness. In particular, light quarks and leptons are almost elementary fields.
This explains why all the standard bounds on the compositeness of these particles can be easily evaded in the present framework.
Furthermore, the Higgs boson and the longitudinal components of the $W$ and the $Z$ are full composites.
The transverse polarization of the SM gauge fields will be instead partly composites, the degree of compositeness this time being
fixed in term of the ratio of elementary and composite gauge couplings.

Besides those sketched above, this theoretical framework  
has simple and important consequences for the physics at present and future colliders, as well as on
the pattern of deviations expected in precision measurements.
We do not have time here to review all of them,  but the interested reader can find more details in Ref.~[\refcite{Contino:2006nn}].

\subsection{Higgs potential from the top quark}
\label{sec:fermionpot}

So far we have assumed that the Higgs potential at its minimum can  induce the correct amount of electroweak symmetry breaking.
Here we want to show that this can naturally follow due to the contribution of the SM top quark.
Let us assume that the fermionic content of the elementary sector is that of the Standard Model, and that each  
$SU(3)_c \times SU(2)_L \times U(1)_Y$ multiplet couples linearly to a corresponding composite operator.
The composite operators transform as a complete representations of the  global symmetry $\cal G$ of the strong sector,
while, in general, the external fermions will not. This means that the linear couplings violate $\cal G$ explicitly, so that
loops of elementary fermions will induce a Higgs potential.
The dominant contribution will come from the top and bottom quarks, since
heavier fermions have larger couplings to the strong sector.

The calculation can be done by closely following the strategy adopted for the gauge contribution in section~\ref{sec:SO5example}.
As before, we will work out the specific case of the $SO(5)/SO(4)$ model, but our results are generic.
As a first step, one needs to specify how the composite  operators transform under $SO(5)$.
Following Ref.[\refcite{Agashe:2004rs}], we will assume that for each quark generation, $q_L$, $u_R$, $d_R$,
there are three composite operators transforming as  \textit{spinorial} representations of $SO(5)$ with $U(1)_X$ charge $X = 1/6$,
such that (a sum over the three flavors is understood)
\begin{equation} \label{eq:linSO5}
{\cal L} = \lambda_q \,\bar q_L O_q + \lambda_u \,\bar u_R O_u + \lambda_d \,\bar d_R O_d + h.c.
\end{equation}
In general the couplings $\lambda_{q,u,d}$ can be arbitrary matrices in flavor space, but here 
we will assume for simplicity that they are diagonal.
A spinorial representation of $SO(5)$ consists of two spinors of $SO(4)$,  in the same way as the smallest
irreducible  representation of the $SO(4,1)$ Lorentz group in 5 dimension, a Dirac fermion, is made
of two Weyl fermions of $SO(3,1)$.  Hence, a spinor of $SO(5)$ decomposes as $4 = (2,1) + (1,2)$ under $SU(2)_L \times SU(2)_R$.

Similarly to the case of the gauge fields, a useful trick to derive 
the  effective action for the elementary quarks in the Higgs background is
that of uplifting $q_L$, $u_R$ and $d_R$ to complete $SO(5)$ spinorial representations,
\begin{equation}
\Psi_q = \begin{bmatrix} q_L \\ Q_L \end{bmatrix} \, , \qquad
\Psi_u = \begin{bmatrix} q^u_R \\[0.1cm] \begin{pmatrix} u_R \\ d_R^\prime \end{pmatrix} \end{bmatrix} \, , \qquad
\Psi_d = \begin{bmatrix} q^d_R \\[0.1cm] \begin{pmatrix} u_R^\prime \\ d_R \end{pmatrix} \end{bmatrix}\, ,
\end{equation}
where  $Q_L$, $q^{u,d}_R$, $u_R^\prime$ and $d_R^\prime$ are non-dynamical spurions. 
Each of the fields $\Psi$ contains one doublet of $SU(2)_L$ (the two upper components of the multiplet)
and one doublet of $SU(2)_R$ (the two lower components).  Specifically, $q_L$ is the $SU(2)_L$ doublet
inside $\Psi_q$, $u_R$ is the upper component of the $SU(2)_R$ doublet inside $\Psi_u$, while
$d_R$ is the lower component of the $SU(2)_R$ doublet inside $\Psi_d$.
Then, according to the definition $Y = T^{3R} + X$, the hypercharge of $q_L$, $u_R$, $d_R$ is correctly reproduced if 
all the fields $\Psi_{q,u,d}$ are assigned $U(1)_X$ charge $1/6$.
Once written in terms of $\Psi_{q,u,d}$, the couplings of eq.(\ref{eq:linSO5}) formally respect $SO(5)\times U(1)_X$. Hence, the most general
$(SO(5)\times U(1)_X)$-invariant effective action for the elementary quarks, at the quadratic order and in momentum space, is:
\begin{equation}
\begin{split}
{\cal L}_{\rm eff}
 = & \sum_{r=q,u,d}\bar \Psi_r \pslash \Big[\Pi^{r}_0(p) +\Pi_{1}^r(p)\, \Gamma^i\Sigma_i \Big]\Psi_{r} \\
    & +\sum_{r=u,d}\bar \Psi_q \big[M_0^r(p)+M_1^r(p)\, \Gamma^i\Sigma_i\big] \Psi_{r}\, .
\label{efflag}
\end{split}
\end{equation}
As before, we have treated $\Sigma$ as a constant background and encoded the effect of the strong dynamics in the
form factors $\Pi_{0,1}^r$ and $M_{0,1}^r$ ($r=q,u,d$). The poles of these latter give the spectrum of the fermionic resonances
of the strong sector. Using the expression of the gamma matrices $\Gamma^{i}$ of $SO(5)$
given in the Appendix one easily obtains:
\begin{equation} \label{matrixS}
\Gamma^i\Sigma_i= 
  \begin{pmatrix}  
    \mathbf{1}\, \cos (h/f) & \hat\sigma\, \sin (h/f) \\
    \hat\sigma^\dagger\, \sin (h/f) & -\mathbf{1}\, \cos (h/f)
  \end{pmatrix} \, , \qquad\quad
  \begin{aligned}
   \hat\sigma &\equiv \sigma^{\hat a}\, h^{\hat a}/h \\[0.1cm]
   \sigma^{\hat a} &= \{ \vec \sigma,-i\mathbf{1} \}\, .
 \end{aligned} 
\end{equation}

At this point we keep only the top quark  multiplets 
$q_L = (t_L ,b_L)$ and $t_R$ as physical, dynamical fields, and set to zero all the other fields.  
The effect of  the other elementary fermions  in the Higgs potential 
 is negligible due to their small couplings to the strong dynamics  at low energy.
We thus obtain the effective action for $q_L$ and $t_R$ we were looking for:
\begin{equation} \label{eq:fermeffaction}
\begin{split}
{\cal L} = & \bar q_L \pslash \,\big( \Pi_0^q(p) + \Pi_1^q(p) \cos(h/f) \big) q_L \\[0.15cm]
 & + \bar t_R \pslash \, \big( \Pi^u_0(p) - \Pi_1^u(p) \cos(h/f) \big) t_R \\[0.15cm]
 & + \sin(h/f)\, M_1^u(p)\, \bar q_L \hat H^c t_R + h.c.
\end{split}
\end{equation}
Here  $\hat H^c = i\sigma^2 \hat H$ and $\hat H$ has been defined in eq.(\ref{eq:Hhatdef}).
In particular, the top quark mass can be extracted from the Yukawa term between $t_L$ and $t_R$
by taking the low-energy limit $p\simeq 0$:
\begin{equation} \label{eq:mtop}
m_t \simeq \frac{v}{f} \frac{M_1^u(0)}{\sqrt{(\Pi^q_0(0) + \Pi_1^q(0)) (\Pi^u_0(0) - \Pi_1^u(0))}} \, .
\end{equation}
By expanding eq.(\ref{eq:fermeffaction}) around the Higgs vev one also immediately obtains the
expression of the parameter $c$ defined in eq.(\ref{eq:CHLag}):
\begin{equation} \label{eq:SO5c}
c = \sqrt{1-\xi}\, .
\end{equation}

\begin{figure}[t]
\centering
%\fbox{ 
\begin{minipage}[t]{0.16\linewidth}
  \vspace{0pt}
  \centering
  \includegraphics[scale=0.65]{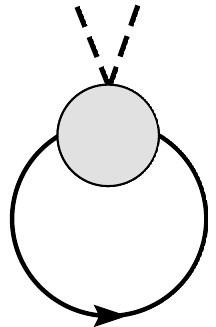} 
\end{minipage} 
%}
\hspace{0.03cm} 
\begin{minipage}[t]{7pt}
  \vspace{1.6cm}
  \centering
  {$\mathbf{+}$}
\end{minipage} 
\hspace{0.03cm} 
\begin{minipage}[t]{0.16\linewidth}
  \vspace{0pt}
  \centering
  \includegraphics[scale=0.65]{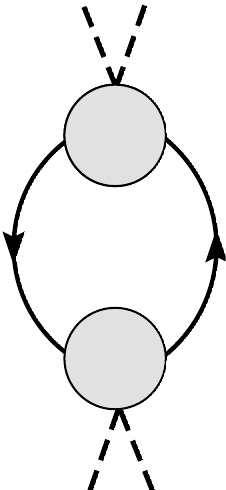} 
\end{minipage} 
\hspace{0.03cm} 
\begin{minipage}[t]{7pt}
  \vspace{1.6cm}
  \centering
  {$\mathbf{+}$}
\end{minipage} 
\hspace{0.03cm} 
\begin{minipage}[t]{0.3\linewidth}
  \vspace{0pt}
  \centering
  \includegraphics[scale=0.65]{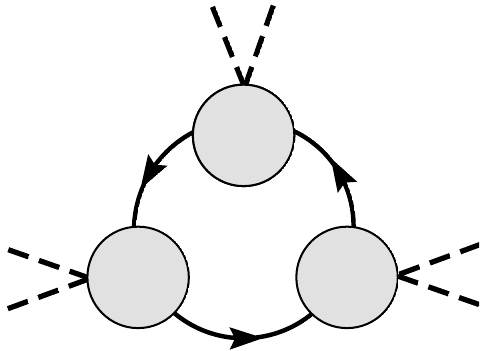} 
\end{minipage} 
\hspace{0.03cm} 
\begin{minipage}[t]{12pt}
  \vspace{1.6cm}
  \centering
  {$\mathbf{ + \; \cdots}$}
\end{minipage} 
\\
\hspace{0.03cm} 
\begin{minipage}[t]{0.35\linewidth}
  \vspace{0.74cm}
  \centering
  \includegraphics[scale=0.65]{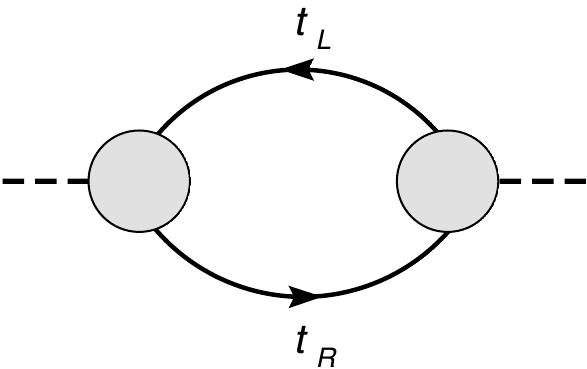} 
\end{minipage} 
\hspace{0.03cm} 
\begin{minipage}[t]{7pt}
  \vspace{1.8cm}
  \centering
  {$\mathbf{+}$}
\end{minipage} 
\hspace{0.03cm} 
\begin{minipage}[t]{0.35\linewidth}
  \vspace{0pt}
  \centering
  \includegraphics[scale=0.65]{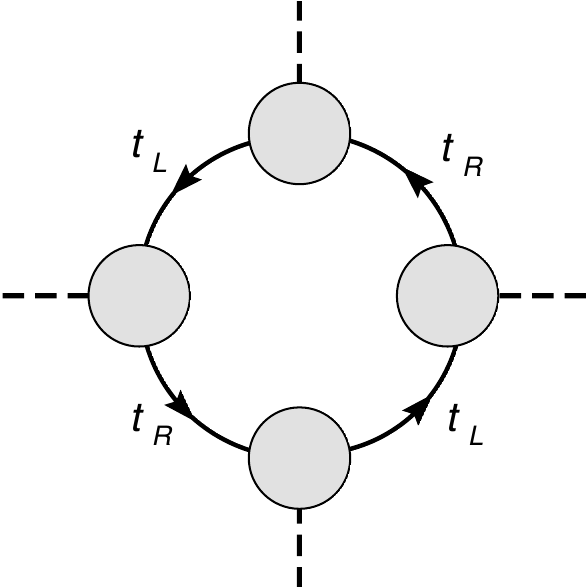} 
\end{minipage} 
\hspace{0.03cm} 
\begin{minipage}[t]{12pt}
  \vspace{1.8cm}
  \centering
  {$\mathbf{ + \; \cdots}$}
\end{minipage} 
\caption{1-loop  contribution of the SM top and bottom quark  to the Higgs potential. 
Upper row: diagrams where the same elementary field, either
$q_L = (t_L , b_L)$  or $t_R$, circulates in the loop with a propagator $i/(\pslash\, \Pi_0)$.
A grey blob denotes the form factor $\pslash \Pi_1$.
Lower row: diagrams where both $t_L $  and $t_R$ circulate in the loop with a Higgs-dependent propagator (see text).
In this case a grey blob denotes the form factor $M_1^u$.
}
\label{fig:HpotdiagsF}
\end{figure}

From the  effective action one  easily derives the  1-loop Coleman-Weinberg potential:
\begin{equation} \label{eq:pot}
V(h) =   - 2  N_c \int\! \frac{d^4p}{(2\pi)^4}\,
 \Big[ \log\Pi_{b_L} + \log\Big(  p^2 \, \Pi_{t_L}\Pi_{t_R}  -\Pi_{t_Lt_R}^2\Big)\Big]\, ,
\end{equation}
where $N_c =3$ and we have defined
\begin{equation}
\label{eq:fff}
\begin{split}
\Pi_{t_L} = \Pi_{b_L} & \equiv \Pi_0^q+\Pi_1^q\cos(h/f) \\[0.1cm]
\Pi_{t_R}     & \equiv \Pi_0^u-\Pi_1^u\cos(h/f) \\[0.1cm]
\Pi_{t_Lt_R} & \equiv M_1^u\sin (h/f)\, .
\end{split}
\end{equation}
The first term in the integral of eq.(\ref{eq:pot}) is the contribution of $b_L$, while the second is due to the top quark ($t_L$ and $t_R$).
The potential can also be conveniently rewritten (up to terms that do not depend on the Higgs field) as
\begin{equation}
\begin{split}
V(h) =  & -2 N_c \int\! \frac{d^4p}{(2\pi)^4}\, \bigg\{  
 2 \log\left( 1+ \frac{\Pi^q_1}{\Pi^q_0} \cos\frac{h}{f} \right)   + \log\left( 1- \frac{\Pi^u_1}{\Pi^u_0} \cos\frac{h}{f} \right) \\[0.15cm]
& \qquad \quad+ \log\left( 1 - \frac{(M_1^u \sin(h/f))^2}{p^2 (\Pi_0^q + \Pi_1^q \cos(h/f)) (\Pi_0^u - \Pi_1^u \cos(h/f)) } \right)
\bigg\} \, ,
\end{split}
\end{equation}
where this time the first two terms in the integral can be thought of as due to the resummation of 1-loop diagrams where only $q_L$
or $t_R$ are exchanged, see  Fig.~\ref{fig:HpotdiagsF} (upper row).
%%%%%%%%%%%%%%%%%%%%%%%%%%%%%%%%%%%%
%%%%%%%%%%%%%%%%%%%%%%%%%%%%%%%%%%%%%%%%%%
The last term, instead, comes from resumming the diagrams where both $t_L$ and $t_R$ circulate in the loop with a 
Higgs-dependent propagator, respectively
\begin{equation*}
\frac{i}{\pslash \left( \Pi_0^q + \Pi_1^q \cos(h/f) \right)} \, , \qquad \text{and} \qquad 
\frac{i}{\pslash \left( \Pi_0^u - \Pi_1^u \cos(h/f) \right)}\, ,
\end{equation*}
see Fig.~\ref{fig:HpotdiagsF} (lower row).
As for the case of the gauge fields, the finiteness of the integral is guaranteed by the convergence of the form factors
$M_1^u$ and $\Pi_1^{u,q}$ at large Euclidean momenta.
Provided these decrease fast enough, the potential can be reasonably well approximated by expanding the logarithms
at first order. This gives:
\begin{equation} \label{eq:potab}
V(h) \simeq \alpha\, \cos\frac{h}{f} - \beta\, \sin^2\frac{h}{f} \, ,
\end{equation}
where the coefficients $\alpha$ and $\beta$ are defined in terms of integrals of the form factors.
Including the contribution of the gauge potential (\ref{eq:gaugepotential}) to  $\beta$, one has:
\begin{equation} 
\begin{split}
\alpha =& \, 2N_c \int\! \frac{d^4p}{(2\pi)^4}\, 
 \left( \frac{\Pi^u_1}{\Pi^u_0} -2 \frac{\Pi^q_1}{\Pi^q_0} \right) \\[0.2cm]
\beta  =& \int\! \frac{d^4p}{(2\pi)^4}\,
 \left( 2N_c \frac{(M_1^u)^2}{(-p^2)\,(\Pi^q_0+\Pi_1^q)(\Pi^u_0-\Pi_1^u)} 
        - \frac{9}{8}\frac{\Pi_1}{\Pi_0} \right)\, .
\end{split}
\end{equation}

We see that even though the gauge contribution to $\beta$ is negative, the EWSB can still be triggered by the top contribution if
$\alpha \leq 2\beta$. In this case the potential has a minimum at 
\begin{equation}
\xi = \sin^2\frac{\langle h \rangle}{f} = 1 - \left( \frac{\alpha}{2\beta} \right)^2\, .
\end{equation}
This shows immediately that small values of $\xi$ require a fine tuning between $\alpha$ and $\beta$.
In fact, this is a general feature of composite Higgs models: the  misalignment of the vacuum comes from the interplay 
of different terms in the potential (specifically, $\sin^2$ and $\cos$ in eq.(\ref{eq:potab})), each of which is a periodic function of $\theta = h/f$.  
One thus naturally expects large values of the angle $\theta$ at the minimum ($\xi \sim 1$), or no symmetry breaking at all ($\xi = 0$).
Small values of $\theta$ are unnatural and can arise only through a fine-tuned cancellation among different terms of the potential.
\footnote{An interesting exception is when one term in the potential starts at order $h^4$ (for example
a $\sin^4$ term~[\refcite{Agashe:2004rs}]), thus contributing only to the quartic coupling and not to the Higgs mass term. 
If the coefficient of such term is slightly larger than that of the other terms in the potential, then a small value of $\theta$ 
naturally follows at the minimum.   This way of getting naturally a large gap between $f$ and $v$ is analogous to the mechanism 
at work in Little Higgs theories, where the large quartic follows from collective breaking (see [\refcite{Schmaltz:LHreview}] and references therein). 
Unfortunately,  no fully natural mechanism
have been found so far (other than collective breaking), to make the coefficient of the $h^4$ term parametrically larger than
that of the remaining terms in the potential.
}
Therefore, the value of $\xi$ gives a rough estimate of the tuning of the theory. In particular, models where $\xi \sim 0.1$ is required
to pass the LEP precision tests are tuned at the level of $10\%$.

As a final exercise, it is instructive to derive the expression of the physical Higgs mass that follows from eq.(\ref{eq:potab}).
Taking the second derivative of the potential at its minimum one has $m_h^2 = 2\beta \xi/f^2$.
It is convenient to define
\begin{equation}
F(Q^2) = \frac{(M_1^u)^2}{(\Pi^q_0+\Pi_1^q)(\Pi^u_0-\Pi_1^u)} , , 
\end{equation}
so that (neglecting for simplicity the gauge contribution)
\begin{equation}
\beta  = \frac{N_c}{8\pi^2} \, F(0)\! \int \! dQ^2 \, \frac{F(Q^2)}{F(0)} \equiv \frac{N_c}{8\pi^2} \, F(0)\, m_*^2 \, ,
\end{equation}
where the last equality defines $m_*$. This is the scale at which the top loop is cut off, and is naturally expected
to be of the order of the lightest fermionic resonance of the strong sector.
Using the fact that $m_t = \xi \sqrt{F(0)}$, see eq.(\ref{eq:mtop}), one finally obtains ($y_t = m_t/v$):
\begin{equation}
m_h^2 = 2 N_c \frac{y_t^2}{8\pi^2} \, m_*^2 \, \xi \, .
\end{equation}
This result could have been guessed simply by  naive dimensional analysis: 
the Higgs mass is one loop suppressed compared to the scale of the heavy resonances $m_*$, 
and the SM coupling responsible for the explicit breaking of the Goldstone symmetry
is the top Yukawa coupling  in this case.
A further suppressing factor $\xi$ comes from the tuning among different terms in the potential.
One can also use the NDA estimate $m_*^2 \approx f^2 N/16\pi^2$, where $N$ is the number of `colors' of the strong dynamics, to 
rewrite $m_h$ as follows:
\begin{equation}
m_h^2 \sim \frac{4 N_c}{N}\, m_t^2 \, .
\end{equation}
This shows that the Higgs mass is naturally expected to be $\lesssim m_t$, 
and that it remains constant in the limit $\xi\to 0$ with $v$ fixed.

%%%%%
%% Section 3
\section{The holographic Higgs}
\label{sec:holoHiggs}

So far we have discussed the phenomenology and the predictions of composite Higgs models assuming that \textit{some}
dynamics exists which forms the Higgs as a bound state at low energy.
Here I want to give one example of such dynamics, illustrating an extremely fascinating possibility:
the composite pNG Higgs might be identified with the fifth component of a gauge field living in a 5-dimensional (5D) spacetime.
Theories of this kind are not just beautiful because of their profound implications on our understanding of Nature, but  
they are also extremely predictive: we will show, although in the context of a simplified abelian model, how the form factors 
of the $SO(5)/SO(4)$ example discussed previously can be computed analytically.

\begin{figure}[t]
\centering
%\hspace*{-0.3cm}
\includegraphics[scale=0.65]{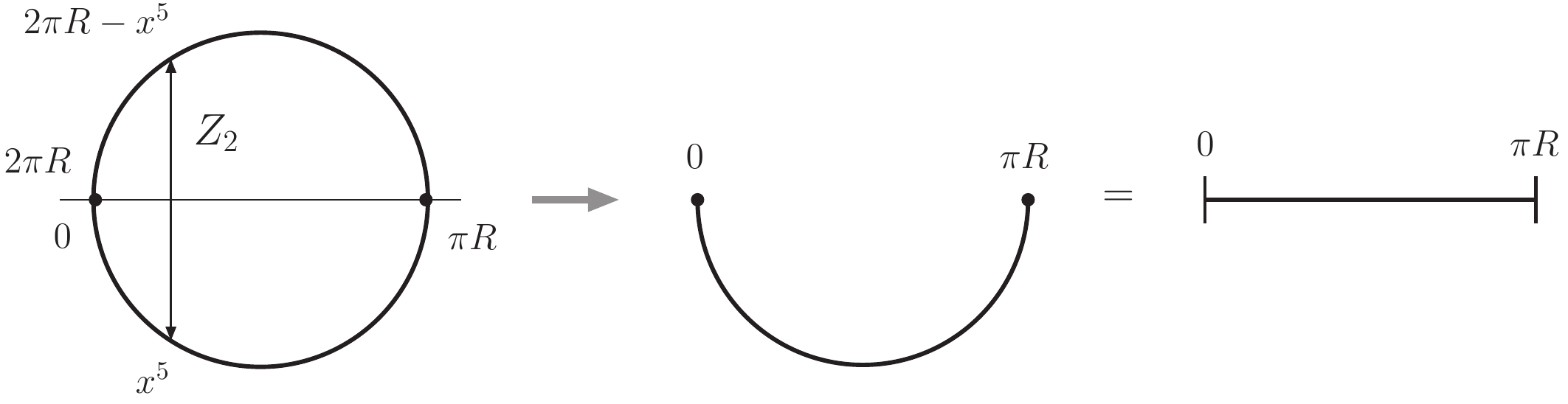} 
\caption{ \label{fig:orbifold}
The orbifold construction:  opposite points on the circle are identified by a $Z_2$ symmetry. The resulting space is equivalent to a 
segment of length $L =\pi R$. 
}
\end{figure}

Let us start by considering a gauge theory on a flat 5-dimensional interval:~\footnote{I will assume 
that the reader is familiar with the formalism of field theories in higher-dimensional spacetimes.
Excellent introductions to the subject are Refs.[\refcite{Cheng:TASI2009,Csaki:TASI2002,Kribs:TASI2004,Rattazzi:Cargese2003}].
Some of the topics discussed in this section are also  introduced in the TASI lectures 
by R. Sundrum~[\refcite{Sundrum:TASI2004}] and the review [\refcite{Serone:2009kf}] by M. Serone, to which I refer for further details and 
a list of references on models of gauge-Higgs unification.
}
the  metric is that of 5D Minkowski spacetime, $\eta_{MN} = (+,-,-,-,-)$ ($M,N = \mu, 5$), 
and the fifth spatial coordinate runs from $0$ to $L$, where $L$ is the dimension of the extra dimension: $x^5 \in [0,L]$. 
One can also start  with $x^5$ defined on a circle of radius $R$ and identify opposite points by means of a $Z_2$ symmetry:
\begin{equation} \label{eq:ident}
x^5 \, \sim \, 2\pi R - x^5\, ,\qquad\quad  x^5 \in [0,2\pi R]\, .
\end{equation}
The spacetime obtained in this way is called \textit{orbifold}, and denoted with $S^1/Z_2$. As a consequence of the 
identification (\ref{eq:ident}), only half of the points are physically inequivalent, for example those lying between $0$ and $\pi R$.
The orbifold is thus equivalent to a segment of length $L=\pi R$, see Fig.~\ref{fig:orbifold}.~\footnote{Although trivial at the level of spacetime,
the equivalence between field theories on the orbifold and on the interval is valid also at the level of field configurations. See
for example the discussion in Ref.~[\refcite{Barbieri:2002ic}].}
%%%%%%%%%
%%%%%%%%%

The action describing  a (non-abelian) gauge field $A_M$ and a fermion field  $\Psi$ living on the 5-dimensional interval is:
\begin{equation} \label{eq:5Daction}
S = \int\! d^4 x \int^L_0 \!\! dx^5 \left[ -\frac{1}{g_5^2}\, F_{MR} F_{NS} \,\eta^{MN} \eta^{RS} + 
\bar \Psi \left( i D_M \Gamma^M - m_\Psi \right) \Psi \right]\, ,
\end{equation}
where $\Gamma^M$ are the 5-dimensional gamma matrices,
\begin{equation}
\Gamma^M = \left\{  \gamma^\mu , -i \gamma^5 \right\}\, , \qquad \quad \left\{ \Gamma^M, \Gamma^N \right\} = 2\, \eta^{MN}\, .
\end{equation}
The smallest irreducible representation of the 5-dimensional Lorentz group $SO(4,1)$ is a Dirac fermion, so that the bulk fermion $\Psi$
has both a left-handed and a right-handed component~\footnote{Here `left' and `right' refer to the chirality in 4 dimensions, that is:
$\gamma_5 \Psi_{R,L} = \pm \Psi_{R,L}$.}
\begin{equation}
\Psi(x,x^5) = \begin{bmatrix} \Psi_L(x,x^5) \\[0.1cm] \Psi_R(x,x^5) \end{bmatrix}\, .
\end{equation}
A gauge-invariant mass $m_\Psi$ for the fermion field is thus allowed in the bulk.
Notice that the 5D gauge coupling has dimension of mass$^{-1/2}$, $[1/g_5^2] = 1$, 
and this is a sign that the theory described by the action (\ref{eq:5Daction}) is non-renormalizable.
Indeed, it is valid up to energies of the order $\Lambda_S \approx 16\pi^2/g_5^2$, below which it can be
considered as the low-energy effective description of some more fundamental theory.
In spite of the non-renormalizability,  there are important physical observables
--  we will see that the Higgs potential is one of those -- which are UV finite and thus calculable.

Since the spacetime has boundaries, the action  (\ref{eq:5Daction}) alone does not completely define 
the theory: one has  to specify the fields' boundary conditions at $x^5=0$ and $x^5=L$.
These must be chosen so that the variation of the action vanishes, upon evaluation on the equations of motion,
both in the bulk and on the boundaries.
For example, in the case of the fermion $\Psi$, the variation of the action reads
\begin{equation} \label{eq:deltaS}
\begin{split}
\delta S 
 & = \frac{\delta S}{\delta \Psi} \, \delta \Psi + \delta \bar\Psi \, \frac{\delta S}{\delta \bar \Psi}  \\[0.1cm]
 & = \int\! d^4 x \!\int^L_0 \!\! dx^5 \left[ \delta \bar\Psi  \, D\Psi + \overline{D\Psi} \, \delta\Psi \right]
 + \frac{1}{2}  \int\! d^4 x \left[ \bar\Psi \gamma^5 \delta\Psi - \delta\bar\Psi \gamma^5 \Psi \right]^L_0\, .
\end{split}
\end{equation}
The first term on the second line of the previous formula vanishes upon evaluation on the bulk equations of motion,
\begin{equation} \label{eq:eom}
D\Psi \equiv \left( i \partial_M \Gamma^M - m_\Psi \right) \Psi= 0 \quad \longrightarrow  \quad
 \begin{cases} i\dslash \Psi_R = \left( \partial_5 + m_\Psi \right) \Psi_L \\[0.1cm]
  i\dslash \Psi_L = \left( -\partial_5 + m_\Psi \right) \Psi_R\, .  \end{cases}
\end{equation}
The second term instead,
\begin{equation*}
\frac{1}{2}  \int\! d^4 x \left[ \bar\Psi_L \delta\Psi_R -\bar\Psi_R \delta\Psi_L  
- \delta\bar\Psi_L\Psi_R +  \delta\bar\Psi_R\Psi_L  \right]^L_0\, ,
\end{equation*}
must vanish when the boundary conditions at $x^5=0,L$ are imposed.
As implied by the coupled system of equations of motion (\ref{eq:eom}), the boundary conditions of $\Psi_L$ and $\Psi_R$
are not truly independent: fixing one determines automatically also the other.
Thus, at each boundary $x_i^5=0,L$ there are two possible choices of boundary conditions: either
\begin{align} 
\label{eq:bc1}
&\Psi_L(x_i^5) = 0  \qquad \text{and thus} \qquad \partial_5 \Psi_R(x_i^5) = m_\Psi \Psi_R(x_i^5)\, , \\
\intertext{or}
\label{eq:bc2}
&\Psi_R(x_i^5) = 0  \qquad \text{and thus} \qquad \partial_5 \Psi_L(x_i^5) = -m_\Psi \Psi_L(x_i^5)\, .
\end{align}
In the particular case of vanishing bulk mass, $m_\Psi =0$,  the above conditions simplify to
\begin{equation} \label{eq:bc3}
\begin{cases} \Psi_L(x^5_i) = 0 & \text{(Dirichlet, $-$)} \\[0.1cm] \partial_5 \Psi_R(x^5_i) = 0  & \text{(Neumann, $+$)} \end{cases}
\quad \text{or} \quad
\begin{cases} \partial_5 \Psi_L(x^5_i) = 0 & \text{(Neumann, $+$)}  \\[0.1cm] \Psi_R(x^5_i) = 0 & \text{(Dirichlet, $-$)} \end{cases}
\end{equation}
Similar conditions (of Neumann or Dirichlet type) also apply for the gauge field $A_M$.

Since the spacetime is  compact, each 5D field $\Phi$ can be decomposed in Fourier harmonics,
\begin{equation}
\Phi(x,x^5) = \sum_n \phi^{(n)}(x) \xi_{n}(x^5)\, ,
\end{equation}
where the $\xi_n(x^5)$ form a complete set of orthogonal functions on the interval. The $\phi^{(n)}(x)$ (\textit{i.e.} 
the Fourier harmonics of $\Phi(x,x^5)$) are called Kaluza-Klein (KK) modes and behave like 4-dimensional massive fields with masses increasing with $n$.
Consider for example the case of a fermion field: each chiral component has a decomposition  in terms of  Kaluza-Klein
modes $\psi_{L,R}^{(n)}(x)$:
\begin{equation}
\Psi_L (x,x^5) = \sum_n \psi^{(n)}_L(x) \xi^{L}_n(x^5) \, , \quad \Psi_R (x,x^5) = \sum_n \psi^{(n)}_R(x) \xi^{R}_n(x^5) \, .
\end{equation}
For $m_\Psi=0$, a complete set of orthogonal wave functions $\xi_n$  is given by ($n=0,1,2,\dots$)
\begin{equation} \label{eq:wfbasis}
\begin{aligned}
&\xi^{(++)}_n(x_5) = \left\{  \cos\left[ \frac{2n\pi x^5}{L} \right]     \right\} \, , 
&&\xi^{(+-)}_n(x_5)  = \left\{  \cos\left[ \frac{(2n+1)\pi x^5}{L} \right]     \right\} \\[0.2cm]
& \xi^{(--)}_n(x_5)  = \left\{  \sin\left[ \frac{2n\pi x^5}{L} \right]     \right\}  \, , 
&&\xi^{(-+)}_n(x_5)  = \left\{  \sin\left[ \frac{(2n+1)\pi x^5}{L} \right]     \right\}\, ,
\end{aligned}
\end{equation}
where $\xi^{(s_0,s_L)}_n$ satisfies a condition of type $s_0$ at $x^5=0$ and type $s_L$ at $x^5=L$, and $s_i = +$ ($s_i = -$) means Neumann
(Dirichlet).  As implied by eq.(\ref{eq:bc3}), if $\xi^L$ has $(s_0,s_L)$ boundary conditions, then  $\xi^R$ will
have $(-s_0, -s_L)$ conditions.  In the case of $(\pm,\mp)$ fields, the  Kaluza-Klein modes form a tower of
four-dimensional Dirac fermions with mass $m_n = (2n+1)\pi/L$ ($n=0,1,2,\dots$).  In the case of $(\pm,\pm)$ fields,
the massive levels are at $m_n = 2n\pi/L$. In addition to those, there is also a `zero mode'  ($n=0$) corresponding to  
a massless chiral fermion. Chirality comes  from the fact that for $n=0$ only the $\xi^{(+,+)}$ wave function admits a non-trivial
solution (with a constant profile), see eq.(\ref{eq:wfbasis}). These considerations remain valid even for 
$m_\Psi \not = 0$, although the value of the masses of the non-zero KK modes will change.

\begin{figure}[t]
\centering
\includegraphics[scale=0.55]{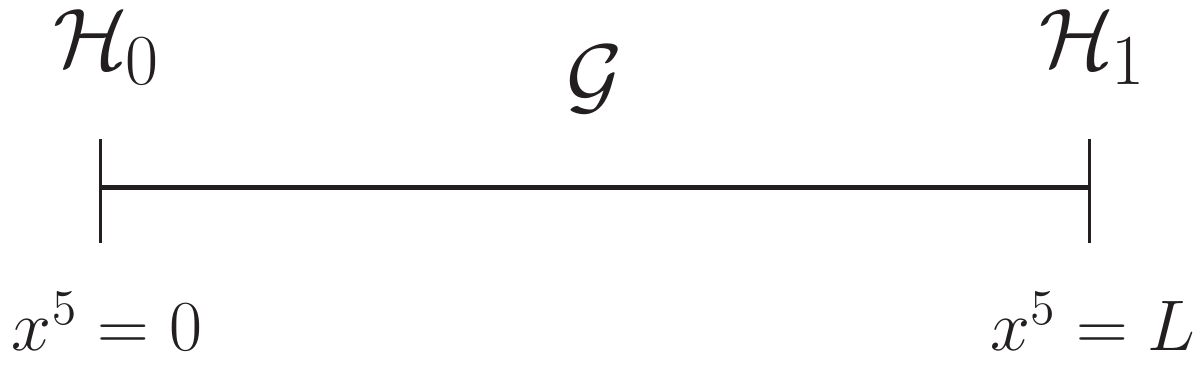} 
\caption{ \label{fig:interval}
The bulk gauge symmetry ${\cal G}$ is reduced to the subgroup ${\cal H}_0$ at the boundary $x^5=0$, and to ${\cal H}_1$ at $x^5=L$.
}
\end{figure}

The possibility of obtaining a  spectrum of chiral fermions at low energy is in fact one of the motivations to consider the interval
rather than other compact spaces, like for example the circle.
There is  another reason however:  the boundary conditions imposed on the gauge field can lead to an elegant
mechanism of symmetry reduction at low energy. Let us see how.
In general, a 5D gauge transformation  on $A_M$ has the form
\begin{equation}
\begin{split}
& A_M \to  \Omega A_M \Omega^\dagger - i\, \Omega \partial_M \Omega^\dagger \\[0.2cm]
& \Omega (x,x^5) = {\cal P} \exp\left\{ i T^A \alpha^A(x,x^5) \right\}\, ,
\end{split}
\end{equation}
where $T^A$ are the generators of the bulk gauge group ${\cal G}$, and ${\cal P}$ represents the path-ordering of the exponential.
An  infinitesimal transformation with $M=\mu$, in particular, transforms $A_\mu^A \to A_\mu^A - \partial_\mu \alpha^A$,
which implies that  each of the gauge parameters
$\alpha^A$ must respect the same boundary conditions of $A^A_\mu$.  Since the gauge field is part of a covariant
derivative and has a geometrical meaning, it follows that $A^A_5$ must have opposite boundary conditions compared to $A^A_\mu$.
The most generic set of boundary conditions that can be consistently applied on the various components of the gauge field is 
then~\footnote{Here we neglect the effect of possible mass terms localized on the boundaries.}
\begin{equation} \label{eq:bc4}
\begin{split}
& A_\mu^a (+,+)\, , \quad A_5^a(-,-) \quad \qquad T^a \in \text{Alg}\left\{  {\cal H} = {\cal H}_1 \cap {\cal H}_0  \right\}  \\[0.2cm]
& A_\mu^{\bar a} (+,-)\, , \quad A_5^{\bar a}(-,+) \quad\qquad T^{\bar a} \in \text{Alg}\left\{  {\cal H}_0/{\cal H}   \right\}  \\[0.2cm]
& A_\mu^{\dot a} (-,+)\, , \quad A_5^{\dot a} (+,-) \quad\qquad T^{\dot a} \in \text{Alg}\left\{  {\cal H}_1/{\cal H}   \right\}  \\[0.2cm]
& A_\mu^{\hat a} (-,-)\, , \quad A_5^{\hat a}(+,+) \quad\qquad T^{\hat a} \in 
  \text{Alg}\left\{  {\cal G}/{\cal H}_0  \right\}  \cap \text{Alg}\left\{  {\cal G}/{\cal H}_1  \right\} \, ,
\end{split}
\end{equation}
where ${\cal H}_{0,1}$ are subgroups of ${\cal G}$.  One can notice two important facts: First, the set of  generators corresponding to 
the gauge fields that do not vanish at $x^5 =0$ ($A_\mu^a$ and $A_\mu^{\bar a}$) form the subgroup ${\cal H}_0$, while those associated 
to the fields that do not vanish at $x^5 =L$ ($A_\mu^{\dot a}$ and $A_\mu^{\hat a}$) form the subgroup ${\cal H}_1$.
In other words, the bulk gauge symmetry ${\cal G}$ is reduced to ${\cal H}_0$ (${\cal H}_1$) on the boundary $x^5=0$ ($x^5=L$),
see Fig.~\ref{fig:interval}.  
%%%%%%%%%
%%%%%%%%%
Second, the gauge invariance  at low energy is $ {\cal H} = {\cal H}_1 \cap {\cal H}_0 $.
Indeed, the basis of wave functions relevant for the Kaluza-Klein decomposition of the fields in (\ref{eq:bc4}) is that of eq.(\ref{eq:wfbasis}).
Since only $(+,+)$ fields have  (massless) zero-modes, this means that the low-energy spectrum comprises
a set of gauge fields $A_\mu^a \, T^a \in \text{Alg}\left\{  {\cal H} \right\}$, and a set of 4D scalars $A_5^{\hat a}$ living in
$\text{Alg}\left\{  {\cal G}/{\cal H}_0  \right\}  \cap \text{Alg}\left\{  {\cal G}/{\cal H}_1  \right\}$

In addition to these massless fields, there is a tower of massive spin-1 fields  transforming under the adjoint representation
of ${\cal G}$.  Level by level, each of the  modes $A_5^{(n)}$ is eaten  in a Higgs mechanism 
to form massive vectors together with $A^{(n)}_\mu$.
In fact,  there is a gauge transformation that eliminates the $x^5$ dependence of $A_5$ from the very beginning, and leaves only its zero mode
(which has a constant wave function).  It can be constructed as follows:
In a 5D Minkowski spacetime one can go to an axial gauge where $A_5=0$ by performing
the following gauge transformation:
\begin{equation}
\Omega(x,x^5) = {\cal P} \exp\left\{  i \int^{x^5} \!\!\! dy\; A_5(x,y) \right\} \, .
\end{equation}
This is  not an allowed gauge transformation on the interval, since if $A_5$ has a constant profile, the gauge parameter
does not satisfy the correct boundary conditions.
However, one can obtain a proper gauge transformation by simply subtracting the zero mode of $A_5$ in the exponent:
\begin{equation}
\Omega(x,x^5) = {\cal P} \exp\left\{  i \int^{x^5}_0 \!\!\! dy\; A_5(x,y) \right\} \exp\left\{ -i \frac{x^5}{\sqrt{L}}\, A_5^{(0)}(x) \right\}\, .
\end{equation}
The factor $1/\sqrt{L}$  comes from the normalization of the zero-mode (constant) wave function,
\begin{equation}
1 = \int_0^L \! dy\; \left(\xi^{(0)}(y) \right)^2 \, .
\end{equation}
The existence of an axial gauge where $A_5$ does not depend on $x^5$ thus shows that only the zero mode $A_5^{(0)}$  is physical,
all the other modes can be gauged away.

\subsection{$A_5$ as a pseudo Nambu-Goldstone boson}

The existence of a massless  scalar field in the spectrum, the zero mode of $A_5$, 
should have raised a crucial question from the reader: is its lightness just an accident, perhaps only valid at tree level,
or there is a more profound reason ? In this latter case $A_5$ would be a natural candidate to play the role of the Higgs boson.
There is in fact  a simple reason why $A_5$ is massless:
locality and the 5D gauge invariance forbid a potential for $A_5$ at tree-level. This is because the only gauge invariant,
local operators one can write in the 5D theory involve the antisymmetric tensor $F_{MN}$, and  
since $F_{\mu5} = \partial_\mu A_5 - \partial_5 A_\mu - i[A_\mu , A_5]$, there is no way to form terms with only $A_5$ and no 
four-dimensional derivative.~\footnote{Notice, on the other hand, that in six dimensions a quartic coupling for $A_5$ and $A_6$ arises 
from the non-abelian structure of the kinetic term. 
For an example of a 6-dimensional theory exploiting this tree-level potential see [\refcite{Csaki:2002ur}] and references therein.}

At the 1-loop level, on the other hand, a potential for $A_5$ can arise from non-local operators.
One can indeed construct a gauge covariant variable, the Wilson line $W(x)$, which is a non-local function of $A_5$:
\begin{equation} \label{eq:WL}
W(x) \equiv \exp\bigg\{i\int_0^L \!\! dx^5 \, A_5 (x,x^5) \bigg\} \equiv \exp\left\{ i\theta(x) \right\} \, .
\end{equation}
After canonically normalizing the kinetic term of $A_5$  one has 
\begin{equation} \label{eq:WLtheta}
\theta(x) =  (g_5 \sqrt{L}) A_5^{(0)}(x)\, ,
\end{equation}
so that $W(x)$ is just the exponent of the zero mode of $A_5$.
According to its definition, under a gauge transformation the Wilson line transforms as
\begin{equation}
W(x) \to \Omega(x,L)\, W(x)\, \Omega^\dagger(x,0) \, , 
\end{equation}
which intuitively suggests that a non-vanishing potential for $A_5$ can arise from 5-dimensional loops that stretch
from one boundary to the other. We will show in the following that this intuition is indeed correct. Here we just want to notice that a possible
potential for $A_5$ must be of the form:
\begin{equation}
V(\theta) = \frac{1}{L^4} f(\theta)\, ,
\end{equation}
where $f(\theta)$ is a periodic function of $\theta$. Indeed, if $V$ arises at the 1-loop level 
as the effect of non-local operators, it must be finite, since there are no local counterterms which could
cancel possible divergences.  Furthermore, it can depend on $A_5$  only though the Wilson line $W$, which is a periodic function of $\theta$.
This implies the periodicity of $f$, and the fact that the overall  dimension of $V$ is  set by the length $L$ of the extra dimension.
In other words, the potential for $A_5$ is a finite-volume effect, very much similarly to the Casimir effect.

Thus, $A_5$ is massless at the tree level and acquires a finite (\textit{i.e.} non-divergent) mass radiatively.
This should sound familiar to the reader as the usual situation for  a pseudo-NG boson.  Indeed, $A_5$ \textit{is}
a pseudo Nambu-Goldstone boson of the 5D theory. 
The easiest way to show it is by adopting the  point of view of a 4-dimensional observer located  on
one of the two boundaries of the extra dimension, for example at $x^5=0$.
From her/his local perspective, the values of the bulk fields at the $x^5=0$ boundary, $\Phi_0(x) = \Phi(x,x^5=0)$, as well as  
possible additional localized fields, act like a 4D sector with local invariance ${\cal H}_0$.
The dynamics associated with the degrees of freedom living in the bulk and at $x^5=L$, on the other hand, 
is  interpreted as  a 4D strongly interacting sector with a global invariance ${\cal G}$ broken down to 
${\cal H}_1$. As we now want to show, this breaking is spontaneous rather than explicit, and the associated
NG bosons are the zero modes of $A_5$.
The operative definition of the strong sector is through the 5D functional integral, which can be performed in two steps:
In the first step one integrates over the bulk fields $\Phi(x,x^5)$  while
keeping their value at $x^5=0$ fixed:
\begin{equation} \label{eq:holoproced}
\begin{split}
Z = & \int \! d\Phi \; e^{i S[\Phi]+iS_0[\Phi]} \\[0.1cm]
   = & \int \! d\Phi_0 \; e^{i S_0[\Phi_0]} \int_{\Phi_0} \!\! d\Phi \; e^{i S[\Phi]}  \\[0.1cm]
   \equiv & \int \! d\Phi_0 \; e^{i S_0[\Phi_0]+ i S_{eff}[\Phi_0]} \, .
\end{split}
\end{equation}
This is equivalent to integrating out the degrees of freedom in the bulk and at the $x^5=L$ boundary,
and defines a 4-dimensional effective action,
\begin{equation}
i S_{eff}[\Phi_0] \equiv \log \int_{\Phi_0} \!\! d\Phi \; e^{i S[\Phi] }  \, ,
\end{equation}
which encodes their dynamics. In eq.(\ref{eq:holoproced}) we have singled out possible terms in the action localized at $x^5=0$,
\begin{equation}
S_{0}[\Phi] = \int \! d^5x\; \delta(x^5) {\cal L}_0 \, ,
\end{equation}
which thus depend only on $\Phi_0$.
As a second step, one  integrates over all values of $\Phi_0$.\footnote{At this level one can make use of
Lagrange multipliers to ensure that bulk fields with Dirichlet boundary conditions at $x^5=0$ vanish, see 
for example ref.[\refcite{Contino:2004vy}].}

\begin{figure}[t]
\begin{center}
\includegraphics[scale=0.55]{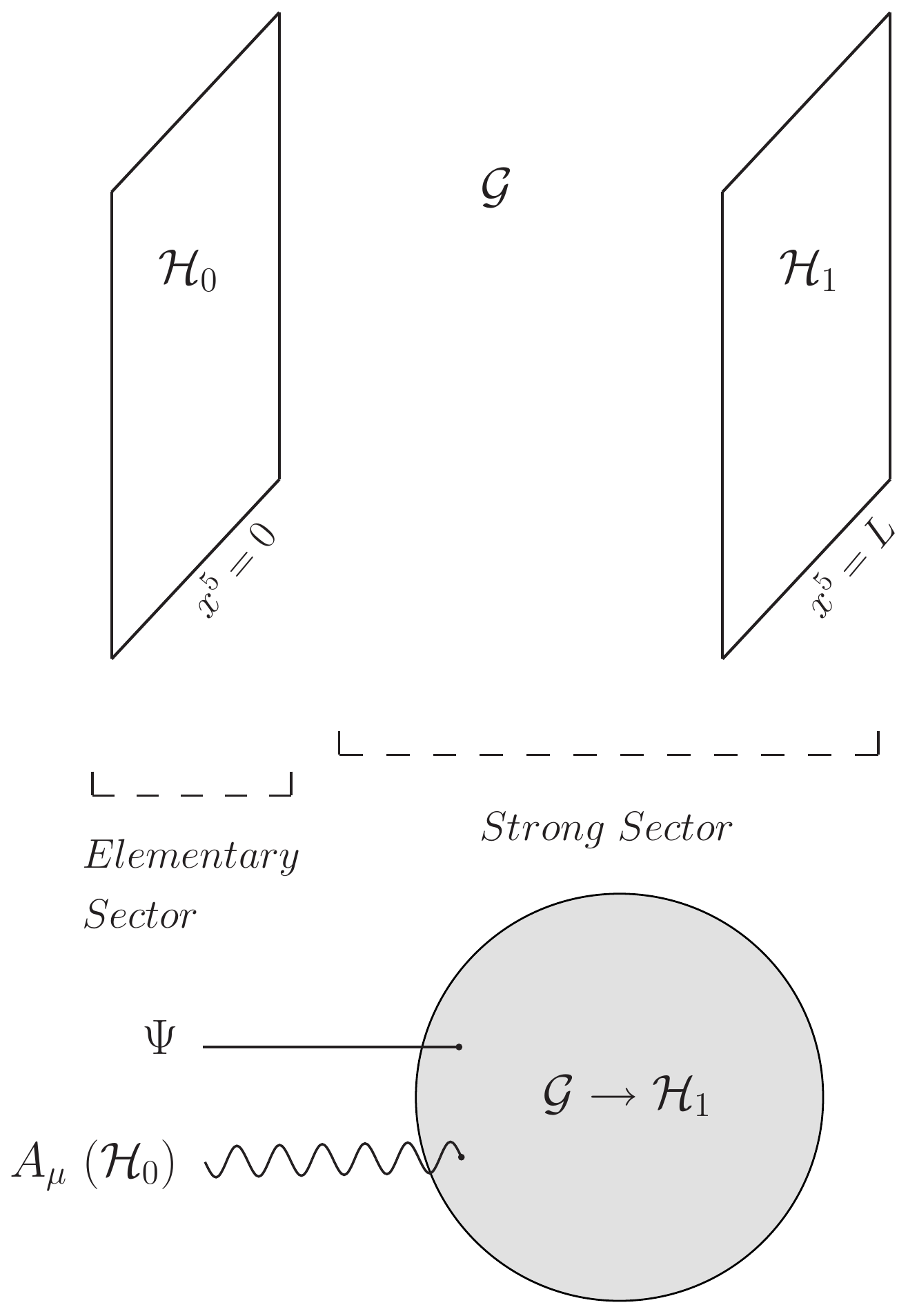} 
\end{center} 
\caption{ \label{fig:holography}
The holographic dictionary.
}
\end{figure}

This defines a correspondence, or rather a `holographic dictionary', which allows one to 
translate the 5D theory into a 4-dimensional one defined in terms of an `elementary' weakly-interacting
sector (the boundary degrees of freedom at $x^5=0$) coupled to a strongly-interacting one
(the dynamics of the bulk and of the $x^5=L$ boundary), see Fig.~\ref{fig:holography}.
%
%%%%%%%%%%%
%%%%%%%%%
%
In this perspective, the Kaluza Klein modes of the 5D theory must be interpreted as the mass eigenstates
resulting from the admixture of the massive resonances of the strong sector with the fields of the elementary sector.
This is in fact in complete analogy with the discussion on partial compositeness of section~\ref{sec:CHandEWPT}.
The above holographic description is clearly inspired by the notorious AdS/CFT 
correspondence~[\refcite{AdSCFT},\refcite{ArkaniHamed:2000ds,Rattazzi:2000hs,PerezVictoria:2001pa}],
but it does not mean to be an exact duality. Rather, it is a way to \textit{define}
the 4D strong dynamics.  Still, such 4D description of the 5D theory turns to be extremely
useful in order to  get a quick qualitative understanding of its  physics and to devise models.
At the same time, it indicates a different procedure, alternative to a Kaluza-Klein reduction,
to extract the 4D low-energy theory starting from the full 5D one (see for 
example~[\refcite{ArkaniHamed:2000ds,Rattazzi:2000hs,PerezVictoria:2001pa,Luty:2003vm,Barbieri:2003pr},
\refcite{Contino:2004vy,Panico:2007qd}] and the reviews~[\refcite{Serone:2009kf},\refcite{Gherghetta:LesHouches2005}]).

To prove that the strong dynamics, as defined above via the holographic dictionary, 
has a global symmetry $G$ spontaneously  broken to ${\cal H}_1$, one can argue as follows.\footnote{See for example
Refs.~[\refcite{Contino:2003ve},\refcite{Panico:2007qd},\refcite{Serone:2009kf}]}
Let us consider first the case in which  ${\cal H}_0$ is a subgroup of ${\cal H}_1$,
so that  there is a zero mode of $A_5$ for each of the ${\cal G}/{\cal H}_1$ generators, see eq.(\ref{eq:bc4}).
As shown above, it is always possible to choose an axial gauge where $A_5$ does not depend on $x^5$, so that
only its zero modes are non vanishing.
Then,  saying that the strong dynamics has a global symmetry ${\cal G}$ means
that if we rotate the boundary fields $\Phi_0$ by a global transformation $g \in {\cal G}$, 
we expect $S_{eff}[\Phi_0]$ to be invariant:
\begin{equation} \label{eq:Ginv}
\Phi_0 \to g\, \Phi_0\, ,  \quad\quad  S_{eff}[g\, \Phi_0] = S_{eff}[\Phi_0]\, , \qquad g\in {\cal G} \, .
\end{equation}
In this sense, rotating the external fields $\Phi_0$ coupled to the strong sector is a way to probe its
global symmetries.  
Intuitively, we expect $S_{eff}[\Phi_0]$ to be at least invariant under global ${\cal H}_1$ transformations,
since the degrees of freedom in the bulk and on the $x^5=L$ boundary certainly respect this symmetry. 
In fact,  $S_{eff}[\Phi_0]$ is invariant under \textit{local} ${\cal H}_1$ transformations,
\begin{equation} \label{eq:H1inv}
\Phi_0 \to h(x) \Phi_0\, ,  \quad\quad  S_{eff}[h(x) \Phi_0] = S_{eff}[\Phi_0]\, , \qquad h\in {\cal H}_1 \, ,
\end{equation}
since even after choosing the axial gauge, the action in the bulk and at  $x^5=L$  is invariant under 
5D gauge transformations that do not depend on $x^5$. The boundary conditions at $x^5=L$ imply that such
$x^5$-independent transformations must belong to~${\cal H}_1$.

Starting from the axial gauge, one can further perform a \textit{field redefinition} of the form
\begin{equation} \label{eq:fred}
\begin{split}
& \Phi \to \Phi^\prime = \Omega \Phi \\[0.1cm]
& A_M \to A_M^\prime = \Omega A_M \Omega^\dagger - i \Omega \partial_M \Omega^\dagger\, , 
\end{split}
\end{equation}
with 
\begin{equation}
\Omega(x,x^5) = \exp\left\{ i (x^5-L) A_5(x)  \right\}\, .
\end{equation}
This sets $A_5$ to zero everywhere in the extra dimension ($A_5^\prime =0$) except at $x^5=0$.
Notice indeed that eq.(\ref{eq:fred}) is \textit{almost} a gauge transformation: it would be so if its exponent vanished at $x^5=0$, 
as required for the parameter of a genuine gauge transformation along the ${\cal G}/{\cal H}_1$ direction.
This means that  away from $x^5=0$  the redefinition (\ref{eq:fred}) acts like a real gauge transformation and leaves
the action in the bulk and at $x^5=L$ invariant. Its only additional effect 
is that of changing all the boundary conditions   at $x^5=0$ by an $A_5$-dependent  factor,
\begin{equation} \label{eq:newbc}
\Phi_0(x) \to \Phi^\prime_0(x) = \Omega(x,x^5=0) \Phi_0(x) = e^{-i\theta(x)} \Phi_0(x) \, ,
\end{equation}
where $\exp(i \theta(x))$ is the Wilson line from $x^5=0$ to $x^5=L$ defined by eqs.(\ref{eq:WL}) and (\ref{eq:WLtheta}).
This shows that the theory with $A_5$ and boundary conditions $\Phi_0$ is equivalent to a theory with
vanishing $A_5$ and boundary conditions $\Phi^\prime_0 = e^{-i\theta(x)}\Phi_0$.

The invariance under a global ${\cal G}$ rotation, eq.(\ref{eq:Ginv}), then follows for $\exp(i\theta(x))$  transforming 
according to the usual transformation rule of a NG field~[\refcite{CCWZ}],
\begin{equation} \label{eq:CCWZrule}
e^{i\theta(x)} \to g \, e^{i\theta(x)} h^\dagger(\theta(x),g) \equiv e^{i\theta'(x)}\, ,
\end{equation}
where $h(\theta(x),g)$ is an element of ${\cal H}_1$. Indeed, from eqs.(\ref{eq:CCWZrule}) and (\ref{eq:newbc}) 
one has that $\Phi_0^\prime\to h(\theta,g)\Phi_0^\prime$,
and from the invariance of the effective action under local ${\cal H}_1$ transformations, eq.(\ref{eq:H1inv}), one deduces the invariance under
global ${\cal G}$ transformations.
This proves that the global symmetry ${\cal G}$ of the strong sector is non-linearly realized, and that the zero modes of $A_5$
are the associated NG bosons. In the general case in which the local invariance of the $x^5=0$ boundary, ${\cal H}_0$, is not a subgroup
of ${\cal H}_1$, part of the NG bosons are eaten to give mass to the corresponding elementary gauge fields. 
The uneaten NG bosons are exactly $\text{dim}({\cal G}/{\cal H}_1)-\text{dim}({\cal H}_0/{\cal H})$, and correspond to the zero modes
of $A_5$, see eq.(\ref{eq:bc4}).

At this point it should be clear that the same pattern of global and local symmetries realized in
4-dimensional composite Higgs models can be obtained in a 5D gauge theory on the interval.
For example, the $SO(5)/SO(4)$ model discussed in the previous section  can be obtained 
from a 5D theory with bulk gauge symmetry ${\cal G} = SO(5)\times U(1)_X$ reduced to ${\cal H}_0 = SU(2)_L \times U(1)_Y$ at $x^5=0$ 
and to ${\cal H}_1 = SO(4)\times U(1)_X$ at $x^5=L$: 
%%
%%%%%%%%%
\vspace{0.2cm}
\begin{center}
\includegraphics[scale=0.6]{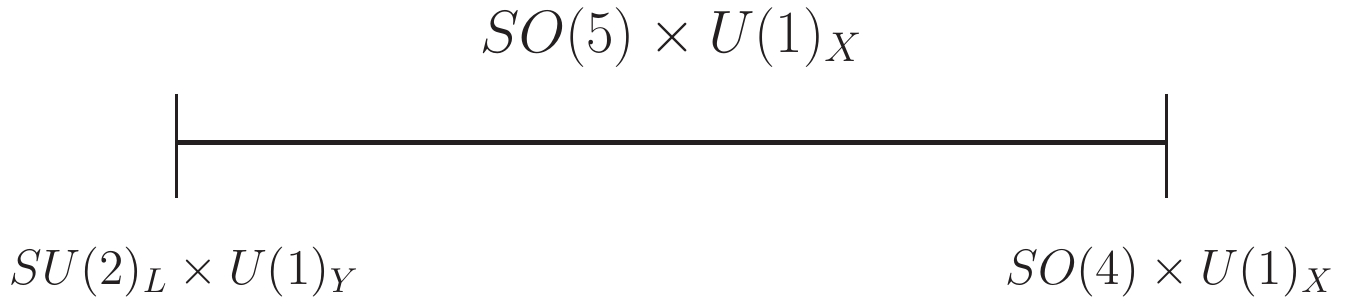} 
\end{center} 
%%%%%%%%%
%%
The  spectrum of zero-modes consists of  the gauge fields of $SU(2)_L \times U(1)_Y$, $A_\mu^a$, 
and four massless scalars $A_5^{\hat a}$ transforming as a $4$~of $SO(4)$ or, equivalently, as a complex doublet of $SU(2)_L$.
In addition to these, there is a tower of massive spin-1 states transforming as  adjoints of $SO(5)$.
Comparing with the $SO(5)$ composite Higgs theory discussed in the previous section, it is clear that the $A_\mu^a$'s
play the role of the SM gauge fields, while the massive KK  vectors correspond to the spin-1 resonances of the strong sector.
From the viewpoint of an observer on the $x^5=0$ boundary, the dynamics of the bulk and of the boundary at $x^5=L$
act like a strongly-interacting sector with global invariance $SO(5)\times U(1)_X \to SO(4)\times U(1)_X$.
The associated NG bosons, our \textit{holographic Higgs},  are the zero modes of $A_5$~[\refcite{Contino:2003ve,Agashe:2004rs}].
As a consequence, $S_{eff}$ has the same structure and the same global symmetry of
the effective action (\ref{eq:holoL}) obtained in the previous section by integrating out the strong dynamics.
By perturbatively solving the 5D theory one can thus compute the analog of the form factors which were previously introduced 
to encode the 4D strong dynamics. The 5D theory, in other words, provides a calculable model for the 4D strong dynamics.

It has to be remarked that the above distinction between a strongly coupled sector (corresponding to the  dynamics
in the bulk and at $x^5=L$) and the `elementary' sector living at $x^5=0$ truly makes sense only if the latter is  weakly
coupled (within itself and to the strong sector). In the case of a 5D theory on a flat extra dimension 
this can be ensured by introducing large kinetic terms for the fields at $x^5=0$, as part of $S_0$~[\refcite{Barbieri:2003pr}].    
In this way the two boundaries are treated differently, and the equivalence between observers at $x^5=0$ and $x^5=L$ is lost.
Remarkably,  5D theories defined on a warped extra dimension~[\refcite{RS}]  automatically satisfy the above requirement,
provided the elementary sector is identified with the degrees of freedom living on the so-called UV brane.
Most importantly, in that case the theory can be extrapolated up to the Planck scale, and the Planck-electroweak hierarchy
is generated by the 5-dimensional geometry.

If one is interested in observables saturated in the infrared, on the other hand, the exact ultraviolet completion of the
theory is not important.
The effective potential of the holographic Higgs, as we discussed above, is one of these calculable quantities independent of the
UV physics. For its calculation we can thus consider a flat extra dimension with no loss of generality.

\subsection{Effective Potential of the Holographic Higgs}

\begin{figure}[t]
\centering
\hspace{0.03cm} 
\begin{minipage}[c]{23pt}
  \vspace{0.5cm}
  \centering
  {\large $\displaystyle \sum_n$}
\end{minipage} 
\begin{minipage}[c]{0.16\linewidth}
  \vspace{0.18cm}
  \centering
  \includegraphics[scale=0.45]{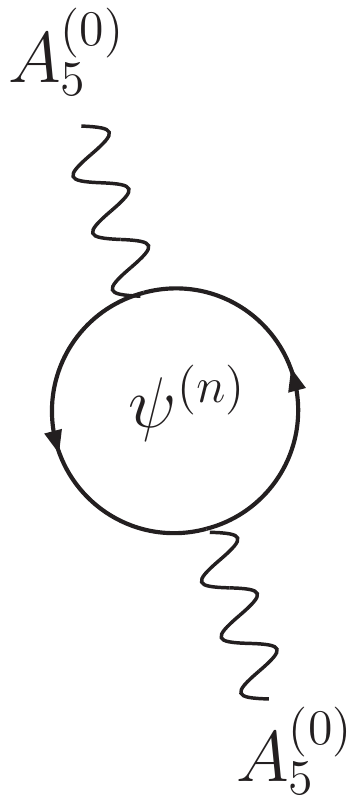} 
\end{minipage} 
\hspace{0.03cm} 
\begin{minipage}[c]{33pt}
  \vspace{0.5cm}
  \centering
  {\large $\displaystyle + \ \ \ \sum_n$}
\end{minipage} 
%
%
%\hspace{0.03cm} 
\begin{minipage}[c]{0.16\linewidth}
  \vspace{0pt}
  \centering
  \includegraphics[scale=0.45]{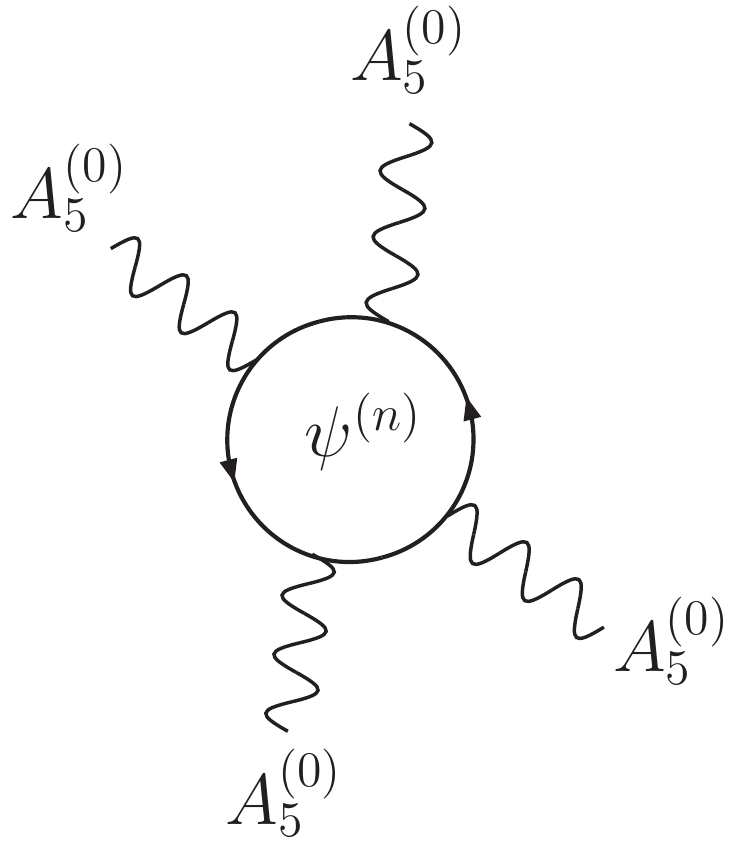} 
\end{minipage} 
\hspace{1.3cm} 
\begin{minipage}[c]{40pt}
  \vspace{0.24cm}
  \centering
  {\large $\displaystyle + \ \cdots$}
\end{minipage} 
\caption{ \label{fig:KKpot}
1-loop contribution to the effective potential of $A_5$  from the KK modes of a bulk fermion.
}
\end{figure}

A straightforward way to compute the potential for $A_5$, our holographic Higgs from the 5D  theory, 
is that of performing a Kaluza-Klein decomposition of  the bulk fields and resumming the series of 1-loop diagrams
induced by the virtual exchange of the KK modes. For example, Figure~\ref{fig:KKpot} shows the diagrams corresponding to the
1-loop contribution of  a bulk fermion. 
%%
%%%%%%%%%
%%%%%%%%%
%%
As a consequence of locality and 5D gauge invariance, we expect the final result to be finite, although the  contribution
of  single modes is divergent.~\footnote{Obviously, for the cancellation to happen properly it is crucial to choose a regulator
that respects the 5D gauge invariance. See for example~[\refcite{Contino:2001uf}] and references therein.}

There is  another way to proceed, however, which is more natural from the holographic point of view: one can first derive
the effective holographic action $S_{eff}[\Phi_0]$ at  $x^5=0$ in the background of $A_5$.
Using that action one then computes the contribution to the potential of $A_5$ that comes from loops of the elementary
fields $\Phi_0$.
This way to proceed is equivalent to the calculation in the Kaluza-Klein basis, since we know that the potential
is a non-local finite-volume effect, which means that any 5D loop has to stretch from one boundary to the other
in order to give a non-vanishing contribution:
%%%%%%%%%%%
\begin{center}
\includegraphics[scale=0.79]{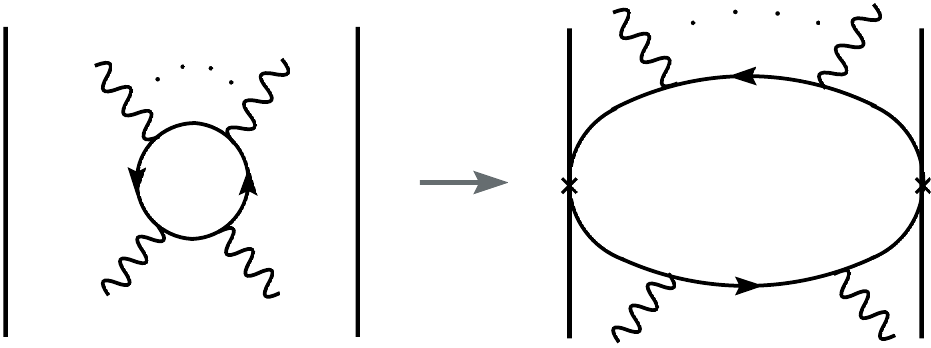} 
\end{center} 
%%%%%%%%%
This is again  simply understood from the holographic viewpoint, where the boundary at $x^5=0$ is interpreted as the
source of the explicit breaking of the global symmetry ${\cal G}$, while the symmetry reduction  ${\cal G}\to {\cal H}_1$ 
at $x^5=L$ corresponds to a spontaneous breaking.
The above argument thus shows that any loop contributing to the potential of $A_5$ must involve the virtual propagation
of the boundary degrees of freedom $\Phi_0$:
%%%%%%%%%%%
\vspace{0.1cm}
\begin{center}
\hspace{0.03cm} 
\begin{minipage}[c]{25pt}
  \vspace{0.1cm}
  \centering
  {\large $\displaystyle \Phi_0$}
\end{minipage} 
%\hspace{0.03cm} 
\begin{minipage}[c]{0.28\linewidth}
  \vspace{0pt}
  \centering
  \includegraphics[scale=0.69]{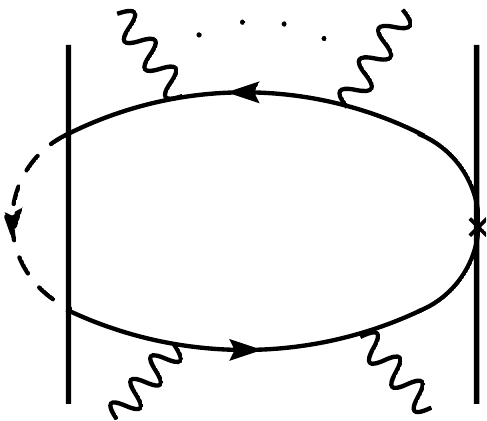} 
\end{minipage} 
\vspace{0.1cm}
\end{center}
%%%%%%%%%
This means that one can perform the calculation by first deriving the effective action for the  $\Phi_0$'s, and then
use that to make 4-dimensional loops. In full analogy with the 4D composite Higgs model of the previous section,
the finiteness of the result is  ensured in this case by the momentum dependence of the form factors that describe the interactions
of the holographic Higgs with the boundary fields. In this sense, even in this case we can speak of the  Higgs
as a composite particle.

A crucial simplification in the explicit computation of the boundary  action
comes from performing the field redefinition (\ref{eq:fred}), which moves the dependence upon $A_5$
to the fields' boundary conditions at $x^5=0$.
This means that we can derive the holographic action at $x^5=0$ simply by setting $A_5=0$ when solving the
equations of motions in the bulk, and  using the new boundary conditions $\Phi^\prime_0= e^{-i\theta(x)}\Phi_0$.

All the above considerations can be more concretely illustrated by means of an explicit example.
We will consider a simplified abelian theory with an  $SO(2)$ bulk gauge invariance fully broken on the
boundaries. We introduce one bulk fermion transforming as a doublet of $SO(2)$.  The fields' 
boundary conditions are thus as follows:
\begin{equation} \label{eq:SO2bc}
\left[ A_\mu (-,-) \ \  A_5 (+,+) \right] \, , \qquad\quad  
\begin{bmatrix}  \Psi_L^1 (+,+) & \Psi_R^1 (-,-) \\[0.15cm] \Psi_L^2 (-,-) & \Psi_R^2 (+,+) \end{bmatrix}\, ,
\end{equation}
and the 5D action reads:
\begin{equation}
S = \int \! d^4x \int_0^L \!\! dx^5 \, \left[ -\frac{1}{4g_5^2} F_{MN}^2 + \frac{1}{g_\Psi^2}  \bar\Psi \left( i D_M \Gamma^M -m \right) \Psi \right]\, .
\end{equation}
For later convenience, and in analogy with the gauge field, 
we have introduced a parameter $g_{\Psi}$ with dimensions of mass$^{-1/2}$, so that the dimension of the
fermion field $\Psi$ is 3/2 as in 4 dimensions. The value of $g_{\Psi}$ can be for example chosen so as to canonically normalize the 
kinetic term of the holographic fermion. We will see that, in absence of localized kinetic terms for $\Psi$,  $g_{\Psi}$ does not enter 
the expression of the potential.

As a first step to compute the potential of $A^{(0)}_5$ we need to derive the boundary action at $x^5=0$. 
As already stressed, in the case of a fermion field the bulk equations of motion connect its chiral components $\Psi_L$
and $\Psi_R$, and one cannot fix simultaneously the value of both at $x^5=0$.
We will thus fix that of $\Psi_L$,
\begin{equation}
\Psi_L (x,x^5=0) \equiv \Psi_L^0\, ,
\end{equation}
and let $\Psi_R$ be free to vary. At $x^5=L$ we instead impose the boundary conditions specified by eq.(\ref{eq:SO2bc}):
\begin{equation} \label{eq:condatL}
\Psi_R^{(1)}(x,x^5=L) = 0\, , \qquad \Psi_L^{(2)}(x,x^5=L) = 0\, .
\end{equation}
A consequence of demanding a fixed (non-zero) boundary value $\Psi^0_L$ 
is that the variation of the action does not vanish anymore at $x^5=0$ (see eq.(\ref{eq:deltaS})):  
\begin{equation}
\delta S = \frac{1}{2}  \int\! d^4 x \int^L_0 \! dx^5 \; \delta(x^5) \left[ \bar\Psi^0_L \delta\Psi_R +  \delta\bar\Psi_R\Psi^0_L  \right]\, .
\end{equation}
We can however solve this problem by introducing  an extra boundary action of the form
\begin{equation}
S_0 = \frac{1}{2 g_\Psi^2} \int \! d^4x \! \int^L_0 \!\! dx^5\; \delta(x^5) \left[ \bar \Psi_L \Psi_R  + \bar \Psi_R \Psi_L \right]\, , 
\end{equation}
so that the variation of the total action is zero: $\delta S + \delta S_0 =0$~[\refcite{Contino:2004vy}].

At this point we are ready to solve the bulk equations of motions and derive the boundary action for $\Psi_L^0$.
Let us work in mixed  momentum-coordinate space and look for a solution of the system of equations 
\begin{equation}
\begin{cases}
- \pslash \, \Psi_L + \left( \partial_5 - m \right) \Psi_R = 0 \\[0.1cm]
- \pslash \, \Psi_R + \left( -\partial_5 - m \right) \Psi_L = 0
\end{cases}
\end{equation}
of the form
\begin{equation} \label{eq:ansatz}
\Psi_{L,R} (p,x^5) = f_{L,R}(p,x^5) \Psi_{L,R}(p) \, .
\end{equation}
We require
\begin{equation}
\begin{cases}
 \left( \partial_5 - m \right) f_R(p,x^5) = \alpha \, p\,  f_L(p,x^5) \\[0.1cm]
 \left( -\partial_5 - m \right) f_R(p,x^5) = \beta \, p\,  f_R(p,x^5)\, ,
\end{cases}
\end{equation}
where $\alpha$ and $\beta$ are numerical coefficients.
If these conditions are satisfied, the Dirac equations  become
\begin{equation}
\begin{cases}
- \pslash \, \Psi_R(p) = \beta \, p\,  \Psi_L(p) \\[0.1cm]
- \pslash \, \Psi_L(p) = \alpha\, p\, \Psi_R(p)\, , 
\end{cases}
\end{equation}
thus requiring $\alpha = 1/\beta$ for consistency.
Without loss of generality we can choose $\alpha = 1 = \beta$, so that
\begin{equation}
\pslash \, \Psi_R(p) = p \, \Psi_L(p)\, .
\end{equation}
Using $\Psi_{L,R}(p) = \Psi^0_{L,R}(p)/f_{L,R}(p,x^5=0)$, which follows from eq.(\ref{eq:ansatz}) 
upon defining $\Psi_{L,R}^0(p) \equiv \Psi_{L,R}(p,x^5=0)$,  we finally get
\begin{equation} \label{eq:solR}
\pslash \, \Psi^0_R (p) = p \, \frac{f_R(p,x^5=0)}{f_L(p,x^5=0)} \, \Psi_L^0(p)\, ,
\end{equation}
where $f_{L,R}$ satisfy
\begin{equation}
\left( - \partial_5^2 + m^2 \right) f_{L,R}(p,x^5) = p^2\,  f_{L,R}(p,x^5) \, .
\end{equation}
Depending on whether $\Psi_L$ (hence $f_{L}$) satisfies Neumann  or Dirichlet boundary conditions at $x^5=L$,
there are two possible solutions to the above equations (respectively dubbed as $L_+$ and $L_-$):
\begin{align}
\label{eq:solLP}
{\large L_+:} \qquad 
\begin{split}
f_R(p,x^5) =& \sin\left[ \omega (x^5-L) \right] \\[0.1cm]
f_L(p,x^5) =& \frac{1}{p} \Big( \omega  \cos\left[ \omega (x^5-L) \right]  -m\, \sin\left[ \omega (x^5-L)\right]   \Big)
\end{split}  \\[0.5cm]
\label{eq:solLM}
{\large L_-:} \qquad 
\begin{split}
f_L(p,x^5) =& \sin\left[ \omega (x^5-L) \right] \\[0.1cm]
f_R(p,x^5) =& -\frac{1}{p} \Big( \omega  \cos\left[ \omega (x^5-L) \right]  + m\, \sin\left[ \omega (x^5-L)\right]   \Big)\, ,
\end{split} 
\end{align}
where we have defined $\omega\equiv \sqrt{p^2-m^2}$.
Evaluating the 5D action on the above solutions leads to the boundary action.
Since the 5D part of the action vanishes once evaluated on the equations of motions, the only contribution
to $S_{eff}(\Psi^0_L)$ comes from the boundary term $S_0$.
It is useful to define
\begin{equation}
\begin{split}
\Sigma^{(+)}(p) =& \frac{1}{p L} \, \frac{f^{(+)}_R(p,x^5=0)}{f^{(-)}_L(p,x^5=0)} = - \frac{1}{L} \, \frac{1}{m+\omega \cot(\omega L)} \\[0.2cm]
\Sigma^{(-)}(p) =& \frac{1}{p L} \, \frac{f^{(-)}_R(p,x^5=0)}{f^{(+)}_L(p,x^5=0)} = \frac{1}{p^2 L}\, \big(  \omega \cot(\omega L) - m \big)
\end{split}
\end{equation}
where the superscripts $(+)$, $(-)$ refer on whether the corresponding left-handed field satisfies Neumann or Dirichlet
conditions at $x^5=L$.
From eqs.(\ref{eq:ansatz},\ref{eq:solR},\ref{eq:solLP},\ref{eq:solLM}) and eq.(\ref{eq:condatL}) one then obtains the
following boundary action:
\begin{equation} \label{eq:SO2Seff}
 \frac{L}{g_\Psi^2} \int \! d^4x\;  \begin{pmatrix} \bar\Psi^{0\, (1)}_L & \bar\Psi^{0\, (2)}_L \end{pmatrix}
 \pslash \, K(p)
 \begin{pmatrix} \Psi^{0\, (1)}_L \\ \Psi^{0\, (2)}_L \end{pmatrix} \, ,
\end{equation}
where
\begin{equation}
K(p) = \begin{bmatrix} \Sigma^{(+)}(p) & 0 \\ 0 &  \Sigma^{(-)}(p) \end{bmatrix}   \, .
\end{equation}

Until this point we have proceeded as if $A_5$ was vanishing. To reintroduce its dependence back into the action we
just have to adopt the following new boundary conditions at $x^5=L$:
\begin{equation} \label{eq:newbc2}
\Psi_L^{0 \, (i)} \to \left[ \exp(- i\theta(x))  \right]^{ij} \Psi_L^{0 \, (j)} \, .
\end{equation}
According to its definition (\ref{eq:WLtheta}),  the expression of the Wilson line, in the case of our $SO(2)$ model, is
\begin{equation}
e^{ -i \theta(x)} = e^{  -i A  h(x)/f } = \begin{pmatrix} \cos (h/f) & \sin(h/f) \\ -\sin(h/f) & \cos(h/f)   \end{pmatrix} \, ,
\quad A = \begin{bmatrix} 0 & -1 \\ +1 & 0 \end{bmatrix}\, , 
\end{equation}
where $A$ is the generator of $SO(2)$ rotations, and we have knowingly defined
\begin{equation} \label{eq:deff}
h(x) \equiv A_5^{(0)}(x)\, , \qquad f \equiv \frac{1}{g_5 \sqrt{L}}\, .
\end{equation}
After performing the redefinition of eq.(\ref{eq:newbc2}), one obtains a boundary action of the same form of  eq.(\ref{eq:SO2Seff}), 
but where now ($\Delta\Sigma(p) \equiv \Sigma^{(+)}(p) -\Sigma^{(-)}(p)$)
\begin{equation} \label{eq:K}
K(p) = \begin{bmatrix}
\Sigma^{(+)} - \sin^2(h/f)\, \Delta\Sigma &  \quad \sin(h/f)\cos(h/f)\,  \Delta\Sigma \\[0.1cm]
 \sin(h/f)\cos(h/f)\,  \Delta\Sigma  & \quad \Sigma^{(-)} + \sin^2(h/f)\, \Delta\Sigma  \end{bmatrix}\, .
\end{equation}
The last step before obtaining the final expression of $S_{eff}$ is setting $\Psi_L^{0\, (2)}=0$. This is required  to
reproduce the Dirichlet condition of $\Psi_L^{(2)}$ at $x^5=0$ in eq.(\ref{eq:SO2bc}), since 5D fields which vanish on the
holographic boundary  do not have any corresponding elementary field. 
In other words,  the 4D elementary sector consists of one single left-handed field, $\Psi_L^{0\, (1)}$.
Its effective action is given by eqs.(\ref{eq:SO2Seff}) and (\ref{eq:K}) with 
$\Psi_L^{0\, (2)}$ set to zero:
\begin{equation} \label{eq:Sefffinal}
S_{eff} =  \frac{L}{g_\Psi^2} \int \! d^4x\;  \bar\Psi^{0\, (1)}_L \pslash \left[ 
\Sigma^{(+)}(p) - \left( \Sigma^{(+)}(p) -\Sigma^{(-)}(p) \right) \sin^2\frac{h}{f} \right] \Psi^{0\, (1)}_L \, .
\end{equation}
It closely resembles the expression of the effective action of the composite Higgs, eq.(\ref{eq:fermeffaction}), and in fact it 
is has exactly the form that one would have obtained, following the procedure of section~\ref{sec:CHM}, in the case
of a composite Higgs theory where the strong dynamics has an $SO(2)$ global symmetry fully broken down
in the infrared, and an elementary sector consisting of one left-handed fermion.

It is at this point clear that  $h$ (hence $A_5$) can be fully considered as a composite scalar from the point of view of 
a low-energy 4D observer. Compositeness, indeed, is experienced (and can be thus defined) as a non-trivial dependence of
the couplings on the 4D momentum. In the case of the holographic scalar $h$, such momentum dependence
is a consequence of the 5-dimensional profile of  the corresponding bulk field.
Also, $h$ is truly a NG boson with decay constant $f$, as the periodic dependence of $S_{eff}$ upon $h/f$ testifies.
Its Coleman-Weinberg potential can be easily derived starting from the effective action (\ref{eq:Sefffinal}): the series
of 1-loop diagrams to resum is of the form showed in the upper row of Fig.~\ref{fig:HpotdiagsF}.  After rotating to
Euclidean momenta, $Q^2= -p^2$, one obtains:
\begin{equation}
V(h) = -2 \int \! \frac{d^4Q}{(2\pi)^4}\;  \log\left[  1 -  \frac{\Sigma^{(+)}(Q)-\Sigma^{(-)}(Q)}{\Sigma^{(+)}(Q)} \, \sin^2\frac{h}{f} \right]\, .
\end{equation}
Similarly to the case of 4D composite Higgs theories, the convergence of the integral is related to the behavior of the form
factor $(\Sigma^{(+)} - \Sigma^{(-)})$ at large virtual momenta. In the Euclidean one has
\begin{equation}
\frac{\Sigma^{(+)}(Q) -\Sigma^{(-)}(Q) }{\Sigma^{(+)}(Q) } = 1 - \frac{1}{Q^2} \left( \omega_E^2 \coth^2 (\omega_E L) - m^2 \right)\, , 
\end{equation}
where $\omega_E \equiv \sqrt{Q^2+m^2}$. At large momenta, $QL , Q/m \gg 1$, one has
\begin{equation}
\frac{\Sigma^{(+)}(Q) -\Sigma^{(-)}(Q) }{\Sigma^{(+)}(Q)}   = -4 \left( 1 + \frac{m^2}{Q^2} \right) e^{-2QL} + \cdots 
\end{equation}
This means that the 5D bulk dynamics leads to an \textit{exponential} convergence of the integral. According to the discussion 
at the end of section~\ref{sec:SO5example},
this is equivalent to saying, in the 4D language, that the operator responsible for the symmetry breaking in the infrared (\textit{i.e.} the
order parameter of the spontaneous breaking) has infinite dimension.
A good approximation of the potential thus comes by expanding the logarithm at leading order, so that:
\begin{equation}
\begin{split}
& V(h) \simeq -\frac{1}{4\pi^2} \frac{1}{L^4} f(m) \sin^2\frac{h}{f}\, ,  \\[0.3cm]
& f(m) \equiv \int_m^\infty \! d y_E\;   y_E^3 \left( \coth^2(y_E) -1 \right) \, . 
\end{split}
\end{equation}
For example $f(0) = 3 \zeta(3)/2 \simeq 1.8$.

All the above results and considerations clarify the meaning and the utility of the holographic description of the 5D theory.
We have already said that, by virtue of the holographic interpretation, the 5D theory gives a model for the strong dynamics 
of composite Higgs theories. Here we want to stress that such model of the strong dynamics is especially interesting
because it  admits a perturbative expansion, thus allowing one to compute a large class of infrared-saturated
quantities.  Calculability in the 5D theory requires that the 5D expansion parameter be small
\begin{equation*}
\frac{g_5^2 L^{-1}}{16\pi^2} \ll 1\, .
\end{equation*}
The corresponding perturbative parameter in the 4D picture is the number of `colors' of the strong sector, 
which can thus be defined as:
\begin{equation}
\frac{1}{N} \equiv \frac{g_5^2 L^{-1}}{16\pi^2}  \, .
\end{equation}
This is completely consistent with the NDA expectation
\begin{equation*}
\frac{N}{16\pi^2} \approx \frac{f^2}{m_\rho^2}\, , 
\end{equation*}
where now $f$ is  defined by eq.(\ref{eq:deff}), and $m_\rho\sim 1/L$ sets the scale 
of the lightest  resonances of the strong sector, whose spectrum is given by the poles of  $K(p)$ in eq.(\ref{eq:K}),
see for example~[\refcite{Contino:2004vy}].   As already mentioned,  
the KK modes are to be identified with the mass eigenstates obtained from the mixing of  elementary 
and  composite states ($m_{KK} \sim m_\rho \sim 1/L$). They are thus partial composites in the sense of section~\ref{sec:CHandEWPT}.

%%%%%

\section{Epilogue}

In these lectures I tried to give an overview of the basic mechanisms behind the idea of composite Higgs and 
of the central qualitative features a  model have to incorporate to be compatible with the present experimental data.
Mastering the general mechanisms should make the reader well equipped to
go through the vast literature on the subject and build
her/his own model.   Several realistic constructions have been proposed so far whose phenomenological implications will
soon be tested at the LHC.  These include $SO(5)$ `minimal' models in 5D 
warped~[\refcite{Agashe:2004rs,Contino:2006qr},\refcite{Carena:2006bn,Medina:2007hz,Carena:2009yt},\refcite{Panico:2008bx}]  
and flat [\refcite{Panico:2005dh}] spacetimes, 
as well as modern 4D composite Higgs models [\refcite{Evans:2010bp}]. 
Although I cannot discuss the detailed predictions of each of these models, I would like to conclude
by spending a few words on some general aspects of the phenomenology of a composite Higgs
at  present  and future colliders. 
(Ref.[\refcite{Giudice:2007fh}] is an excellent place to start to learn more on this subject, see also [\refcite{Espinosa:2010vn},\refcite{stronghh}].)

If  a light Higgs boson is discovered at the LHC or at Tevatron, the most important questions to address
will be: what is its role in the mechanism of electroweak symmetry breaking~? Is it an elementary or a composite scalar~?
Crucial evidence will come from a precise measurement of the parameters $a$, $b$, $c$ 
in the effective Lagrangian (\ref{eq:CHLag}):  any deviation from the unitary point $a=b=c=1$ will 
be the sign of a departure from the simple SM description and will give hints on the nature
of the symmetry breaking dynamics.
A first determination of $a$ and $c$ will come from the measurement 
of the couplings of the Higgs to the SM fermions and vectors. This requires disentangling possible
modifications of both the Higgs production cross sections and decay rates. Preliminary
studies have shown that the LHC should be eventually able to extract the individual
Higgs couplings with a $\sim 10-20\%$ precision~[\refcite{Duhrssen:2004cv}], though much will depend on the value of
its mass. This would imply a sensitivity on $(1-a^2)$ up to $0.1-0.2$~[\refcite{Giudice:2007fh}]. 
While the determination of the Higgs couplings will give a first hint on its nature,  a more direct probe of the
symmetry-breaking dynamics will  come only from a precise study of the scattering processes
that the exchange of the Higgs is assumed to unitarize.
A smoking gun of the compositeness of the light Higgs would  be finding an excess of events in 
$V_LV_L\to V_LV_L$  at the LHC compared to the SM expectation. Another important though difficult
process to monitor is $V_L V_L \to hh$~[\refcite{stronghh}].

Determining $a$, $b$, $c$  gives direct information on the symmetry breaking
structure associated to the light composite Higgs.  Indeed,
while  these three parameters are  independent for a generic composite scalar, we have seen that they are  related 
to each other in specific models where the Higgs is a pseudo Nambu-Goldstone boson.
For instance, the $SO(5)/SO(4)$ coset  implies $a = \sqrt{1-\xi}$, $b = 1-2\xi$, while the value of $c$ depends on how
 the composite operators coupled to the SM fermions transform under $SO(5)$. We have seen that in the case of spinorial representations of $SO(5)$
one has $c=\sqrt{1-\xi}$.
Different curves in the $(a,b)$ plane are thus associated to different symmetry-breaking cosets.
It has been also shown by the authors of  Ref.~[\refcite{Giudice:2007fh}] that if the light composite Higgs belongs to an 
$SU(2)_L$ doublet, regardless of whether it has a NG interpretation, the parameters $a$ and $b$ follow a universal trajectory
in the vicinity of the unitary point (\textit{i.e.} for small $\xi = v^2/f^2)$:  $a \simeq 1- \xi/2$, $b \simeq 1-2 \xi$.
Any deviation from this curve would be the signal of a different origin for the light scalar $h$.
It is  possible, for example, that a light dilaton arises from the spontaneous breaking of the 
scale invariance of the strong sector~[\refcite{dilaton}].
In that case conformal invariance requires $a^2 = b = c^2$, with the Lagrangian~(\ref{eq:CHLag}) exactly truncated
at quadratic order in $h$. For this choice one can define the dilaton decay constant by $v/a \equiv f_D$,
and the dilaton field as $\exp (\phi(x)/f_D) = 1+ h(x)/f_D$.  

Along with the study of the  phenomenology of a light composite Higgs, crucial information on the symmetry breaking
dynamics will come from the production of resonances of the strong sector, in particular the  fermionic resonances coupled to the top quark.
Extracting their masses and couplings by measuring their production cross section and decay fractions
will give the unique opportunity to understand the mechanism by which the Higgs is light and
identify the global symmetries of the strong dynamics.

%%%
%%%  Acknowledgments
%%%

\section*{Acknowledgments}

I would like to thank Csaba Csaki, Scott Dodelson and K.T. Mahanthappa for inviting me to TASI 2009, ``Physics of the Large and the Small'',
to give these lectures. I thank all the students for the great atmosphere and for their many stimulating questions.
I am also grateful to Andrea Wulzer for discussions and to David Marzocca, Marco Matassa and Natascia Vignaroli for reading the 
manuscript and pointing out several errors.

%%%
%%%  Appendix
%%%

\addcontentsline{toc}{section}{\protect\numberline{}Appendix}
\section*{Appendix}
%\appendix{SO(5) group theory details}
\label{app:SO5}

We collect here some useful group theory results and formulas.

The isomorphism between $SO(4)$ and $SU(2)_L\times SU(2)_R$ mentioned at the beginning of section~\ref{sec:SO5example} can be shown 
by associating to any 4-dimensional  vector $v^{\hat a}$ a  matrix $V \equiv \sigma^{\hat a} v^{\hat a}$ ($\sigma^{\hat a} \equiv (\vec \sigma , -i 1) $, 
with $\hat a = 1,2,3,4$).  The group $SO(4)$ acts on the vector  $v$ as a rotation, preserving its norm:
\begin{equation}
SO(4) : \qquad v^{\hat a} \to S^{\hat a \hat b} v^{\hat b}\, , \qquad |v| = \text{constant} \, .
\end{equation}
The action of $SU(2)_L\times SU(2)_R$   can then be defined on the matrix $V$ as the left multiplication by  $L\in SU(2)_L$
and right multiplication by $R\in SU(2)_R$, so that the determinant of $V$ is unchanged:
\begin{equation}
SU(2)_L\times SU(2)_R : \qquad V \to L\, V\, R^\dagger \, , \qquad \text{det}(V) = - |v|^2 = \text{constant}\, .
\end{equation}
Then, for each $SO(4)$ matrix $S$ there are  \textit{two}  $SU(2)_L\times SU(2)_R$ transformations that 
act in the same way on $V$,
\begin{equation}
S \to \{\,  (L,R) \; , \; (-L,-R)  \, \}\, ,
\end{equation}
and that differ by a sign.  At the level of group elements such correspondence  implies the following exact equivalence:
\begin{equation}
SO(4) = \frac{SU(2)_L \times SU(2)_R}{Z_2} \, .
\end{equation}

A suitable basis of $SO(5)$ generators for the fundamental representation is the following
\begin{equation}
\begin{split}
 T^{a_{L,R}}_{ij} =& -\frac{i}{2}
  \left[ \frac{1}{2}\, \epsilon^{abc} \left(\delta^b_i \delta^c_j - \delta^b_j \delta^c_i \right)
         \pm \left(\delta^a_i \delta^4_j-\delta^a_j \delta^4_i \right) \right]\\
 T^{\hat a}_{ij}  =& -\frac{i}{\sqrt{2}}
  \left( \delta^{\hat a}_i \delta^5_j - \delta^{\hat a}_j \delta^5_i \right)\, ,
 \end{split}
 \end{equation}
where $i,j=1,\dots,5$ and   $T^{a_{L,R}}$ ($a_{L,R}=1,2,3$) are  the generators of $SO(4)\sim SU(2)_L \times SU(2)_R$.
The spinorial representation of SO(5) can be defined in terms of the Gamma matrices
\begin{equation}
\Gamma^{\hat a} =
 \begin{bmatrix} 0 & \sigma^{\hat a} \\ \sigma^{\hat a\,\dagger} & 0 \end{bmatrix}\, , \qquad
\Gamma^5 = \begin{bmatrix} \mathbf{1} & 0 \\ 0 & -\mathbf{1} \end{bmatrix}\, ,
 \qquad\qquad \sigma^{\hat a}=\{\vec\sigma,-i \mathbf{1}\} \, ,
\end{equation}
as
\begin{equation}
T^{a_{L,R}} = -\frac{i}{2\sqrt{2}}
 \left[ \frac{1}{2}\, \eps^{abc} [\Gamma^b,\Gamma^c] \pm [\Gamma^a,\Gamma^4] \right]\, , \qquad
T^{\hat a}  = -\frac{i}{4\sqrt{2}} [\Gamma^{\hat a},\Gamma^5]\, ,
\end{equation}
so that
\begin{equation}
T^{a_L} = \frac{1}{2} \begin{bmatrix} \sigma^a & 0\\ 0 & 0 \end{bmatrix}\, , \qquad
T^{a_R} = \frac{1}{2} \begin{bmatrix} 0 & 0\\ 0 & \sigma^a \end{bmatrix}\, , \qquad
T^{\hat a} = \frac{i}{2\sqrt{2}}
 \begin{bmatrix} 0 & \sigma^{\hat a} \\ -\sigma^{\hat a\, \dagger} & 0 \end{bmatrix}\, .
\end{equation}
%

%%%%%
%% References

%%%
%%%  References
%%%

%%%%%

\end{document}